\documentclass[aps,prd,preprintnumbers,groupedaddress,nofootinbib,amssymb,eqsecnum,notitlepage]{revtex4-2}
\usepackage{graphicx}
\usepackage{dcolumn}
\usepackage{bm}
\usepackage{graphicx}
\usepackage{dcolumn}
\usepackage{amsmath}
\usepackage{amssymb}
\usepackage{comment}
\usepackage{epsf}
\usepackage{epsf}
\usepackage{psfrag}
\usepackage{epsfig}
\usepackage{epstopdf}
\usepackage{threeparttable}
\usepackage{natbib}
\usepackage{bm}
\usepackage[usenames,dvipsnames]{color}
\usepackage[colorlinks=true, linkcolor=BrickRed, citecolor=Blue, urlcolor=Blue, filecolor=Blue]{hyperref}
\usepackage{booktabs}
\usepackage{hhline}
\usepackage{empheq}
\usepackage{enumerate}
\usepackage{tikz}
\usetikzlibrary{arrows.meta,automata,positioning,shadows}

\begin{document}
\title{Constraints on Coupled Dark Energy in the DESI Era}
\author{Adrià Gómez-Valent$^{a}$, Ziyang Zheng$^{b}$, Luca Amendola$^{b,c}$}\email{agomezvalent@icc.ub.edu}\email{zheng@thphys.uni-heidelberg.de}\email{l.amendola@thphys.uni-heidelberg.de}
\affiliation{}
\affiliation{$^1$Departament de Física Quàntica i Astrofísica (FQA) and Institut de Ciències del Cosmos (ICCUB), Universitat de Barcelona (UB), c. Martí i Franqués, 1, 08028 Barcelona, Catalonia, Spain} 

\affiliation{$^b$Institut f\"{u}r Theoretische Physik, Ruprecht-Karls-Universit\"{a}t Heidelberg, Philosophenweg 16, D-69120 Heidelberg, Germany} 
\affiliation{$^c$New York University Abu Dhabi, PO Box 129188, Abu Dhabi,
UAE, and Center for Astrophysics and Space Science (CASS), New York University Abu Dhabi, UAE}

\begin{abstract}
We investigate the current viability of a well-known coupled dark energy scenario in which cold dark matter (DM) interacts with a spin-0 dark energy component through a non-trivial field dependence of the DM mass. This ultra-light scalar mediates a fifth force between DM particles, which can leave signatures on cosmological scales. We use state-of-the-art data on the cosmic microwave background from Planck's CamSpec likelihood, baryon acoustic oscillations from the second DESI data release as well as the supernovae of Type Ia (SNIa) from Pantheon+ and DES-Dovekie. We perform the analysis considering both a  flat potential and a Peebles-Ratra (PR) potential for the scalar field in order to assess the impact of the potential slope on the fitting performance of the model. While for a constant potential the scalar field dynamics is insensitive to the sign of the coupling parameter $\beta$, the PR potential breaks the existing symmetry in the solutions at late times and could induce a difference at the phenomenological level between positive and negative values. We study for the first time if it is actually the case, finding no important asymmetry in the fitting results, essentially due to shifts in other parameters, which can lead to significant compensating effects. In the light of the aforesaid datasets, we find in all cases a peak at $|\beta|\sim 0.03$ -- less pronounced than reported in some recent works --, excluding the no-coupling scenario at $\sim 95\%$ CL at most. The model is able to explain an effective crossing of the phantom divide, with the equation-of-state parameter lying within the $2\sigma$ bands of model-agnostic reconstructions, albeit with important differences in the shape of the central curves. Our results are very robust under changes in the SNIa sample used in the analysis and is not significantly altered when we replace a constant potential with the PR one, although the latter is crucial to produce the aforesaid crossing. In passing, we also provide constraints obtained with the PR potential in the uncoupled case.
\end{abstract}
\maketitle

\section{Introduction}

Recent data on baryon acoustic oscillations (BAO) from the Dark Energy Spectroscopic Instrument (DESI) \cite{DESI:2025zgx}, combined with a host of additional datasets including cosmic microwave background (CMB) and supernovae of Type Ia (SNIa), have shown significant evidence for a cosmology that differs from what has been so far denoted as standard cosmological model, i.e. $\Lambda$CDM. This result has sparked an intense debate both for as concerns its robustness \cite{Patel:2024odo,Cortes:2025joz,Gonzalez-Fuentes:2025lei,Keeley:2025rlg,Ong:2025utx,Hergt:2026moc,Ong:2026tta,Gonzalez-Fuentes:2026rgu, Afroz:2025iwo}, and for what it entails in terms of new physics, see \cite{Chakraborty:2024xas,Gomez-Valent:2024tdb,Benisty:2024lmj,Poulin:2024ken,Ye:2024ywg,Wolf:2024eph,Wolf:2024stt,Park:2024vrw,Gomez-Valent:2024ejh,Odintsov:2024woi,Giare:2025pzu,Khoury:2025txd,Pan:2025psn,Wolf:2025jed,Chaussidon:2025npr,Chakraborty:2025syu,Yang:2025mws,Chen:2025wwn,Giani:2025hhs,Cai:2025mas,Braglia:2025gdo,Poulin:2025nfb,Camarena:2025upt,Gomez-Valent:2025mfl,Wang:2025znm,Yang:2025uyv,Yao:2025wlx,Nojiri:2025low,Artola:2025zzb,Mishra:2025goj,Goh:2025upc,Tsujikawa:2025wca,Wolf:2025acj,Adi:2025hyj,Sharma:2025iux,Efstratiou:2025iqi,Cheng:2025yue,Ghedini:2025epp,deCruzPerez:2025dni,Li:2026xaz,LaPenna:2026avs,Ibarra-Uriondo:2026zbp,Akarsu:2026anp,Park:2026iqa,Jhaveri:2026bla,Wang:2026wrk} for a non-exhaustive list of proposals. It supports a dynamical form of dark energy, which might require an additional weakly-clustered field to be present, and actually dominating, in the cosmic recipe. Dark energy dynamics could also be induced by the quantum-field-theory running of the cosmic vacuum; see \cite{SolaPeracaula:2022hpd} and references therein.

When reported in terms of a varying dark energy equation of state (EoS) in the so-called Chevallier-Polarski-Linder (CPL) \cite{Chevallier:2000qy,Linder:2002et}  form $w(a)=w_0+w_a(1-a)$, the current data excludes the $\Lambda$CDM model (i.e. $w_0=-1,w_a=0$) at $\sim 3\sigma$ CL with a preference for a peak in the dark energy density at $z\sim 0.3-0.6$, which has also been corroborated in more model-agnostic analyses \cite{DESI:2024aqx,Jiang:2024xnu,Berti:2025phi,DESI:2025fii,Li:2025ops,Gonzalez-Fuentes:2025lei,Gonzalez-Fuentes:2026rgu}. The fact that the degeneracy curve on the $w_0,w_a$ plane points almost exactly towards  the $\Lambda$CDM is of course puzzling, because it implies that any issue with the data might move the best fit back towards the standard cosmology. Nevertheless,  the deviation from $\Lambda$CDM is  quite impressive and should be taken into serious consideration.

Another puzzling aspect stems from the fact that the best fit $(w_0,w_a)$ values imply the existence of a phantom phase $(w(z)<-1)$ in the past. This is perplexing from a theoretical perspective since a phantom phase might require a fundamental instability of the physical processes because the Hamiltonian becomes unbounded from below. However, this must not necessarily be the case. It has long been known, in fact, that if dark energy interacts with matter or if gravity is not-Einsteinian, one can have a positive definite Hamiltonian and still generate an epoch in which dark energy appears to go phantom, just due to neglecting the additional interaction \cite{Sola:2005et,Gannouji:2006jm,Das:2005yj}.

Several papers in the last couple of years attempted therefore to explain the DESI data with an interacting dark energy-dark matter model, see, e.g., \cite{Chakraborty:2024xas,Chakraborty:2025syu,deCruzPerez:2025dni,Li:2026xaz,LaPenna:2026avs}. Modified-gravity models leading to interactions of this sort were introduced in the cosmological literature even before the discovery of acceleration \cite{Wetterich:1987fm} and studied as forms of dynamical dark energy ever after (see e.g. \cite{Amendola:1999er,Uzan:1999ch,Chiba:1999wt,Amendola:1999qq,Perrotta:1999am,Holden:1999hm,Bartolo:1999sq}). A popular class of coupled dark energy (CDE) models introduces a fifth force between matter particles, mediated by a scalar field. In these models, it is common to assume that baryons are uncoupled, so as to avoid the strong constraints from local gravity. In its simplest incarnation, interacting models just require a scalar field $\phi$ representing dark energy coupled to dark matter (DM) through a single extra parameter, the coupling $\beta$. Adding a potential for the field requires of course more parameters, for instance the potential slope. For the sake of simplicity, one typically assumes both parameters, coupling and slope, to be constant. 

Some works in the past focused on constraining this model, already with CMB data from the WMAP satellite and the South Pole Telescope \cite{Pettorino:2012ts}, or considering past releases of Planck CMB data in combination with low-redshift datasets, as e.g. from BAO and SNIa \cite{Pettorino:2013oxa,Planck:2015bue,Gomez-Valent:2020mqn,Gomez-Valent:2022hkb,Gomez-Valent:2022bku,Goh:2022gxo}. See also \cite{Barros:2018efl,Archidiacono:2022iuu,Bottaro:2023wkd,Bottaro:2024pcb}. Interestingly, these pre-DESI works found a peak in the likelihood at a non-vanishing value of the coupling constant, at around 2$\sigma$ CL\footnote{We refer the reader to Refs. \cite{Amendola:2017xhl,Savastano:2019zpr,Goh:2023mau} for possible implications of a large coupling leading to long-range attractive forces stronger than gravity in the radiation-dominated epoch.}. 

More recently, in \cite{Costa:2025kwt} the authors also found a $\sim 2\sigma$ preference for a non-null coupling with CMB and DESI DR2 data, but excluding SNIa and considering a flat potential. They either fixed the sum of the neutrino masses to zero or allowed it to vary freely in the Monte Carlo analysis, obtaining in the latter case slightly larger upper bounds due to the positive correlation between $\beta$ and $M_\nu$\footnote{Their coupling is defined as two times our $\beta^2$.}.

Instead, using the combination Planck 2018+DESI DR2+PantheonPlus, the authors of
Ref.~\cite{Chakraborty:2024xas} found a preference for a negative coupling at more than
$3\sigma$ CL\footnote{Notice that in their convention, our $\beta$ corresponds to their coupling divided by $\sqrt{8\pi}$, with a relative minus sign.} (see also \cite{Wang:2025znm} for similar results with a positive coupling, but using the supernovae from DES and with a different treatment of the initial conditions). However, the Akaike information criterion did not provide significant
evidence for a departure from $\Lambda$CDM and even mildly favored the latter, which appears in
tension with the inferred posterior constraints on $\beta$. 

In \cite{Li:2026xaz}, the authors reported evidence for a non-zero coupling at more than $5\sigma$
CL using CMB data from Planck PR4, ACT, and SPT-3G, in combination with DESI DR2 BAO
measurements and several SNIa compilations. Despite obtaining minimum $\chi^2$ values similar to
those found for the CPL parametrization, and considering the same number of free parameters, their
reported preference for the CDE model corresponds to a much stronger statistical signal when
expressed in terms of the coupling parameter. This is substantially larger than the $\sim 3\sigma$
evidence for dark-energy dynamics typically obtained within the CPL framework. Moreover, the coupling signal remains above $5\sigma$ independently of the specific SNIa sample
adopted, which is somewhat unexpected. Finally, one may raise concerns regarding the use of high
CMB multipoles to constrain CDE models, since non-linear effects at those scales could be
non-negligible and may require a more careful treatment of the model-specific phenomenology in
order to avoid potential biases. Recent developments in non-linear perturbation theory have begun to address these non-linearities in CDE models and other modified gravity scenarios, providing a framework for more robust constraints from current and forthcoming large-scale surveys \cite{Tudes:2024jpg, Zheng:2025owb, Silva:2025bnn}.

In view of the situation described above, we find it timely to revisit our previous constraints on
CDE, taking into account the most up-to-date datasets, but without including data on large-scale structure nor 
high-$\ell$ CMB data from ACT and SPT. This is a more conservative approach, which also facilitates the comparison of our results with those available in the literature. Relative to earlier analyses, the present work implements
several important changes. First, we explore both a constant and a power-law potential in order to assess the role of the
scalar-field self-interaction in the ability of the model to accommodate current observations.
Second, we consider both negative and positive couplings. While for a constant potential the
equations of motion are symmetric under a sign change of the coupling, this symmetry is broken once
a non-trivial potential is introduced. In principle, this could lead to late-time differences when
the potential starts to dominate the expansion history. We explicitly show, however, that such
asymmetries are largely compensated by shifts in other cosmological parameters, yielding very
similar results for either sign of the coupling. 
Third, we employ the most recent observational datasets. In particular, we analyze the combination
of Planck PR4, DESI DR2, and several SNIa compilations, including Pantheon+ and the newly
recalibrated SNIa from DES, contained in the so-called DES-Dovekie sample. Overall, our results indicate a statistical significance not exceeding the $2\sigma$ CL for a non-zero coupling. In contrast to previous claims in the literature
\cite{Chakraborty:2024xas,Li:2026xaz}, this finding is fully consistent with the conclusions drawn
from information criteria and with the relative performance of the CPL parametrization. Our results are more aligned with \cite{Costa:2025kwt}, but in our analysis we take into account the data on SNIa and study also the effect of a non-flat potential.

This work is organized as follows. In Sec. \ref{sec:Model} we explain the most relevant features of the CDE model studied in this work, including its background dynamics -- initial conditions, scaling solutions, role of the scalar field potential in the late universe -- and the effect of the fifth force at the linear perturbations level. We also review how to construct the EoS parameter of the effective dark energy component, which encapsulates the effect of both the scalar field dynamics and the non-null variation of the DM mass. In Sec. \ref{sec:method}, we describe the methodology  followed to perform the fitting analyses as well as some technical details about the implementation of the model in the Einstein-Boltzmann solver. We devote Sec. \ref{sec:results} to present our results, and Sec. \ref{sec:conclusions} for the conclusions. Some appendices complement the main content of the paper.


\section{Models of coupled dark energy}\label{sec:Model}
\subsection{Background and perturbation equations}
We consider a flat, perturbed Friedmann-Lema\^{i}tre-Robertson-Walker (FLRW) metric and work in the conformal Newtonian gauge, with the two scalar degrees of freedom encoded in the two gravitational potentials $\Psi$ and $\Phi$,
\begin{equation}
\mathrm{d}s^2 = -a^2(1 + 2\Psi)\mathrm{d}\tau^2 + a^2(1 - 2\Phi)\delta_{ij} \mathrm{d}x^i \mathrm{d}x^j  \;,
\end{equation}
where $a$ is the scale factor and $\tau$ is the conformal time. In our CDE scenario, in addition to the matter species described in the standard model of particle physics (with some extension to explain the neutrino masses), we consider a  dark sector in the form of a scalar field, $\phi$, representing dark energy, and cold DM, which is coupled to $\phi$ through a field-dependent mass. We take the mass to be of the form $m_{\rm dm}(\phi)\propto e^{-\kappa\beta\phi}$, with $\beta$ the  constant and dimensionless coupling strength, and $\kappa=\sqrt{8\pi G}$ the inverse of the reduced Planck mass. This coupling modifies the covariant conservation equations in the dark sector. They read, 
\begin{equation} \label{eq:consereqs}
    \nabla^{\mu}T^{(\phi)}_{\mu\nu}=\beta\kappa T^{(\rm dm)} \nabla_{\nu}\phi; \quad 
    \nabla^{\mu}T^{(\rm dm)}_{\mu\nu}=-\beta\kappa T^{(\rm dm)} \nabla_{\nu}\phi  \,,
\end{equation}
with $T^{(\rm dm)}$ the trace of the dark matter energy-momentum tensor\footnote{These conservation equations can be derived from fundamental actions, e.g. from the action of a Dirac field with mass term $m(\phi)\bar{\psi}\psi$, or in a more general setup (valid for both fermions and bosons) by considering the single-particle action and promoting the mass to a functional of the scalar field, see, e.g., Appendix A2 of Ref. \cite{Costa:2025kwt}. The variation of particle masses could also be triggered by a cosmic evolution of the Higgs field \cite{Sola:2016our} or the running of the vacuum \cite{Fritzsch:2012qc}, without the need of introducing additional degrees of freedom.}. Assuming the DM component behaves as a pressureless perfect fluid, one can derive its modified conservation equation, which at the background level takes the following form, 
\begin{equation}
    \rho_{\rm dm}'+3\mathcal{H}\rho_{\rm dm}=-\beta\kappa\rho_{\rm dm}\phi'     \,,
\end{equation}
where $\mathcal{H}=aH$ is the conformal Hubble function, and $\rho_{\rm dm}$ corresponds to the dark matter energy density. Hereafter, we use a prime to denote a derivative with respect to the conformal time. The solution to this equation is

\begin{equation}\label{eq:DMdensity}
\rho_{\rm dm}(a)=m_{\rm dm}(a)n_{\rm dm}(a)=\rho_{\rm dm}^{0}a^{-3}e^{-\beta\kappa(\phi(a)-\phi^0)}\,,
\end{equation}
where $\rho_{\rm dm}^{0}$ and $\phi^0$ are the current DM energy density and scalar field, respectively. The deviation from the standard dependence on the scale factor is due to the variation of the DM mass, since DM particles are conserved and, hence, its number density scales as $n_{\rm dm }(a)\sim a^{-3}$. Furthermore, the energy density and pressure associated with the scalar field are given by:
\begin{equation}\label{eq:densitypressure}
    \rho_{\phi}=\frac{\phi'^2}{2a^2}+V(\phi); \qquad  p_{\phi}=\frac{\phi'^2}{2a^2}-V(\phi)  \,.
\end{equation}
$V(\phi)$ is the scalar field potential. The evolution of $\phi$ can then be derived from Eq. \eqref{eq:consereqs}, leading to the modified Klein-Gordon equation,
\begin{equation}\label{eq:KG}
    \phi'' + 2\mathcal{H}\phi'+a^2\frac{\partial V}{\partial \phi}=\kappa\beta a^2\rho_{\rm dm} \, .
\end{equation}
Alternatively, one can absorb the coupling term into an effective potential (see, e.g., \cite{Fardon:2003eh,Chakraborty:2025syu}), 
\begin{equation}\label{eq:effV}
V_{\rm eff}(\phi,a) =  V(\phi) + m_{\rm dm}(\phi)n_{\rm dm}(a) = V(\phi) + \rho_{\rm dm}(\phi,a)\,.
\end{equation}
The form of Friedmann equations remains unmodified with respect to the $\Lambda$CDM scenario, but here of course we have a dynamical DE component instead of a cosmological constant as well as mass-varying DM particles. They read,

\begin{equation}\label{eq:Friedmann}
    3\mathcal{H}^2 = \kappa^2a^2\left(\rho_{\rm dm}+\rho_{\rm b}+\rho_{\rm r}+\rho_\phi\right)  \qquad ;\qquad 
    2\mathcal{H}'+\mathcal{H}^2 = -\kappa^2a^2\left(\frac{\rho_r}{3}+p_\phi\right) \, . 
\end{equation}
The subscripts `${\rm r}$' and `${\rm b}$' denote radiation and baryons, respectively. Radiation is uncoupled from $\phi$ because the trace of its energy–momentum tensor vanishes. Baryons are assumed to be uncoupled in order to automatically satisfy the stringent local constraints on fifth forces (see, e.g., Refs.~\cite{Brax:2018zfb,Elder:2019yyp,March:2021mqu,Vagnozzi:2021quy,Tsai:2021irw,Brax:2022olf,Tsai:2023zza,Feleppa:2025clx,Feleppa:2025vop,Yuan:2025twx} for existing solar-system, laboratory or Earth-orbit bounds). Although we do not specify them in the energy and pressure budgets in Eq. \eqref{eq:Friedmann}, in our analysis we treat neutrinos in full detail, considering one massive neutrino of 0.06 eV and two massless neutrinos. This is an implementation of the normal-ordering neutrino mass hierarchy, in its minimal configuration \cite{Esteban:2024eli}.

Considering that recent observational results from collaborations such as DESI have reported hints of dark energy dynamics within the CPL parametrization of the dark energy EoS parameter \cite{DESI:2025zgx}, and that similar indications have also emerged from model-independent reconstructions of the dark energy background evolution \cite{DESI:2024aqx,DESI:2025fii,Gonzalez-Fuentes:2025lei,Gonzalez-Fuentes:2026rgu} under the assumption that dark energy behaves as a self-conserved perfect fluid, it is useful to rewrite the background equations in terms of a self-conserved DM component, with $\tilde{\rho}_{\rm dm}(a)=\tilde{\rho}_{\rm dm}^0 a^{-3}$, and a self-conserved DE fluid with energy density and pressure $\rho_{\rm de}(a)$ and $p_{\rm de}(a)$, respectively. It is the EoS of the latter, $w_{\rm de}(a)$, the one that should be compared to the CPL parametrization, not $w_\phi(a)=p_\phi(a)/\rho_\phi(a)$, cf. Eq. \eqref{eq:densitypressure}. In the late universe, where radiation can be neglected, the total energy density and pressure can be written as follows,
\begin{equation}
\rho_t(a)=\rho_{\rm dm}(a)+\rho_b(a)+\rho_\phi(a)\equiv \tilde{\rho}_{\rm dm}(a)+\rho_b(a)+\rho_{\rm de}(a) \qquad;\qquad p_t(a)=p_\phi(a)\equiv p_{\rm de}(a)\,.
\end{equation}
The effective (self-conserved) DE fluid is therefore characterized by the following effective EoS parameter \cite{Sola:2005et,Gannouji:2006jm,Das:2005yj},
\begin{equation}\label{eq:effEoS}
w_{\rm de}(a) =\frac{p_{\rm de}(a)}{\rho_{\rm de}(a)}= \frac{p_{\phi}(a)}{\rho_{\phi}(a)+\rho_{\rm dm}(a)-\tilde{\rho}_{\rm dm}(a)} = \frac{w_\phi(a)}{1+\frac{\rho_{\rm dm}(a)-\tilde{\rho}_{\rm dm}(a)}{\rho_\phi(a)}}\,,
\end{equation}
There is, however, an ambiguity in this definition of the effective EoS, as the equations of motion can be satisfied for arbitrary values of the constant $\tilde{\rho}_{\rm dm}^0$. The authors of \cite{Das:2005yj} set $\tilde{\rho}_{\rm dm}^0=\rho_{\rm dm}^0$, but this is only one of the infinitely many options, each leading to a different shape of $w_{\rm de}(a)$. Other possibilities include the use of DM density obtained in model-agnostic reconstructions \cite{deCruzPerez:2025dni} or the one from the best-fit CPL model \cite{Wang:2025znm}. However, these choices can lead to inconsistencies in the limit when the coupling tends to zero, since one can artificially find phantom behavior even in pure quintessence models. Here we opt to use the best-fit values of $\tilde{\rho}_{\rm dm}^0$ obtained from the fitting analyses of the various models introduced later, when DM is self-conserved, i.e. when the coupling is switched off. Using Eq. \eqref{eq:effEoS} together with this prescription, we can link our results to those reported in these works, allowing for a direct comparison between the effective dark energy EoS parameter inferred in our model and the one that is actually measured. In any case, our fitting results are independent of this choice; this step is included solely to facilitate their interpretation. Notice that, despite $w_\phi>-1$, since we are considering a canonical -- quintessence -- kinetic term, $w_{\rm de}$ can be below -1. Indeed, as well will see in Sec. \ref{sec:results} (and in Appendix \ref{app:EffEoS}), the effective dark energy component in the CDE model under study can cross the phantom divide, see also \cite{Chakraborty:2025syu}.

At the linear perturbation level, the perturbed Einstein equations remain identical to the GR case, 
\begin{equation}
    2k^2\Phi+6\mathcal{H}\left(\Phi'+\mathcal{H}\Psi\right)=-\kappa^2a^2\left[\delta\rho+\frac{\phi'}{a^2}\delta\phi'+\frac{dV}{d\phi}\delta\phi\right]    \,, 
\end{equation}    
\begin{equation}
    \Phi''+\mathcal{H}\left(\Psi'+2\Phi'\right)+ \left(2\mathcal{H}'+\mathcal{H}^2\right)\Psi +\frac{1}{3}k^2\left(\Phi-\Psi\right) = \frac{1}{2}\kappa^2a^2\left[\delta p+\frac{\phi'}{a^2}\delta\phi'-\frac{dV}{d\phi}\delta\phi\right]  \,, 
\end{equation}    
\begin{equation}
    2k^2\left(\Phi'+\mathcal{H}\Psi\right)=\kappa^2\left[a^2\left(\rho+p\right)\theta+k^2\phi'\delta\phi\right]  \,,
\end{equation}
where $\delta\rho$ and $\delta p$ are the perturbations in the energy density and pressure of all the species except dark energy, and $\theta$ is the velocity divergence  -- notice that in our notation, $(\rho+p)\theta =\sum_i(\rho_i+p_i)\theta_i$. These equations are written in momentum space. The coupling only enters explicitly the perturbed conservation equations, which read,
\begin{equation}
    \delta_{\rm dm}'+\theta_{\rm dm}-3\Phi'=-\kappa\beta\delta\phi' \,,
\end{equation}
\begin{equation}
    \theta_{\rm dm}'+\left(\mathcal{H}-\kappa\beta\phi'\right)\theta_{\rm dm}-k^2\Psi +k^2\kappa\beta\delta\phi=0\,,
\end{equation}
\begin{equation}
    \delta\phi''+2\mathcal{H}\delta\phi'+k^2\delta\phi+a^2m_{\phi}^2\delta\phi-\left(\Psi'+3\Phi'\right)\phi'+2a^2\Psi\frac{dV}{d\phi}=\kappa a^2\rho_{\rm dm}\delta_{\rm dm}\beta\,.
\end{equation}
While baryons do not directly feel the fifth force, since we assume they are uncharged under it, dark matter is subject to an additional pull besides gravity. The equations for their density contrast $\delta_i\equiv\delta\rho_i/\rho_i$ at deep subhorizon scales take the following form, 

\begin{figure*} \label{fig:phi_V_evolution}
\centering
\includegraphics[width=0.45\textwidth]{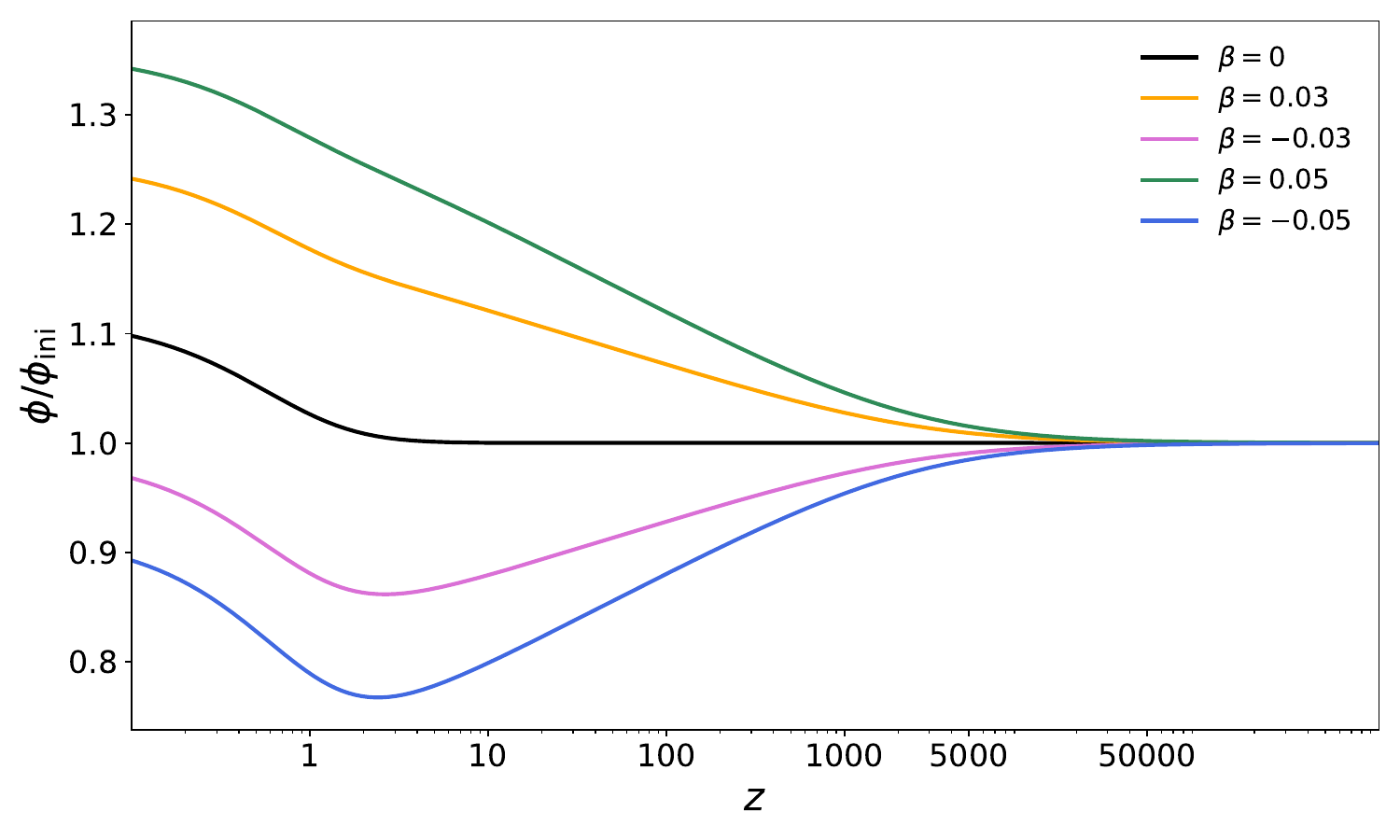}
\hfil
\includegraphics[width=0.45\textwidth]{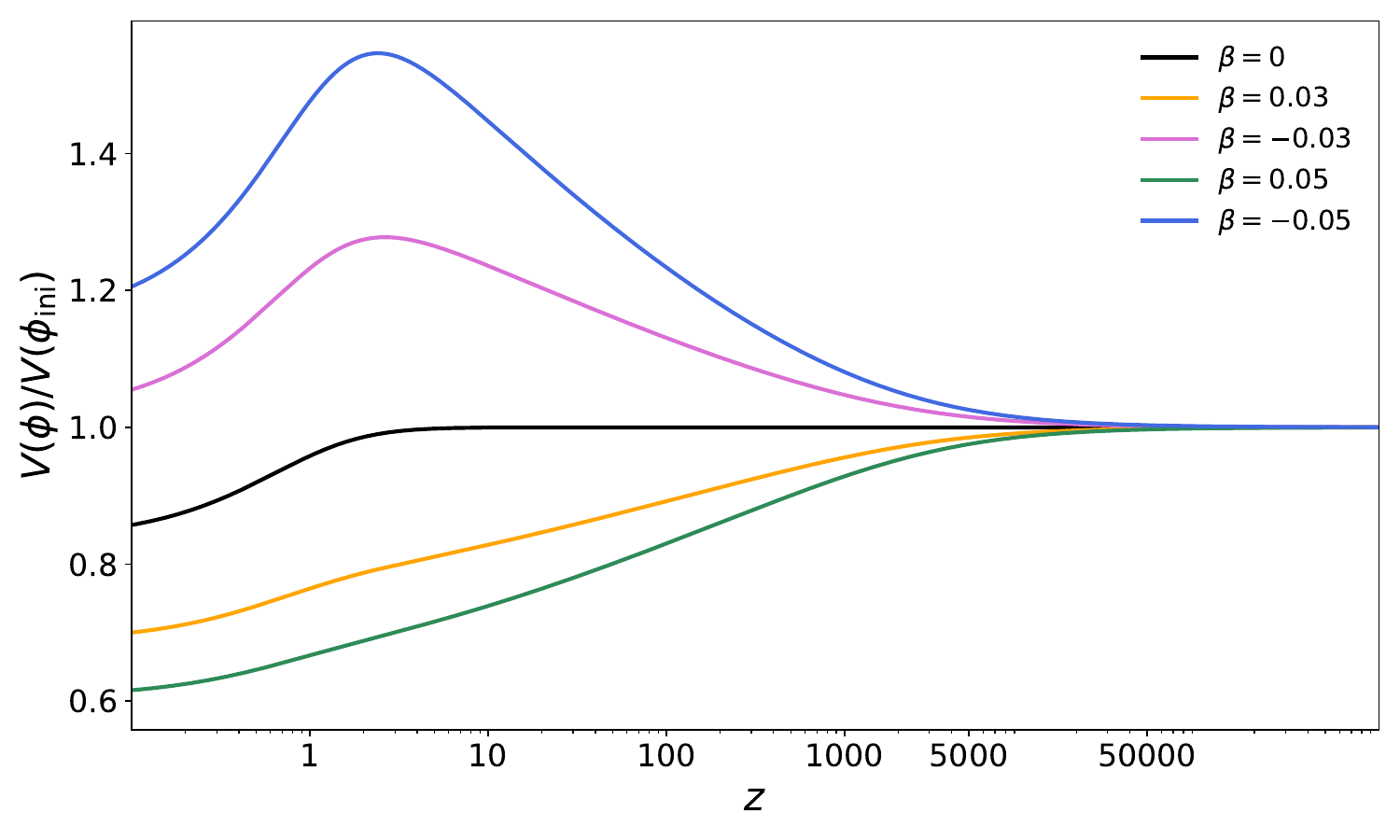}
\medskip
\includegraphics[width=0.45\textwidth]{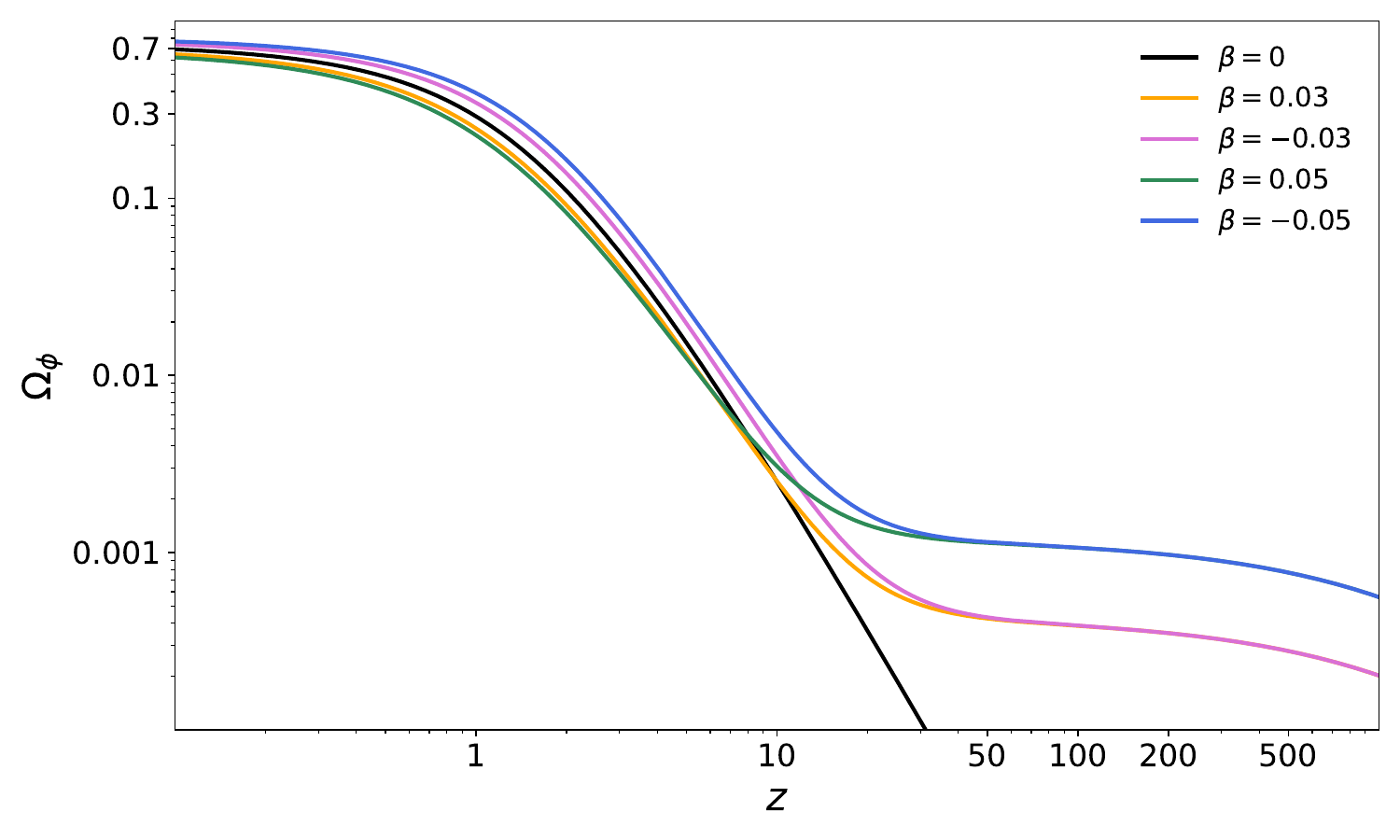}
\hfil
\includegraphics[width=0.45\textwidth]{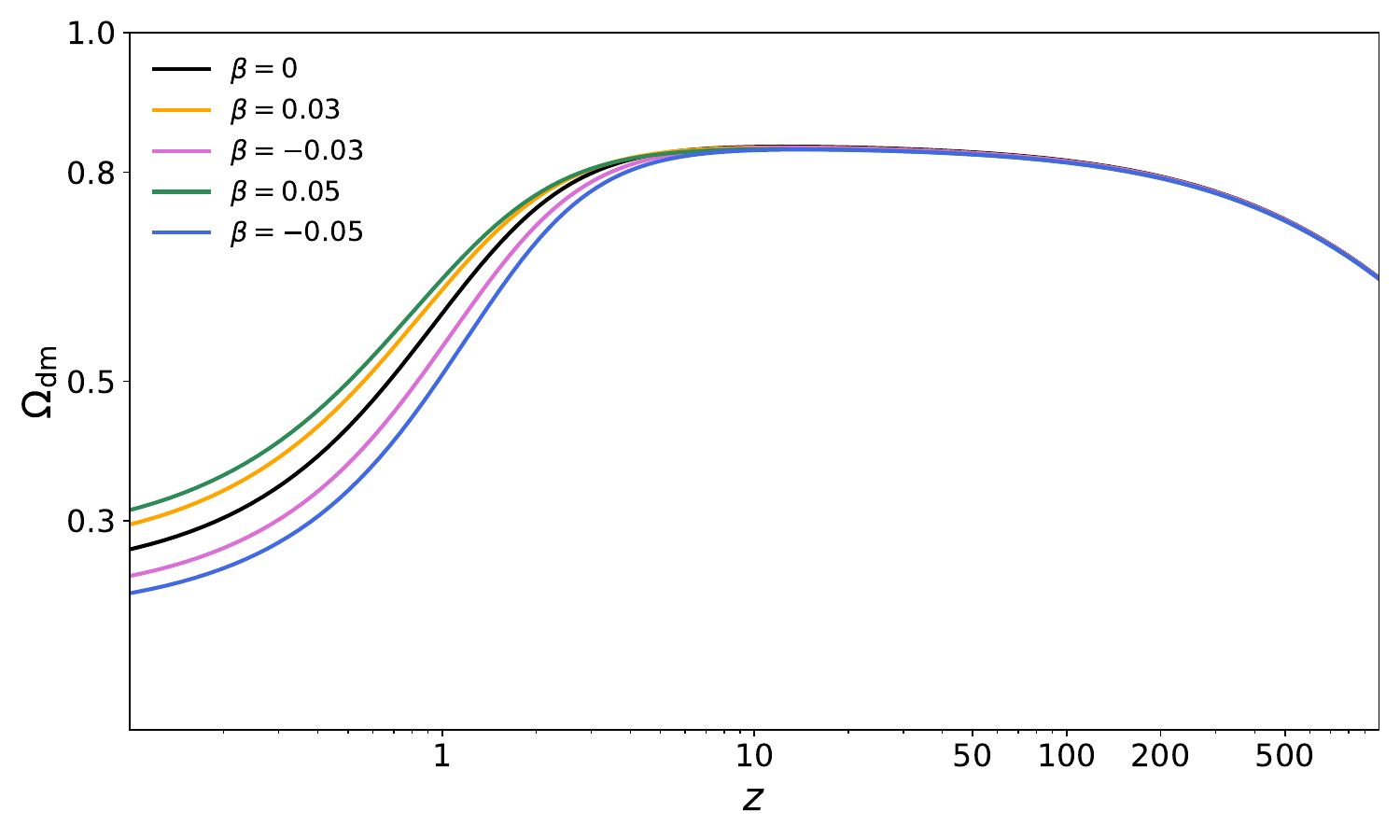}
\medskip
\caption{Evolution of the scalar field (upper-left plot) and its potential energy density (upper-right plot) normalized to their initial values, DE fraction (bottom-left plot) and DM energy fraction (bottom-right plot) obtained with CDE, considering the PR potential (Eq. \eqref{eq:PRpotential}). We have set all parameters except $\beta$ to the best-fit values from the fitting analysis with PlanckPR4+DES-Dovekie+DESI ($\beta>0$ case), cf. Table \ref{tab:BF_DES}. In the bottom-left plot one can clearly appreciate the plateau for $\Omega_\phi=2\beta^2/3$ at high $z$, in the matter-dominated epoch, which corresponds to the so-called $\phi$MDE fixed point of Ref. \cite{Amendola:1999er}.} 
    \label{fig:plots_PR} 
\end{figure*}

\begin{widetext}
\begin{equation}\label{eq:densityContrastB}
\delta_{\rm b}^{\prime\prime}+\mathcal{H}\delta^\prime_{\rm b}-\frac{\kappa^2a^2}{2}[\rho_b\delta_b+\rho_{\rm dm}\delta_{\rm dm}]=0\,,
\end{equation}
\begin{equation}\label{eq:densityContrastDM}
\delta_{\rm dm}^{\prime\prime}+\left[\mathcal{H}-\frac{\beta\kappa\phi^\prime k^2}{k^2+a^2m_\phi^2}\right]\delta^\prime_{\rm dm}+(\delta_b^\prime-\delta^\prime_{\rm dm})\frac{\beta\kappa\phi^\prime a^2m_\phi^2}{k^2+a^2m_\phi^2}-\frac{\kappa^2a^2}{2}\left[\rho_b\delta_b+\rho_{\rm dm}\delta_{\rm dm}\left(1+\frac{2\beta^2k^2}{k^2+a^2m^2_{\phi}}\right)\right]=0\,.
\end{equation}
\end{widetext}
In some previous works (see, e.g., \cite{Castello:2024lhl, Zheng:2025oiq}), phenomenological free functions are introduced to parametrize potential violations of the weak equivalence principle, encoded in the cosmological Euler equation and in the growth of the matter density contrast. More specifically, one can introduce a free function $\Theta$ denoting an additional friction term, while another function $\Gamma$ arises from the Poisson term and is related to the fifth force acting on dark matter. For CDE models, one can map from Eq. \eqref{eq:densityContrastDM} and find the corresponding $\Theta$ and $\Gamma$ in this particular case,
\begin{equation}
\Theta = -\frac{\beta\kappa\phi^\prime k^2}{k^2+a^2m_\phi^2} \qquad ; \qquad 
\Gamma = \frac{2\beta^2k^2}{k^2+a^2m^2_{\phi}}   \,.
\end{equation}
Moreover, as expected, the Yukawa force between dark matter particles is strongly suppressed at distances larger than the mediator's Compton length, i.e., when $a m_\phi\gg k $. In the case under study, the latter is of order the horizon scale, since it corresponds to the inverse of the dark-energy mass, which is extremely small so that DE can trigger the late-time accelerated expansion. Therefore in this paper, as in most previous research, the terms proportional to $m_\phi$ can be neglected in Eq. \eqref{eq:densityContrastDM}. However, in our implementation of the model in the Einstein-Boltzmann solver we take all the scale-dependent corrections fully into account (cf. Sec. \ref{sec:method}).

\subsection{Background dynamics for a constant and Peebles-Ratra potentials} \label{sec:bg_dynamics}

Let us now discuss the evolution of the scalar field and the background energy densities in this CDE model, in which the DM mass evolves with the scalar field as $m_{\rm dm}(\phi)\propto e^{-\kappa\beta\phi}$. The behavior of the model also depends on the shape of the potential $V(\phi)$, so the model is not fully specified until a concrete form for this function is chosen. However, regardless of this choice, if the potential energy is only responsible for driving the late-time acceleration of the universe and is orders of magnitude smaller than the typical energy density of radiation and pressureless matter during the epochs when these components dominate the cosmic expansion -- as we will consider to be in the case under study --, its contribution can be neglected at redshifts much greater than $1$. The potential does not affect the dynamics of the scalar field in these epochs and, hence, the latter will be determined solely by the initial conditions and the value of the coupling strength. 

\begin{figure*}[t!]
\centering
\includegraphics[width=0.48\textwidth]{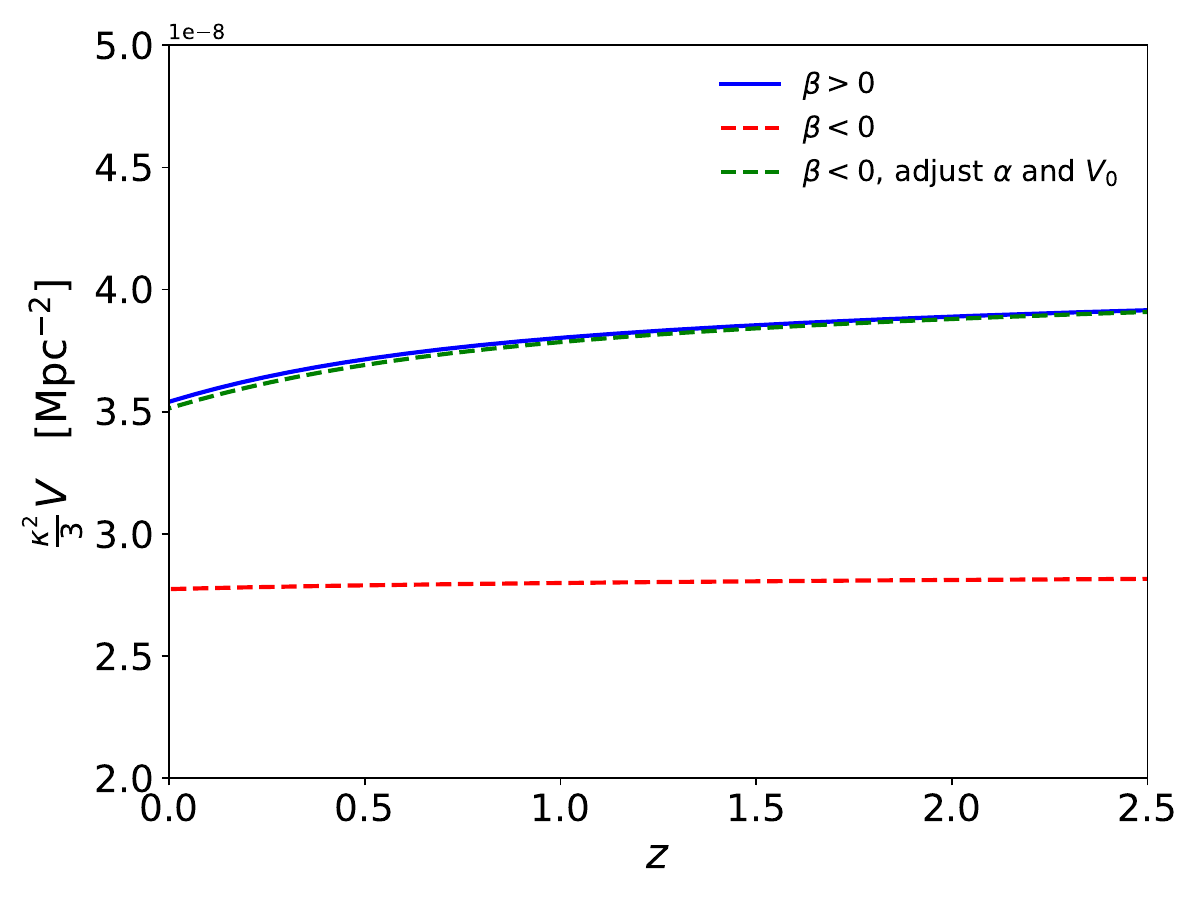}
\hfill
\includegraphics[width=0.48\textwidth]{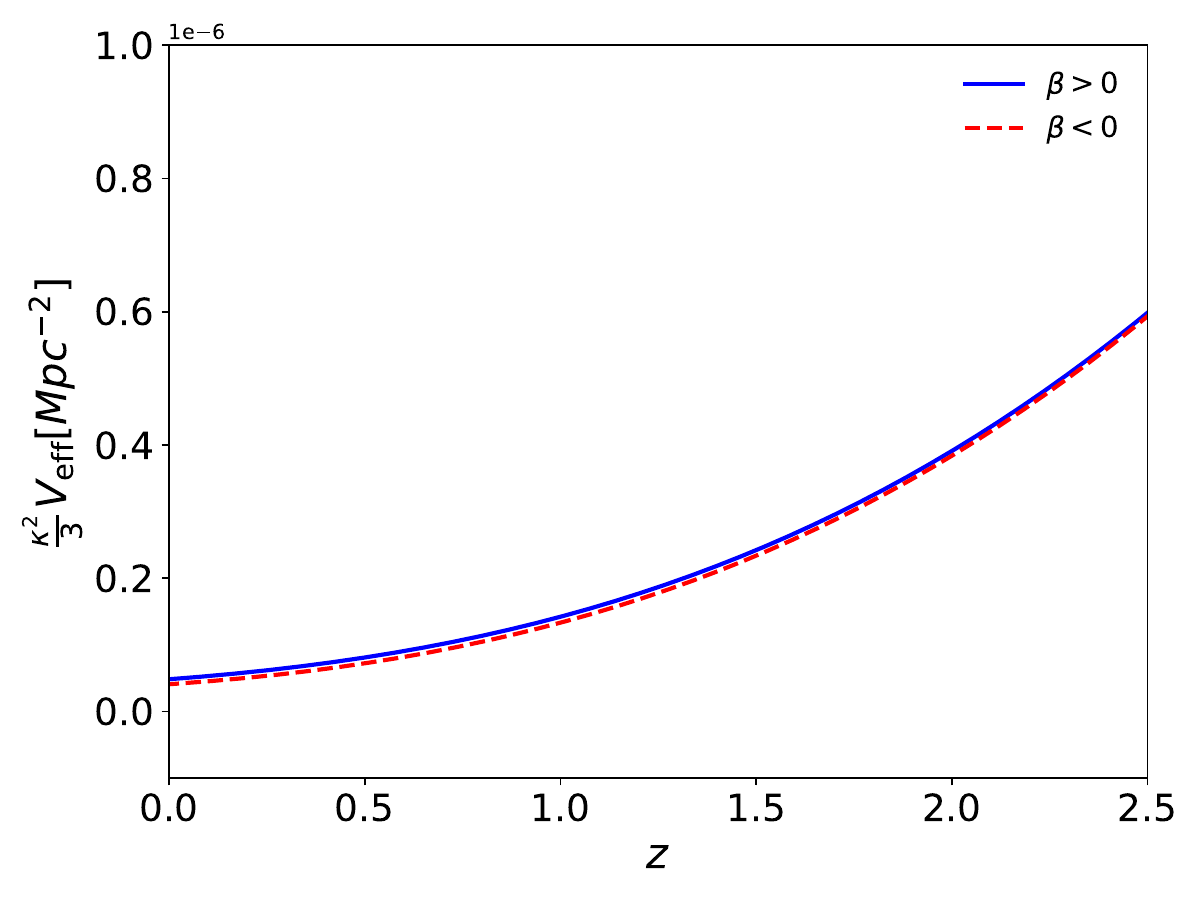}
\caption{Evolution of the PR potential $V$ (left panel) and the effective potential $V_{\rm eff}$ (right panel) within the redshift range $0\leq z \leq 2.5$, for both $\beta>0$ (blue solid line) and $\beta<0$ (red dashed line). Here we show the results rescaled by a factor of $\frac{\kappa^2}{3}$ and in units of Mpc$^{-2}$. All parameters are fixed to the best-fit values obtained from the PlanckPR4+PantheonPlus+DESI dataset (cf. Table \ref{tab:BF_PP}). In the left panel we also show the impact of adjusting $\alpha$ and $V_0$ on the shape of $V$. See Sec. \ref{sec:bg_dynamics} for a more detailed discussion.}
    \label{fig:V_eff} 
\end{figure*}

Deep in the radiation era, the source function entering the right-hand side of the Klein–Gordon equation \eqref{eq:KG} is strongly suppressed, so the scalar field is essentially frozen, as shown in the upper-left plot of Fig. \ref{fig:plots_PR}, and the energy densities evolve as in the standard $\Lambda$CDM model. In the matter-dominated era, the source function -- being proportional to the dark matter density -- becomes non-negligible. As a result, the scalar field becomes dynamical and the system approaches an unstable fixed point where $\Omega_\phi\simeq 2\beta^2/3$ \cite{Amendola:1999er}. This scaling solution is insensitive to the sign of the coupling (cf. the bottom-right plot in Fig. \ref{fig:plots_PR}). For fixed initial conditions, the evolution of the field is completely symmetric under $\beta \rightarrow -\beta$. The only difference is that the field increases if $\beta>0$, whereas it decreases if $\beta<0$, while the total change in absolute value, $|\Delta\phi|$, remains the same. The relative sign change in $\Delta\phi$ induced by reversing the sign of the coupling does not affect the dynamics of the model at any stage of the cosmic expansion if the potential is constant. However, when a non-trivial potential is introduced, differences can arise in the late universe, since the dark-energy density depends on the value of the potential once it begins to rule the scalar-field dynamics after matter domination (see the the upper-right plot in Fig. \ref{fig:plots_PR}).

In this work, we first constrain the model with a constant potential, $V(\phi)=V_0$. This corresponds to considering a cosmological constant in the action, although the scalar field is still dynamical due to the non-zero coupling with matter. For the latter, we only consider positive values, since, as explained above, negative $\beta$ lead to exactly the same dynamics. Our constraints are actually on $|\beta|$ if the potential is constant. 

On the other hand, we also use the Peebles-Ratra potential \cite{Peebles:1987ek,Ratra:1987rm}, 
\begin{equation}\label{eq:PRpotential}
V(\phi)=V_0 \left(\frac{\phi}{\phi_{\rm ini}}\right)^{-\alpha}\,,
\end{equation}
and analyze the model considering both $\beta>0$ and $\beta<0$ to test: (i) to what extent having a non-constant potential influence the description of the cosmological data; and (ii) the impact of the asymmetry induced by the potential in the late universe, and whether it can be indeed compensated by shifts in other parameters or in the initial conditions. We devote Fig.~\ref{fig:V_eff} to showing that such compensations are indeed important for describing the cosmological data. In the left panel, we display the best-fit curves of the PR potential $V(z)$ obtained from analyses with positive and negative $\beta$. In the latter case, $V$ is nearly flat because $\alpha$ is small ($\alpha \approx 0.17$), in contrast to the positive-$\beta$ case, where the best-fit value of the slope is significantly larger ($\alpha \approx 1.08$). By fixing $\beta$ and the remaining parameters while tuning $V_0$ and $\alpha$, one can obtain very similar shapes for $V$ in both cases. However, this is not favored by the data. The right panel of Fig.~\ref{fig:V_eff} shows instead that the best-fit solutions yield remarkably similar shapes for the effective potential $V_{\rm eff}(z)$, defined in Eq.~\eqref{eq:effV}. This is expected, since this quantity enters the right-hand side of the Friedmann equation \eqref{eq:Friedmann}. Regardless of the sign of the coupling strength, the model reproduces similar expansion histories and therefore provides a comparable fit to the data.

\begin{figure*}
\centering
\includegraphics[width=0.6\textwidth]{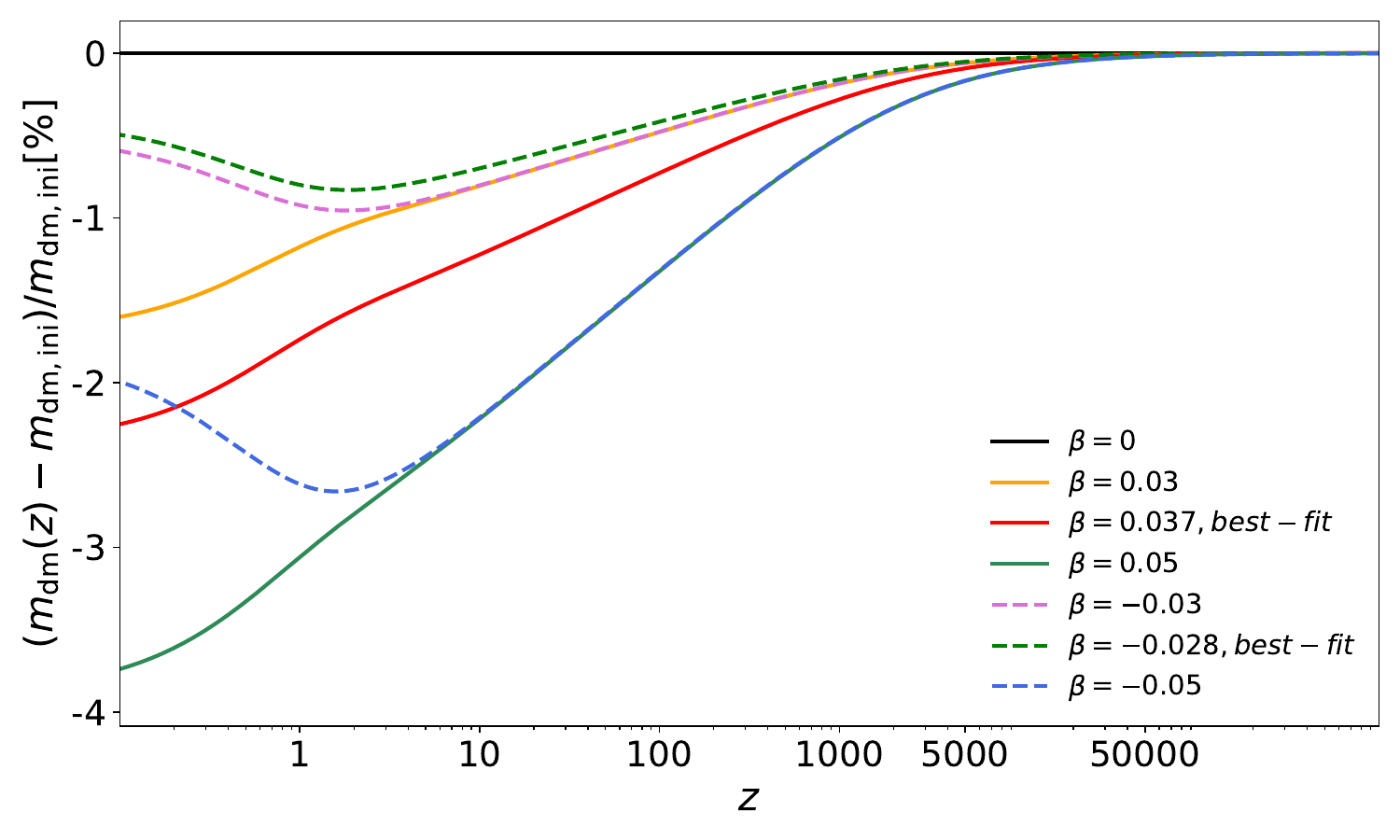}
\caption{Relative change of the DM particle mass over cosmic time for a CDE model with a PR potential. All parameters except $\beta$ are fixed to the best-fit values of the $\beta>0$ and $\beta<0$ cases, obtained using the PlanckPR4+DES-Dovekie+DESI dataset (cf. Table \ref{tab:BF_DES}). Solid and dashed lines represent the positive and negative $\beta$ branches, respectively, cf. Table \ref{tab:tableCDEII}.}
    \label{fig:mass_dm} 
\end{figure*}

From the fitting analyses, we can infer the preferred evolution of the DM density. The relative contributions of the mass and number density to the energy density \eqref{eq:DMdensity} depend on the details of the dark matter decoupling mechanism from the cosmic plasma in the early universe, which is left unspecified and falls beyond the current study. However, we can learn about the relative variation of the mass throughout the cosmic history, $\Delta m_{\rm dm}(a)/m_{\rm dm,ini}$.  One can easily find that $\frac{m_{\rm dm}(z)-m_{\rm dm, ini}}{m_{\rm dm, ini}}=e^{\beta\kappa\left(\phi_{\rm ini}-\phi(z)\right)}-1$. In Fig. \ref{fig:mass_dm}, we can see that for positive values of the coupling, the DM mass is monotone decreasing with the cosmic expansion, since both the fifth force and the PR potential pushes the scalar field to larger positive values. However, once the potential starts to rule the expansion at $z\simeq \mathcal{O}(1)$, the decrease becomes steeper, as expected. If, instead, we take $\beta<0$, the PR potential allows for a stage at which the DM mass increases with $z$ at late times, after passing a minimum. This does not happen for sufficiently flat potentials. In any case, for the best-fit models, we always find that the current value of the DM mass is a few percent smaller than the one in the radiation-dominated epoch.

\section{Methodology and Datasets}\label{sec:method}

We use our own modified version of the Einstein-Boltzmann solver \texttt{CLASS}\footnote{\url{https://github.com/lesgourg/class_public}} \cite{Lesgourgues:2011re,Blas:2011rf}, in which we have implemented the CDE model described in the previous section. We use it to compute the theoretical predictions for the various cosmological observables involved in the fitting analyses. We employ the code \texttt{Cobaya}\footnote{\url{https://github.com/CobayaSampler/cobaya}}
 \cite{Torrado:2020dgo} to perform the exploration of the parameter space with Monte Carlo runs and the \texttt{GetDist}\footnote{\url{https://github.com/cmbant/getdist}} package \cite{Lewis:2019xzd} for the processing of the Monte Carlo Markov chains (MCMC) and the extraction of the  parameter constraints.

  \begin{table*}[t!]
\centering
\begin{tabular}{|c ||c |c | c | c|  c|   }
 \multicolumn{1}{c}{} & \multicolumn{1}{c}{} & \multicolumn{1}{c}{} & \multicolumn{1}{c}{} & \multicolumn{1}{c}{} & \multicolumn{1}{c}{} \\
\multicolumn{1}{c}{} &  \multicolumn{4}{c}{PlanckPR4+PantheonPlus+DESI} 
\\\hline
{\small Parameter} & {\small $\Lambda$CDM} & {\small PR} & {\small CDE\_const}  & {\small CDE\_PR ($\beta>0$)} & {\small CDE\_PR ($\beta<0$)} 
\\\hline
$10^2\omega_b$ & $2.231^{+0.010}_{-0.009}$ & $2.241\pm 0.014$ & $2.222\pm 0.017$  & $2.221^{+0.015}_{-0.016}$ & $2.223^{+0.014}_{-0.016}$ \\\hline
$\omega_{\rm dm}$ & $0.1178\pm 0.0006$ & $0.1174^{+0.0006}_{-0.0007}$ & $0.1173\pm 0.0007$ & $0.1169\pm 0.0007$  & $0.1176\pm 0.0006$ \\\hline
$\ln(10^{10}A_s)$ &  $3.047^{+0.013}_{-0.014}$ & $3.050\pm 0.012$ & $3.040\pm 0.014$ & $3.040^{+0.013}_{-0.014}$  & $3.040^{+0.014}_{-0.015}$ \\\hline
$n_s$ &  $0.968^{+0.003}_{-0.004}$ & $0.969\pm 0.004$ & $0.964^{+0.004}_{-0.005}$ & $0.964^{+0.004}_{-0.005}$ &  $0.964\pm 0.005$\\\hline
$\tau_{\rm reio}$ &  $0.059^{+0.009}_{-0.008}$ & $0.060\pm 0.006$ &$0.055\pm 0.007$ & $0.055\pm 0.007$ & $0.055\pm 0.007$ \\\hline
$H_0$ & $68.12\pm 0.27$ & $67.83^{+0.48}_{-0.38}$ & $68.63^{+0.40}_{-0.43}$ & $68.46^{+0.47}_{-0.46}$ & $68.52^{+0.39}_{-0.44}$ \\\hline
$\beta$ & $-$ & $-$ & $0.033^{+0.017}_{-0.009}$ &  $0.035^{+0.016}_{-0.009}$ & $-0.030^{+0.012}_{-0.018}$ \\\hline
$\alpha$ & $-$ & $<0.11$ & $-$ & uncons. & $<0.86$ \\\hline
${\varphi}_{\rm ini}$ & $-$ & $-$ & $-$ & $<5.44$ & $1.73^{+0.67}_{-0.62}$\\\hline\hline
$r_d$ & $147.76^{+0.17}_{-0.19}$   & $147.75^{+0.25}_{-0.27}$ & $147.54^{+0.26}_{-0.25}$ & $147.55\pm 0.26$   & $147.56^{+0.24}_{-0.25}$ \\\hline
$M$ & $-19.420^{+0.009}_{-0.008}$  & $-19.426^{+0.013}_{-0.010}$ & $-19.405^{+0.012}_{-0.013}$ & $-19.508\pm 0.013$  & $-19.408^{+0.014}_{-0.013}$ \\\hline
$\sigma_{12}$ & $0.794^{+0.005}_{-0.006}$ & $0.791\pm 0.006$ & $0.801^{+0.008}_{-0.009}$ & $0.802^{+0.008}_{-0.009}$ & $0.800^{+0.008}_{-0.009}$ \\\hline
$S_{8}$ & $0.810\pm 0.008$ & $0.805\pm 0.008$ & $0.812\pm 0.009$ & $0.812\pm 0.008$ &  $0.812\pm 0.008$   \\\hline
$\chi^2_{\rm min}$ & $12393.13$ & $12392.65$ & $12392.01$ & $12389.92$  &  $12390.01$
\\\hline
$\Delta{\rm AIC}$ & $-$  & $-1.52$ & $-0.88$  & $-2.79$ & $-2.88$ 
\\\hline
$E_{\Lambda{\rm CDM}}$ & $-$  & $0.69\sigma$ & $1.06\sigma$  & $0.92\sigma$ & $0.89\sigma$ 
\\\hline
\end{tabular}
\caption{Mean and $1\sigma$ uncertainties of the individual parameters of the $\Lambda$CDM, PR and CDE models obtained using the dataset PlanckPR4+PantheonPlus+DESI. The corresponding best-fit values are displayed in Table \ref{tab:BF_PP}. Here, $\omega_{\rm dm}=\Omega_{\rm dm}^0h^2$ denotes the reduced dark matter density parameter. The upper bound on $\alpha$ is provided at 95\% CL. $H_0$ and $r_d$ are expressed in km/s/Mpc and Mpc, respectively. The initial condition for the scalar field is presented in terms of the following dimensionless quantity, $\varphi_{\rm ini}\equiv \kappa\phi_{\rm ini}/\sqrt{3}$. In the last three rows we report the minimum values of $\chi^2$, $\chi^2_{\rm min}$, the differences $\Delta {\rm AIC}\equiv{\rm AIC}_{\Lambda}-{\rm AIC}_{\rm CDE}$ and the level of exclusion of $\Lambda$CDM obtained through a likelihood-ratio test, $E_{\Lambda{\rm CDM}}$.}
\label{tab:tableCDE}
\end{table*}

 We constrain the CDE models with constant and PR potentials, considering for the latter two distinctive cases with positive and negative couplings, and compare our fitting results with those obtained with the $\Lambda$CDM, which we treat as a natural benchmark model, as usual. In passing, we also provide constraints on the original PR model.

 In order to speed up the MCMC runs, we avoid the use of the shooting method by directly sampling over the initial conditions instead of sampling over quantities evaluated at $z=0$; we essentially follow the approach described in Appendix A of Ref. \cite{Gomez-Valent:2020mqn}. In the CDE model with PR potential, the set of parameters that are allowed to vary in the Monte Carlo are: the amplitude and spectral index of the primordial power spectrum, $\ln(10^{10}A_s)$ and $n_s$, respectively, the optical depth to reionization $\tau$, 
 the amplitude and slope of the scalar field potential, $V_0$ and $\alpha$, the coupling strength $\beta$, the reduced baryon density parameter $\omega_b=\Omega_b^0h^2$ (with $h$ the reduced Hubble constant), the initial value of the  DM energy density, and the initial condition of the scalar field $\phi_{\rm ini}$. In CDE, we set $\phi^\prime_{\rm ini}=0$ for the reason explained in Sec. \ref{sec:Model}. The standard model is obviously retrieved in the limit $\alpha\,,\beta\to 0$, i.e., when we switch off the dark sector interaction and consider a cosmological constant. For the CDE model with constant potential, we obviously set $\alpha=0$ and $\phi_{\rm ini}$ plays no role in the fitting results, so we can set it to an arbitrary value. Therefore, CDE with a flat potential has one more free parameter than the $\Lambda$CDM (the coupling $\beta$), whereas in CDE with the PR potential the model has three more parameters, namely, $(\beta,\alpha,\phi_{\rm ini})$. For the runs with the original PR model, we assume that the scalar field is initially already in the tracker solution, deep in the radiation-dominated epoch (see, e.g., \cite{SolaPeracaula:2018wwm}), and we set $\beta=0$.

  For CDE with PR potential, instead of performing a joint exploration of the $\beta>0$ and $\beta<0$ regions of parameter space using nested sampling techniques, we opt to study both signs of the coupling separately sticking to the standard Metropolis-Hastings algorithm. This approach allows us to interpret in a cleaner way the posterior distributions obtained for other parameters of the model (as the power of the PR potential or the initial value of the scalar field), which have quite different shapes for the two signs of $\beta$. These shapes could be otherwise distorted due to enhanced marginalization effects.

In our fitting analyses, we use the following datasets:

\begin{itemize}
\item \textbf{CMB:} We employ the Planck CMB temperature (TT), polarization (EE), and cross-correlation (TE) power spectra. More concretely, we use the \texttt{simall}, \texttt{Commander}, and \texttt{NPIPE PR4} \texttt{CamSpec} likelihoods \cite{Efstathiou:2019mdh,Rosenberg:2022sdy} for multipoles $\ell < 30$ and $\ell \geq 30$, respectively, in combination with the \texttt{NPIPE PR4} CMB lensing likelihood \cite{Carron:2022eyg}.

\item \textbf{BAO:} We use measurements from the DESI Data Release 2 (DR2), accounting for the existing correlations, cf. Table~IV of Ref.~\cite{DESI:2025zgx}.

\item \textbf{SNIa:} We combine the aforesaid CMB and BAO data with the Pantheon$+$ \cite{Scolnic:2021amr} and DES-Dovekie \cite{DES:2025sig} samples, separately. The latter is obtained upon applying an updated SNIa calibration which corrects some issues afflicting the previous DESy5 sample \cite{DES:2024hip,DES:2024jxu}. For completeness, we also present the results obtained with DESy5 in Appendix \ref{app:tables_DESy5}.

\end{itemize}

The triad of observables chosen to constrain our models is robust and widely regarded by the cosmological community as a baseline combination. It also facilitates comparison with previous results in the literature.

In our fitting tables, we do not display the constraints on all the sampled parameters, since some of them --- for instance, the initial DM energy density --- are not very informative. Instead, we report the constraints on the most relevant parameters, such as the Hubble--Lema\^itre parameter $H_0$ and $\omega_{\rm cdm}$. As a consequence of avoiding the shooting method, some of these are derived parameters. We also present constraints on the weak-lensing observable $S_8$ and on the root-mean-square mass fluctuations at a fixed scale of 12\,Mpc, $\sigma_{12}$, which is less affected by some of the issues associated with the more commonly used parameter $\sigma_8$ (see Refs.~\cite{Sanchez:2020vvb,Forconi:2025cwp} for details).

To facilitate the comparison of the fitting performance of the CDE and PR models both among themselves and relative to $\Lambda$CDM, we also report the minimum values of $\chi^2$ and the Akaike information criterion (AIC), defined as ${\rm AIC} = \chi^2_{\rm min} + 2n$ \cite{Akaike}, where $n$ denotes the number of free parameters in the model. The AIC balances goodness of fit against model complexity by rewarding an improved description of the data while penalizing the introduction of additional parameters. It can therefore be regarded as a proxy for more elaborate Bayesian model comparison methods. In addition, we compute the exclusion level of $\Lambda$CDM relative to the models under study using the likelihood-ratio test and assuming the validity of Wilk's theorem\footnote{The posterior distributions in the CDE model suffer from some deviations from Gaussianity mainly affecting the parameters $\alpha$ and $\phi_{\rm ini}$, which may invalidate the assumptions underlying Wilks' theorem. Our estimation of the exclusion level of $\Lambda$CDM must be understood as an approximation. A more accurate estimate of $E_{\Lambda{\rm CDM}}$ would require simulations to determine the exact distribution of $\Delta\chi^2_{\rm min}$. Such an analysis would entail though a substantial computational effort, which we consider unnecessary given the overall consistency between the fitting results reported in the tables, the contour plots, and the information criterion (cf. Sec. \ref{sec:results})}. Instead of displaying the resulting $p$-values in the tables, we express them as an exclusion level, $E_{\Lambda{\rm CDM}}$, translating these $p$-values into a number of standard deviations ($\sigma$), as if they were extracted from univariate Gaussian distributions.

\begin{table*}[t!]
\centering
\begin{tabular}{|c ||c| c|c | c | c|     }
 \multicolumn{1}{c}{} & \multicolumn{1}{c}{} & \multicolumn{1}{c}{} & \multicolumn{1}{c}{} & \multicolumn{1}{c}{} & \multicolumn{1}{c}{} \\
\multicolumn{1}{c}{} &  \multicolumn{4}{c}{PlanckPR4+DES-Dovekie+DESI} 
\\\hline
{\small Parameter} & {\small $\Lambda$CDM} & {\small PR} & {\small CDE\_const}  & {\small CDE\_PR ($\beta>0$)} & {\small CDE\_PR ($\beta<0$)} 
\\\hline
$10^2\omega_b$ & $2.230\pm 0.012$ & $2.240^{+0.014}_{-0.015}$ & $2.223^{+0.016}_{-0.017}$ & $2.222\pm 0.017$ & $2.219\pm0.018$\\\hline
$\omega_{\rm dm}$ & $0.1178\pm 0.0006$ & $0.1173^{+0.0006}_{-0.0007}$ & $0.1174\pm 0.0007$ & $ 0.1168_{-0.0008}^{+0.0009}$ & $0.1176^{+0.0007}_{-0.0006} $ \\\hline
$\ln(10^{10}A_s)$ & $3.046^{+0.013}_{-0.014}$ & $3.050^{+0.014}_{-0.015}$ & $3.040\pm 0.014 $ & $3.041 \pm 0.014$ & $3.038\pm0.015$\\\hline
$n_s$ & $0.968^{+0.003}_{-0.004}$ & $0.969\pm 0.004$ & $0.964^{+0.004}_{-0.005}$ & $0.964\pm 0.005$  & $0.963\pm0.005$ \\\hline
$\tau_{\rm reio}$ & $0.059^{+0.007}_{-0.008}$ & $0.061^{+0.006}_{-0.008}$ & $0.055\pm 0.007$ & $0.055 \pm 0.007$ & $0.054\pm0.007$ \\\hline
$H_0$ & $68.09\pm 0.27$ & $67.71^{+0.46}_{-0.39}$ & $68.55^{+0.38}_{-0.43}$ & $68.32^{+0.55}_{-0.45}$ & $68.48^{+0.66}_{-0.51}$ \\\hline
$\beta$ & $-$ & $-$ & $0.031^{+0.017}_{-0.010}$ & $0.034^{+0.016}_{-0.009}$  & $-0.036^{+0.010}_{-0.019}$ \\\hline
$\alpha$ & $-$ & $<0.13$ & $-$ & uncons. & $<1.46$ \\\hline
${\varphi}_{\rm ini}$ & $-$ & $-$ & $-$ & $4.0^{+1.5}_{-3.6}$ & $1.71^{+0.41}_{-1.10}$ \\\hline\hline
$r_d$ & $147.75\pm 0.18$ & $147.80^{+0.21}_{-0.24}$ & $147.54^{+0.26}_{-0.25}$ & $147.56\pm0.26$ & $147.51\pm0.27$ \\\hline
$\sigma_{12}$ & $0.794\pm 0.006$ & $0.790\pm 0.006$ & $0.801^{+0.007}_{-0.009}$  & $0.801 \pm 0.008$ & $0.803\pm0.009$ \\\hline
$S_{8}$ & $0.810^{+0.008}_{-0.007}$ & $0.805\pm 0.008$  & $0.813\pm 0.009$ & $0.812\pm$ 0.008 & $0.814\pm0.009$  \\\hline
$\chi^2_{\rm min}$ & $12624.17$ & $12622.78$ & $12620.79$ & $12618.92$ & $12616.49$
\\\hline
$\Delta{\rm AIC}$ & $-$  & $-0.61$ & $+1.38$ & $-0.75$ & $+1.68$
\\\hline
$E_{\Lambda{\rm CDM}}$ & $-$  & $1.18\sigma$ & $1.84\sigma$ & $1.42\sigma$ & $1.94\sigma$
\\\hline
\end{tabular}
\caption{As in Table I, but using the SNIa from DES-Dovekie \cite{DES:2025sig}. See Table \ref{tab:BF_DES} for the corresponding best-fit values. We note that the SNIa absolute magnitude is marginalized over analytically, as is done by default in the public likelihood released by the DES collaboration.}
\label{tab:tableCDEII}
\end{table*}


\section{Results and Discussion}\label{sec:results}

Our main results, obtained using PantheonPlus or DES-Dovekie in combination with Planck PR4 CMB data and DESI DR2, are presented in Tables \ref{tab:tableCDE}–\ref{tab:tableCDEII} and Figs. \ref{fig:CP1}–\ref{fig:CP3}. In Appendix \ref{app:breakdown_chi2}, we also provide additional tables with the breakdown of the individual $\chi^2_{\rm min}$ contributions for the various models and datasets as well as tables with the corresponding best-fit values and the resulting effective CPL parameters in order to facilitate the interpretation of the results. Finally, in Appendix \ref{app:tables_DESy5} we report, for completeness, the fitting results obtained with DESy5, allowing the reader to assess the differences induced by the calibration correction in the DES SNIa sample.

The data under consideration leads to very stable constraints on the absolute value of the coupling strength, regardless of the SNIa data employed and the slope of the potential. We find in all cases a peak in its corresponding one-dimensional (1D) posterior distribution located at $|\beta|\sim 0.03-0.04$, $\gtrsim 2\sigma$ away from 0. Volume effects do not play an important role in this signal, cf. \cite{Gomez-Valent:2022hkb}. The constraints on the other parameters are also very robust and not affected by significant shifts when replacing Pantheon+ with DES-Dovekie, which is reassuring.

In Fig. \ref{fig:CP1} we show contour plots in several relevant planes of parameter space for the flat potential case. When CMB data alone are used to constrain the model, there is still room for a mitigation of the Hubble tension, since there exists a strong positive correlation between the coupling and the Hubble parameter, which is even more pronounced at high values of $\beta$. The DM energy density is close to the $\Lambda$CDM value around the decoupling time and, hence, also the angular diameter distance to the last scattering surface remains stable. However, since the DM mass can decrease with the expansion of the universe in a non-negligible way since we are not using low-$z$ data, the cosmological constant is forced to take larger values to keep the location of the first CMB acoustic peak as measured by Planck (see \cite{Costa:2025kwt} for details). This allows for large values of $H_0$, in the ballpark of those measured by SH0ES \cite{Riess:2021jrx,H0DN:2025lyy}, while keeping $\beta$ within the $1-2\sigma$ CL region -- it can easily take values $\beta\sim 0.09$. This feature of the model was already pointed out more than one decade ago in \cite{Pettorino:2013oxa}, and here we confirm it in light of the latest CMB data release from Planck (PR4).

\begin{figure}
    \centering
    \includegraphics[width=0.6\textwidth]{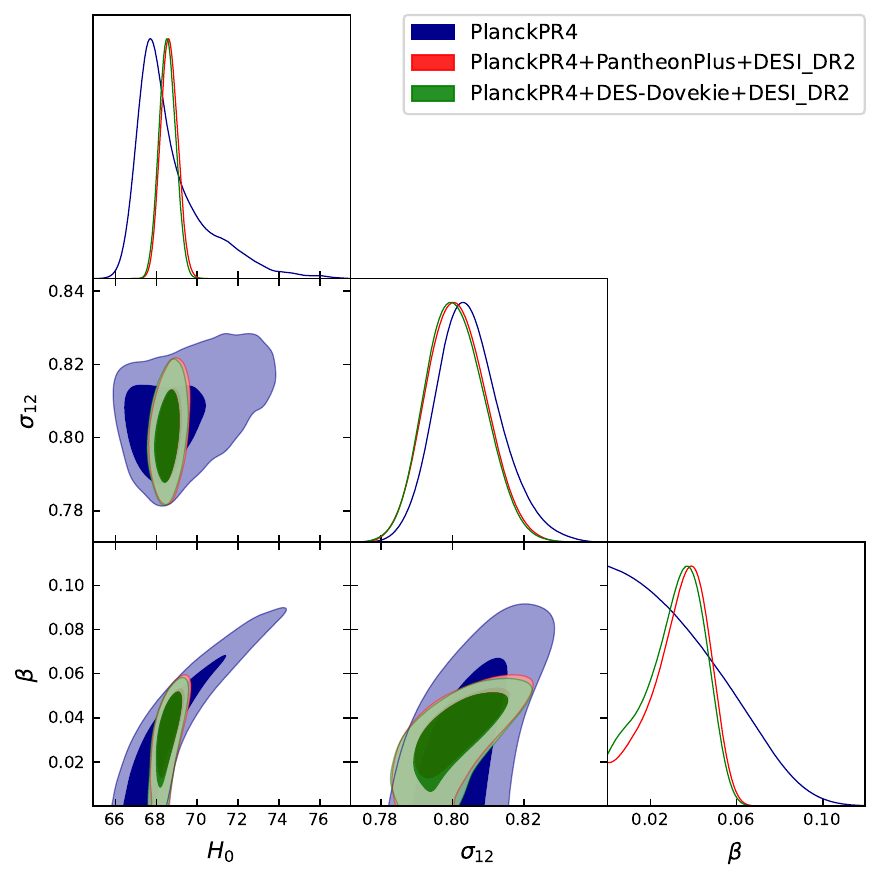}
    \caption{
     One-dimensional posterior distributions for $H_0$ (in units of km/s/Mpc), $\sigma_{12}$ and $\beta$ in the CDE scenario with a constant potential, together with their two-dimensional contour plots at the $68\%$ and $95\%$ CL.}
 \label{fig:CP1}
\end{figure}

However, when low-$z$ data from BAO and SNIa are also included in the analysis, the long tail in the 1D posterior of $\beta$ is unfortunately cut off, limiting $H_0$ to values around $68.5$ km/s/Mpc (cf. again Fig. \ref{fig:CP1}). Hence, the Hubble tension is not substantially alleviated in this CDE model, as already noted in \cite{Gomez-Valent:2020mqn}. The central values of $H_0$ found in CDE are nevertheless $\sim 0.5$ km/s/Mpc larger than in  $\Lambda$CDM and remain fully compatible with the measurements from the Chicago–Carnegie Hubble Program (CCHP), obtained using the Hubble and James Webb Space Telescopes \cite{Freedman:2024eph}. Considering a non-zero slope of the scalar-field potential ($\alpha\ne 0$) does not improve the situation; indeed, it has been shown in many contexts that dynamical dark energy alone cannot produce large values of $H_0$ \cite{Sola:2017znb,Knox:2019rjx,Krishnan:2021dyb,Lee:2022cyh,Gomez-Valent:2023uof,Pedrotti:2025ccw,Bansal:2026axl}, even if it is interacting with dark matter. The future will tell to what extent local measurements of the Hubble constant can act as a discriminator for the CDE model.

The values of the current amplitude of the matter power spectrum at linear scales, which we parametrize with the $\sigma_{12}$ parameter \cite{Sanchez:2020vvb,Forconi:2025cwp}, remains only slightly larger than in the $\Lambda$CDM model, and in any case are compatible with it at 68\% CL. We can therefore conclude that the enhancement induced by the fifth force in the matter clustering is quite mild. This makes complete sense in view of the very small increase of the effective ``gravitational'' coupling felt by the DM particles, which includes the contribution of the fifth force itself, $\Delta G/G_N=2\beta^2\sim 0.2\%$, and resonates perfectly well with the values of  the weak-lensing observable $S_8$ inferred for CDE, which are again fully compatible with the $\Lambda$CDM best-fit value, $S_8\sim 0.81$. They are all in excellent agreement with the measurement by the Kilo-Degree Survey (KiDS),
$S_8 = 0.815^{+0.016}_{-0.021}$ \cite{Wright:2025xka}, although a bit larger (but still within $\lesssim 1.5\sigma$) than the weak-lensing measurements from the Dark Energy Survey (DES),
$S_8 = 0.775^{+0.026}_{-0.024}$ \cite{DES:2021wwk}, and
the Hyper Suprime-Cam (HSC),
$S_8 = 0.763^{+0.040}_{-0.036}$ \cite{Miyatake:2023njf}. See also \cite{Semenaite:2025ohg} for constraints obtained with these weak lensing surveys together with galaxy clustering auto- and cross-correlations from DESI DR1.

We are only able to put upper bounds on the slope of the potential in the original PR model, $\alpha<0.11$ and $\alpha<0.13$ (95\% CL), using Pantheon+ and DES-Dovekie, respectively. They are consistent, but tighter than previous constraints in the literature, see, e.g., \cite{SolaPeracaula:2018wwm,Park:2025fbl}. The constraints on $\alpha$ become much looser in the context of CDE. We refer the reader to the corresponding contour plots in Figs. \ref{fig:CP2} and \ref{fig:CP3} for the results obtained with positive and negative coupling strength. Although the PR potential leads to a decrease of $\chi^2_{\rm min}$ compared to the flat potential case, as expected due to the increase of the parameter space, such a decrease is not very significant from a statistical point of view. We find in all cases values of $\Delta$AIC that are inconclusive, i.e., we do not find a substantial evidence for CDE, even though, as mentioned above, we find an interesting peak in $\beta$ at 95\% CL, similar to that found in \cite{Gomez-Valent:2020mqn} with an older combination of CMB+SNIa+BAO data. The values of $E_{\Lambda{\rm CDM}}$ are in full accordance with those of AIC; $\Lambda$CDM is excluded at $\sim 2\sigma$ level at most. In Tables \ref{tab:chi2_Pan} and \ref{tab:chi2_DESDovekie} of Appendix \ref{app:breakdown_chi2}, we can see that CDE with PR potential is able to describe better the low-$z$ data, at the expense of mildly worsening the fit to the CMB. The improvement in the BAO and SNIa data can be understood by looking at the plots of the effective dark energy EoS parameter in Fig. \ref{fig:EffEoS}. For $\alpha\ne 0$ it is possible to produce an effective crossing of the phantom divide which lies within the $\sim 95\%$ CL band of the model-agnostic reconstruction of Ref. \cite{Gonzalez-Fuentes:2026rgu} -- more details are provided in Appendix \ref{app:EffEoS}. Such a crossing is found, for instance, with the best-fit values obtained in the analysis with PlanckPR4+DES-Dovekie+DESI. We also find that the phantom crossing behavior is significantly more pronounced for negative values of $\beta$ compared to positive ones. However, the shape is quite different from the one found with the CPL and this perhaps explain why the fitting performance of the model is still far from the one offered by that parametrization. The CDE models are unable to deviate from $\Lambda$CDM as dramatically as the CPL parametrization does. To mimic the more extreme $(w_0, w_a)$ values of CPL, the CDE model would require a larger coupling strength $\beta$ to induce the equation of state deeper into the phantom regime at higher redshifts, and a steeper potential slope $\alpha$ to increase the kinetic energy of the scalar field at $z=0$ and thereby pull $w_0$ up to a less negative value today. However, both mechanisms are strongly disfavored by the data: a larger $\beta$ would lead to a stronger fifth force that, besides enhancing structure formation, would make it more difficult to reproduce the late-time expansion history. Consequently, when BAO+SNIa data are added to the CMB, $\beta$ is driven toward smaller absolute values (cf. again our Fig.~4). As a result, the phantom evolution in this CDE model cannot approach that of the preferred CPL model. Consequently, there is no enhancement of the quintessence character at low redshifts, simply because it is not required to preserve the distance to the last scattering surface preferred by the CMB in models with standard pre-recombination physics. We also present the effective CPL parameters obtained from our best-fit values in Tables. \ref{tab:BF_PP} and \ref{tab:BF_DES}. This mapping facilitates direct comparison with current measurements and serves as a useful reference for future observational constraints.

\begin{figure*}
\centering
\includegraphics[width=0.49\textwidth]{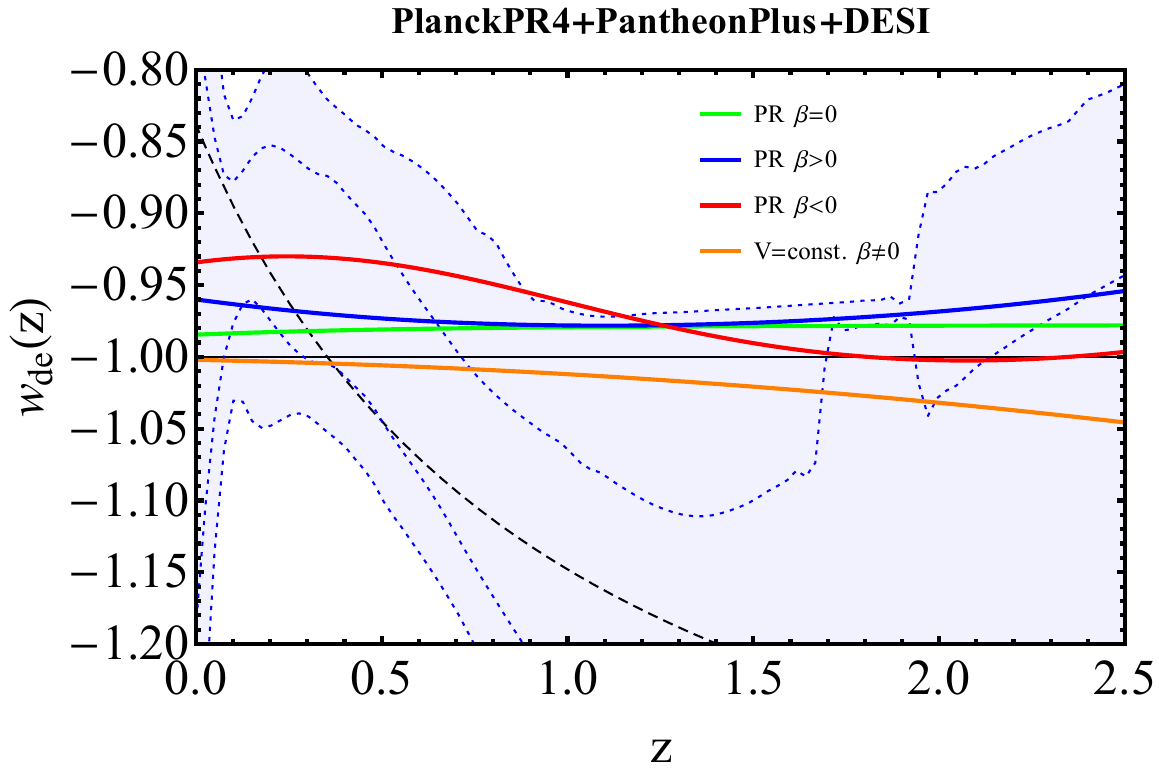}
\hfil
\includegraphics[width=0.49\textwidth]{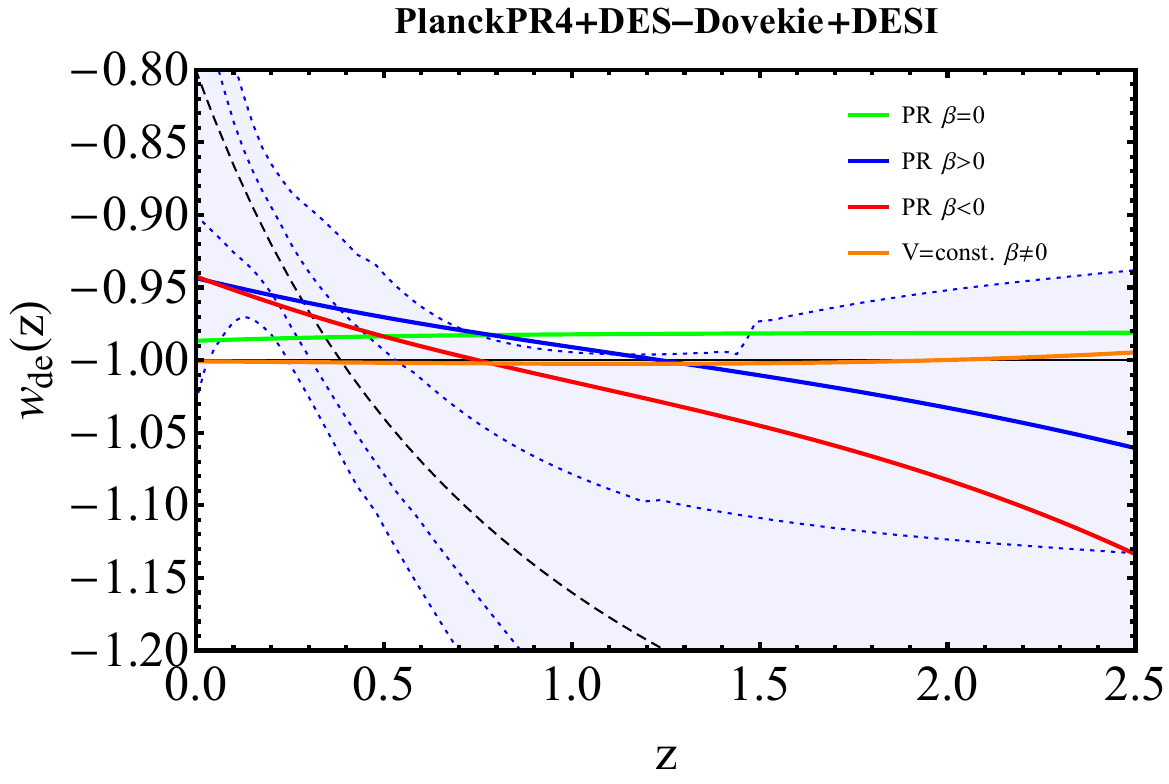}
\medskip
\caption{Effective EoS parameter \eqref{eq:effEoS} for the best-fit CDE models using PlanckPR4+PantheonPlus+DESI (left panel) and PlanckPR4+DES-Dovekie+DESI (right panel), see Tables \ref{tab:BF_PP} and \ref{tab:BF_DES}. The light blue bands show the model-independent reconstruction at 68\% and 95\% CL from \cite{Gonzalez-Fuentes:2026rgu}, while the CPL best-fit curves are indicated by black dashed lines. The best-fit CPL parameters are  $(w_0,w_a)=(-0.84,-0.62)$ \cite{DESI:2025zgx} and $(w_0,w_a)=(-0.80,-0.72)$  \cite{DES:2025sig} for Pantheon+ and DES-Dovekie, respectively. The effective CPL parameters, $\tilde{w}_0\equiv w_{\rm de}(a=1)$ and $\tilde{w}_a\equiv -dw_{\rm de}/da \big|_{a=1}$, for the other models are provided in Tables \ref{tab:BF_PP}-\ref{tab:BF_DES}.}
    \label{fig:EffEoS} 
\end{figure*}

The late-time asymmetry in the background solutions obtained with positive and negative values of $\beta$ that we discussed in Sec. \ref{sec:Model}, which is expected in the presence of a non-flat potential, is in practice very small. As mentioned in Sec. \ref{sec:bg_dynamics}, this is because the model is able to produce similar shapes of the effective potential \eqref{eq:effV} regardless of the sign of the coupling.


\section{Conclusions}\label{sec:conclusions}

In this work, we have revisited constraints on CDE models in light of the most recent cosmological datasets. We include CMB data from the Planck PR4, BAO measurements from  DESI DR2, and the most up-to-date SNIa compilations, such as the Pantheon+ and DES-Dovekie samples. For the CDE models, we have considered a DM field-dependent mass $m_{\rm dm}(\phi)\propto e^{-\kappa\beta\phi}$, both a flat and a Peebles-Ratra (PR) potential, with  positive and negative couplings, and provide also constraints on the original PR model, in which the fifth force is switched off. Our analysis allows us to study the impact of the slope of the scalar-field potential in the late universe and also the sign of the coupling on the ability of the model to explain the data. 

Although we find a persistent peak in the absolute value of the coupling strength, $\sim 2\sigma$ away from 0,  which can be interpreted as a hint of new physics, we find it to be milder than that reported in other recent works \cite{Chakraborty:2024xas,Li:2026xaz}, with the information criteria and likelihood-ratio tests pointing to a similar performance to that offered by  $\Lambda$CDM. While the CMB data alone allow for an important decrease of the Hubble tension in the context of CDE, the addition of SNIa and BAO brings $H_0$ again to values in tension with SH0ES, in the lower range favored by Planck assuming the validity of $\Lambda$CDM. The CDE model is able to accommodate better the low-redshift data than the standard cosmological model, more conspicuously with the PR potential (rather than with the flat one), since in this case it is possible for the effective dark energy to cross the phantom divide. Importantly, our constraints on $|\beta|$ remain stable under both, the shape of the potential and the sign of the coupling. However, the use of additional parameters in the model renders this improvement inefficient from a Bayesian perspective, and the resulting shape of the equation-of-state parameter differs significantly from that obtained with the CPL. This indicates that alternative functional forms for the dark matter mass and the scalar-field potential are needed to achieve a better agreement with current observations \cite{LaPenna:2026avs}. 

It will be of utmost importance to see what future cosmological measurements will reveal, for instance from Euclid or upcoming data releases from DESI. These observations should help us better assess the significance of the evidence for dark energy dynamics and the effective crossing of the phantom divide, as part of the broader effort to determine whether these are genuine features of the underlying cosmological model governing the expansion of the universe. We plan to carry out a detailed forecast of CDE using mock data from these galaxy surveys in future work.


\vspace{0.5cm}
\noindent {\bf Acknowledgments }
\newline
\newline
AGV is funded by “la Caixa” Foundation (ID 100010434) and the European Union's Horizon 2020 research and innovation programme under the Marie Sklodowska-Curie grant agreement No 847648, with fellowship code LCF/BQ/PI23/11970027. He is also supported by projects PID2022-136224NB-C21 (MICIU), 2021-SGR-00249 (Generalitat de Catalunya) and CEX2024-001451-M (ICCUB). AGV acknowledges the
participation in the COST Action CA21136 “Addressing observational tensions in cosmology with systematics and fundamental physics” (CosmoVerse). ZZ acknowledges financial support from the German Research Foundation (DFG) under Germany’s Excellence Strategy – EXC 2181/1 – 390900948 (the Heidelberg STRUCTURES Excellence Cluster) for her research visit to Barcelona, and thanks the University of Barcelona for their hospitality. LA acknowledges support by DFG under Germany’s Excellence Strategy EXC 2181/1 - 390900948 (the Heidelberg STRUCTURES Excellence Cluster) and under Project 554679582 ``GeoGrav: Cosmological Geometry and
Gravity with nonlinear physics''.


\appendix 
 \section{Phantom crossing and the effective dark energy  EoS parameter}\label{app:EffEoS}

Observations are usually interpreted by assuming standard conservation equations for matter and
dark energy, finding an apparent EoS parameter for the effective dark energy fluid $w_{\rm de}(z)$. The relation between
the apparent EoS and the field EoS $w_{\phi}(a)=p_{\phi}(a)/\rho_{\phi}(a)$
is
\begin{equation}\label{eq:effEoS2}
w_{\rm de}(a)=\frac{p_{\phi}(a)}{\rho_{\phi}(a)+\rho_{{\rm dm}}(a)-\tilde{\rho}_{{\rm dm}}(a)}=\frac{w_{\phi}(a)}{1+\frac{\rho_{{\rm dm}}(a)-\tilde{\rho}_{{\rm dm}}(a)}{\rho_{\phi}(a)}}\,,
\end{equation}
where $\tilde{\rho}_{\rm dm}=\tilde{\rho}^{0}_{\rm dm}a^{-3}=\frac{3\tilde{H}_{0}^{2}}{\kappa^2}\tilde{\Omega}_{\rm dm}^0a^{-3}$. In CDE, we have
\begin{equation}
\rho_{\rm dm}(a)=\rho_{\rm dm}^0a^{-3}e^{-\kappa\beta(\phi(a)-\phi^0)}=\frac{3H_{0}^{2}}{\kappa^2}\Omega_{\rm dm}^0a^{-3}e^{-\beta\kappa(\phi(a)-\phi^0)}\,,
\end{equation}
with $\phi^0\equiv\phi(a=1)$, so 
\begin{equation}
\Omega_{\rm dm}(a)=\kappa^2\frac{\rho_{\rm dm}(a)}{3H^{2}(a)}=\frac{\Omega_{\rm dm}^0a^{-3}e^{-\beta\kappa(\phi(a)-\phi^0)}}{E^{2}(a)}\,.
\end{equation}
Using $\rho_{\phi}=3H^{2}\Omega_{\phi}/\kappa^2\approx 3H^{2}(1-\Omega_{\rm dm})/\kappa^2$ we can write Eq. \eqref{eq:effEoS2} as follows, 
\begin{equation} \label{eq: app_w_de}
w_{\rm de}(a) \approx \frac{w_{\phi}(a)}{1+\frac{\Omega_{\rm dm}(a)}{1-\Omega_{\rm dm}(a)}\left(1-\frac{\tilde{\omega}_{\rm dm}}{\omega_{\rm dm}}e^{\beta\kappa(\phi(a)-\phi^0)}\right)}\,,
\end{equation}
with $\omega_{\rm dm}=\Omega_{\rm dm}^0h^2$ the reduced dark matter density parameter and $\tilde{\omega}_{\rm dm} = \tilde{\Omega}_{\rm dm}^0\tilde{h}^2$ its self-conserved counterpart, and $\phi(a)$ is obtained by integrating Eq. \eqref{eq:KG}. By definition, $w_\phi>-1$, but $w_{\rm de}$ could be in principle smaller than -1. This appendix is devoted to understanding what are the required conditions to have a crossing of the phantom divide from phantom (at large $z$) to quintessence (at low $z$), as the one preferred by current observations. 

The last expression shows that in order to have $w_{\rm de}<w_\phi$, which is necessary to get a phantom phase ($w_{\rm de}<-1$), one needs $\beta(\phi-\phi^0) >0$ if $\tilde{\omega}_{\rm dm}=\omega_{\rm dm}$. Note that for a flat potential, $\beta(\phi-\phi^0)$ is always negative, as can be seen by solving the Klein-Gordon equation Eq. \eqref{eq:KG}, since $\beta(\phi-\phi^0)\propto -\beta^2$. Therefore, if $\tilde{\omega}_{\rm dm}=\omega_{\rm dm}$ and for a flat potential, the EoS cannot be phantom, regardless of the sign of the coupling.

\begin{figure*}[t!]
\centering
\includegraphics[width=0.45\textwidth]{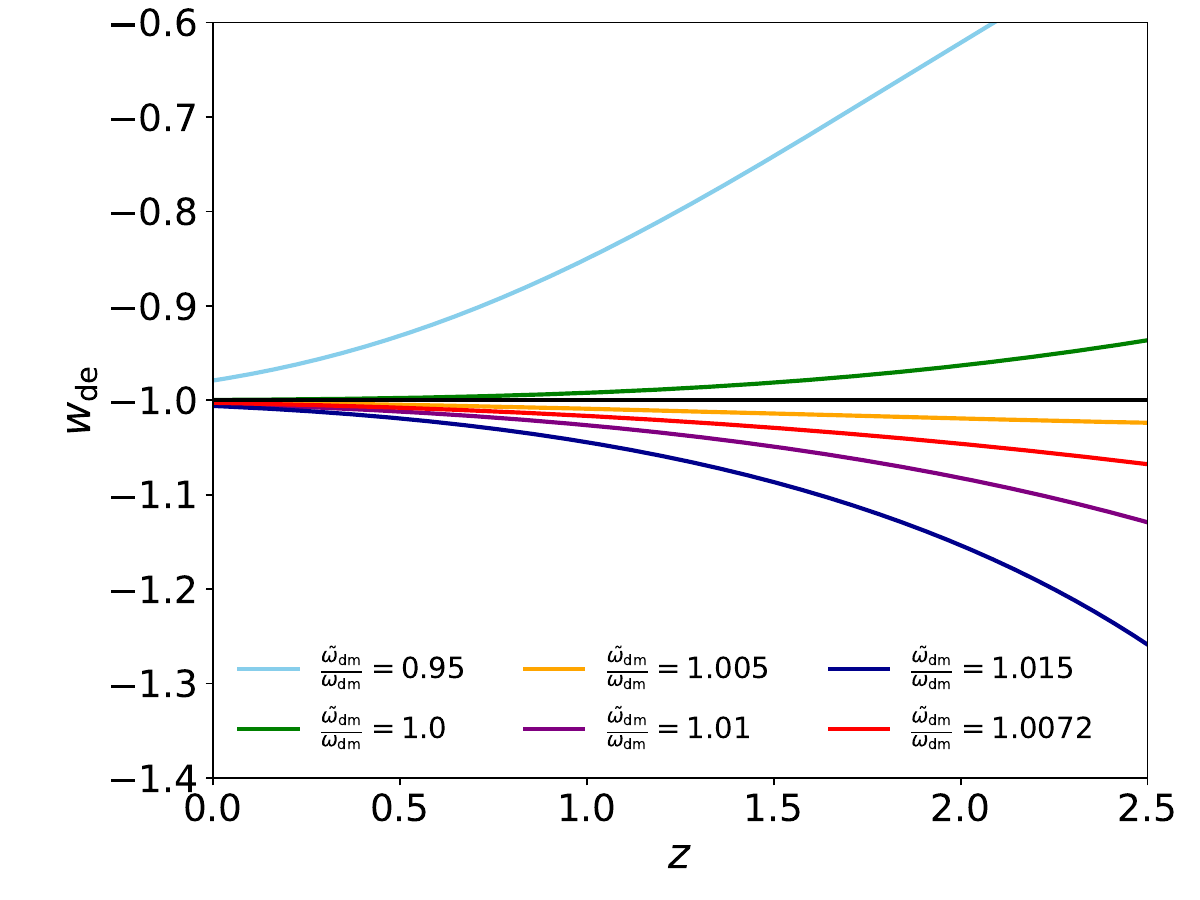}
\hfil
\includegraphics[width=0.45\textwidth]{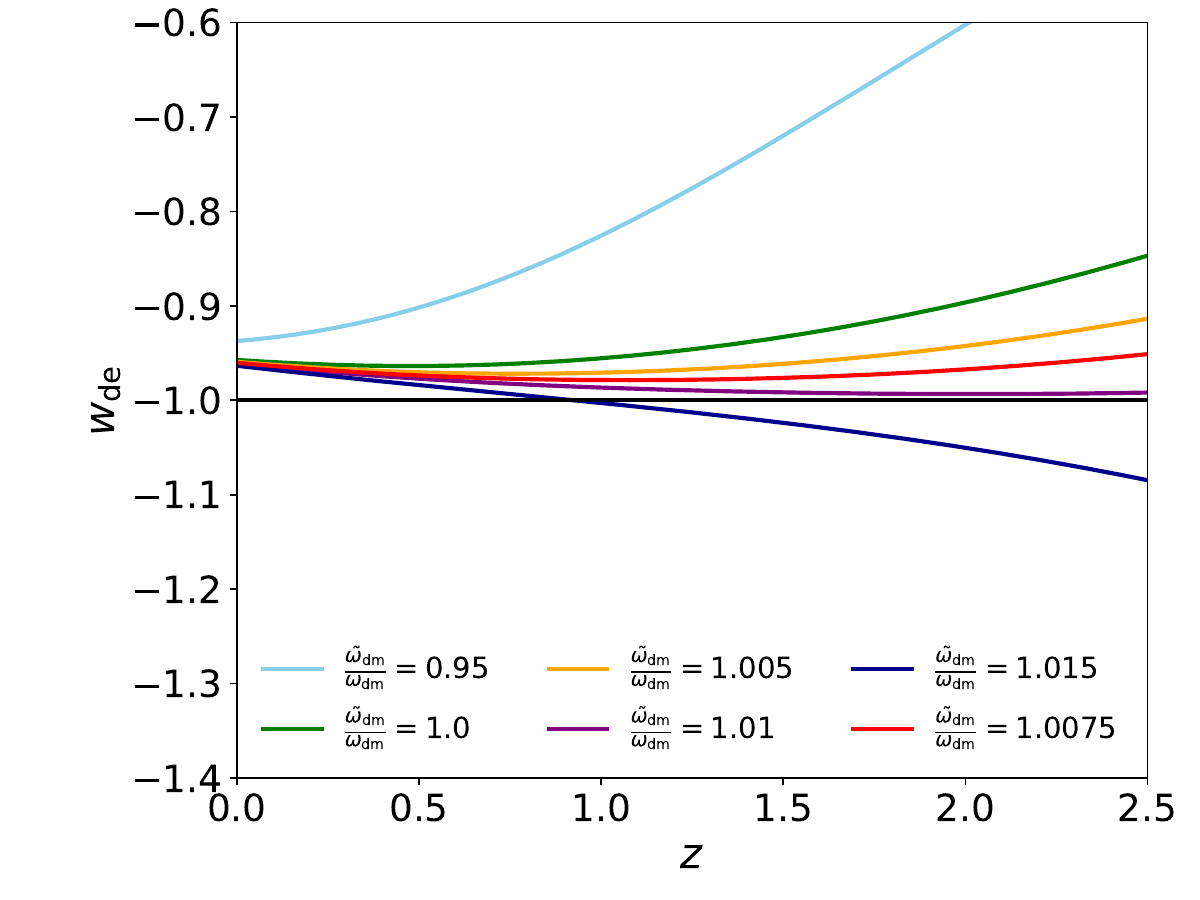}
\hfil
\includegraphics[width=0.45\textwidth]{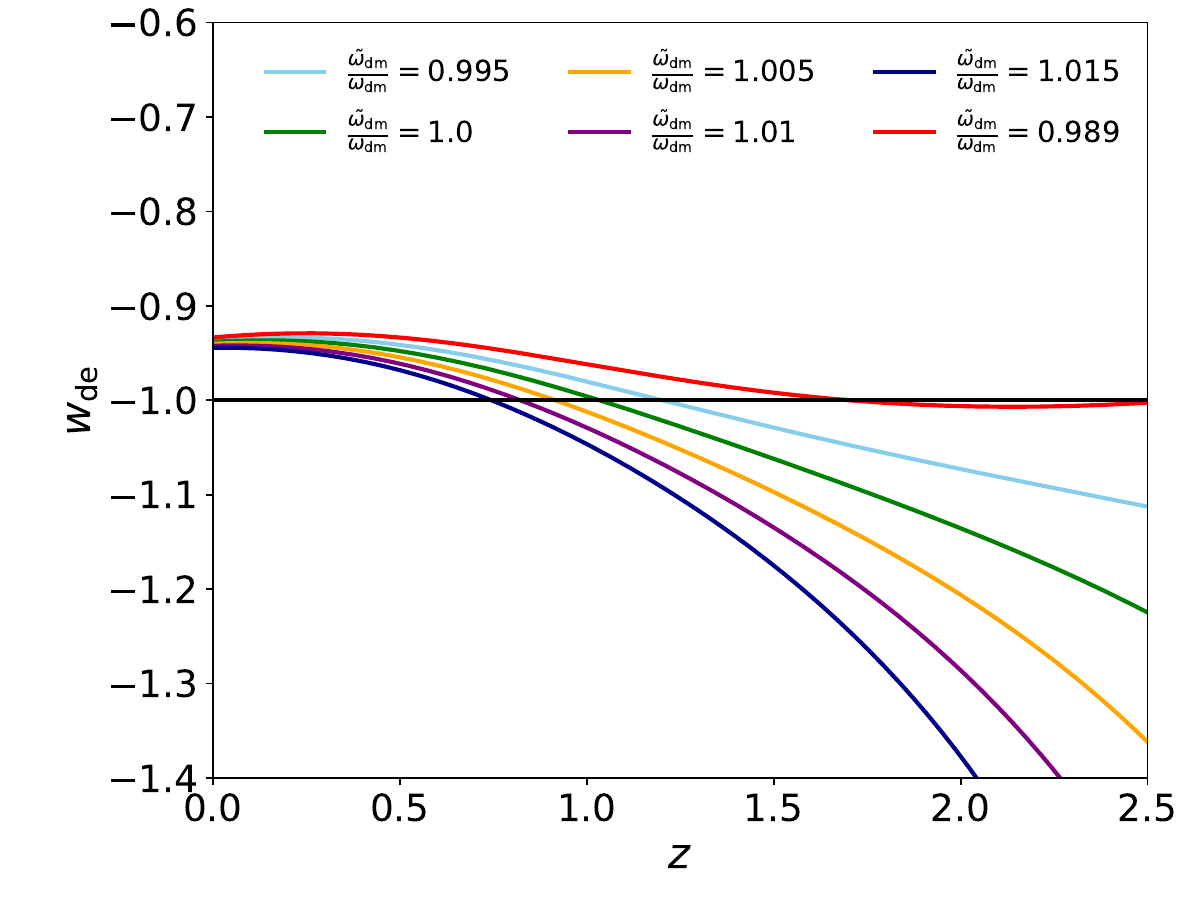}
\medskip
\caption{Evolution of the apparent (effective) dark energy EoS parameter $w_{\rm de}$ in the redshift range $0\leq z\leq 2.5$ within the CDE scenario. Here we show results for all three cases we considered in this work: a flat potential (upper left), a PR potential with $\beta>0$ (upper right), and a PR potential with $\beta<0$ (lower panel). Different curves correspond to different ratios of $\tilde{\omega}_{\rm dm}/\omega_{\rm dm}$. The values of $\tilde{\omega}_{\rm dm}/\omega_{\rm dm}$ taken from the data analysis are shown in red lines (cf. Table \ref{tab:BF_PP}). All parameters are fixed to their best-fit values obtained with the PlanckPR4+PantheonPlus+DESI dataset.}
    \label{fig:phantom_crossing} 
\end{figure*}

In CDE, when using the phase-space variables $x$ and $y$, with $x^{2}=\Omega_{K}$
and $y^{2}=\Omega_{P}$  the energy fraction contained in the scalar-field kinetic and potential terms, respectively, we have
\begin{equation}
w_{\phi}=\frac{x^{2}-y^{2}}{x^{2}+y^{2}}\ge-1\qquad ;\qquad \Omega_{\rm dm}\approx1-x^{2}-y^{2}\,,
\end{equation}
so we can write
\begin{equation} \label{eq: xy_w_de}
w_{\rm de}(a)=\frac{x^{2}-y^{2}}{x^{2}+y^{2}+(1-x^{2}-y^{2})\left(1-\frac{\tilde{\omega}_{\rm dm}^0}{\omega_{\rm dm}^0}e^{\beta\kappa(\phi-\phi^0)}\right)}\,.
\end{equation}
To achieve a phantom-crossing behavior, $w_{\rm de}$ must be larger than $-1$ at $z=0$ and transition to values below $-1$ at higher redshifts. Plugging in these two conditions into Eq. \eqref{eq: app_w_de}, they require:
\begin{equation} \label{eq: B5}
    \frac{\tilde{\omega}_{\rm dm}}{\omega_{\rm dm}} < 1+\frac{1-\Omega_{\rm dm}^0}{\Omega_{\rm dm}^0}\left(w_{\phi}^0+1\right) \quad \text{at} \quad z = 0\,,
\end{equation}
and
\begin{equation} \label{eq: B6}
    \frac{\tilde{\omega}_{\rm dm}}{\omega_{\rm dm}}> e^{-\beta\kappa(\phi(a)-\phi^0)}\left[1+\frac{1-\Omega_{\rm dm}(a)}{\Omega_{\rm dm}(a)}\Big(w_\phi(a)+1\Big)\right] \quad \text{at some} \quad z > 0  \,.
\end{equation}
For a flat potential, it can be found easily that in the absence of coupling, $w_{\rm de}$ would fail to exhibit phantom-crossing since $w_{\phi}\equiv-1$. In the presence of a nonzero coupling, the motion of $\phi$ is sourced only by the coupling term, with $e^{-\beta\kappa(\phi-\phi^0)}>1$. Moreover, $|\phi'|$ monotonically decrease when the dark energy starts to dominate, therefore $-1<w_{\phi}^0 < w_{\phi}(a)$. On the other hand, at $z=0$ we have $\phi'\approx\frac{\kappa\beta a^2\rho_{\rm dm}}{2\mathcal{H}}$, leading to $\frac{x^2}{y^2} \sim \mathcal{O}(\beta^2)\times\frac{(\Omega_{\rm dm}^0)^2}{\Omega^0_P}$. The small ratio then makes $w_{\phi}^0\approx-1$. By comparing Eq. \eqref{eq: B5} with Eq. \eqref{eq: B6}, we conclude that for a flat potential one \textit{cannot} expect a phantom-crossing to occur, since they would lead to opposite conditions; instead, $w_{\rm de}$ either transitions from the quintessence to the phantom regime in the past and remains phantom afterwards, or stays strictly $w_{\rm de} > -1$ throughout cosmic history.  

For a PR potential, the slope given by $\frac{\partial V}{\partial \phi}=-\frac{\alpha}{\phi}V(\phi)$ allows the scalar field to roll down the potential even in the absence of a coupling. In the presence of a coupling with $\beta>0$, the field $\phi$ is initially driven by the coupling term starting from the matter-dominated epoch, before continuing its roll in the same direction (i.e., to larger values of the field) due to the shape of the potential. Consequently, $\phi$ increases monotonically throughout the cosmic expansion, ensuring that $\beta(\phi-\phi^0)$ stays negative. However, $|\phi'|$ depends on both the coupling strength $\beta$ and the slope of the potential. For small $\beta$, it is possible that $|\phi'^0|>|\phi'(a)|$, leading to $w_\phi^0>w_\phi(a)$. In this case, the kinetic energy of the scalar field at $z=0$ is non-negligible due to the steepness of the potential, making it possible to satisfy both of the conditions in Eq. \eqref{eq: B5} and Eq. \eqref{eq: B6} simultaneously - i.e., staying above the phantom divide in the present epoch while achieving a phantom-crossing in the past. We note, however, that in this case it still requires $\frac{\tilde{\omega}_{\rm dm}}{\omega_{\rm dm}}>1$.

For a PR potential with negative $\beta$, $\phi$ is driven against the potential gradient by the negative coupling term before it begins its roll, with $\beta(\phi-\phi^0)>0$ at low redshifts. This allows $w_{\rm de}$ to become phantom even with $\frac{\tilde{\omega}_{\rm dm}}{\omega_{\rm dm}}\leq1$. We present the evolution of $w_{\rm de}$ for different CDE models in Fig. \ref{fig:phantom_crossing}.

\newpage
\section{Breakdown of $\chi^2_{\rm min}$, best-fit values and contour plots of CDE with PR potential}\label{app:breakdown_chi2}

\begin{figure}[h!]
    \centering
    \includegraphics[width=0.9\textwidth]{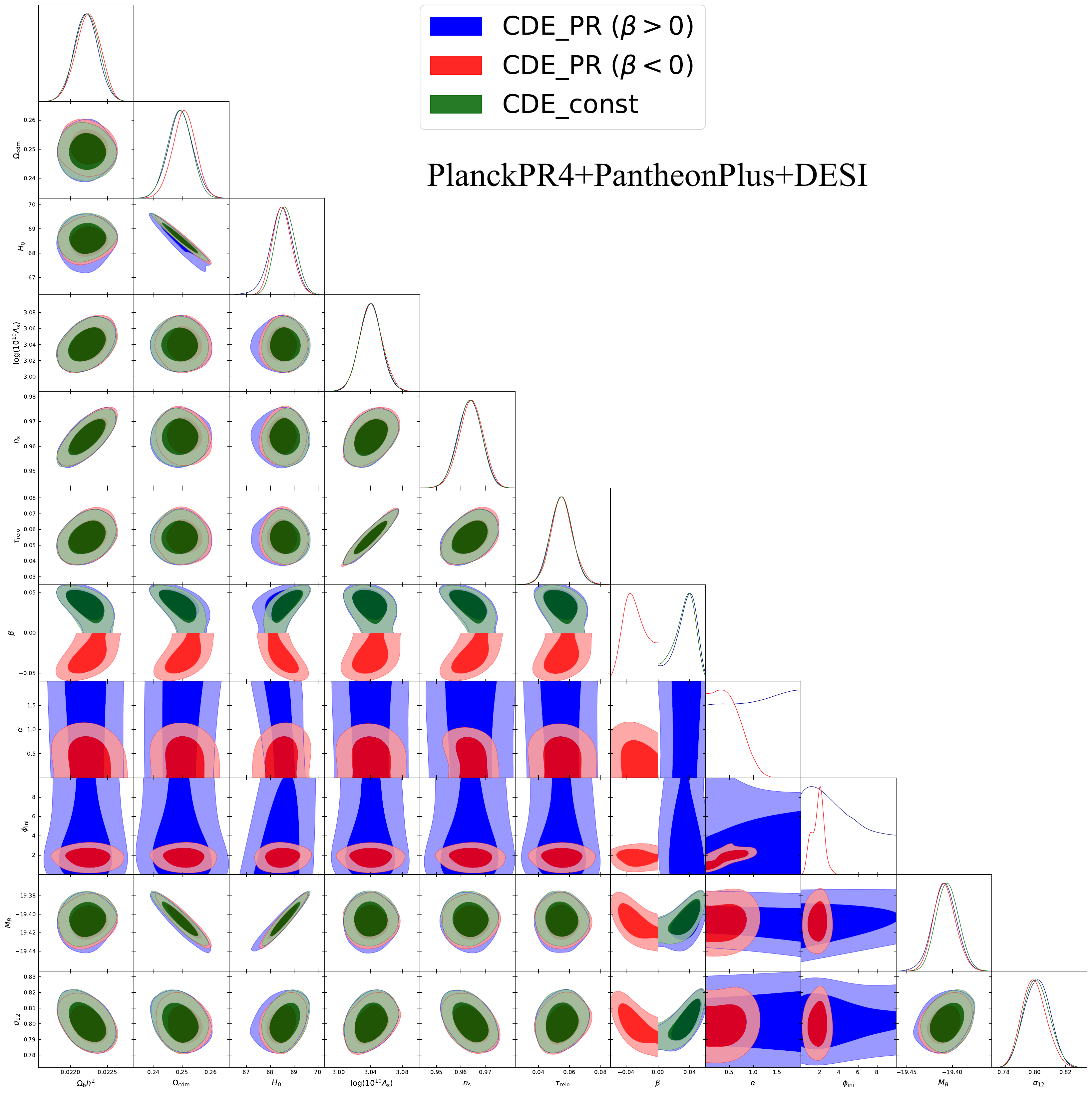}
    \caption{
     Triangle plot for the main parameters of the CDE model with constant potential and with the PR potential both with positive and negative values of the coupling $\beta$, using the data set PlanckPR4+PantheonPlus+DESI.}
 \label{fig:CP2}
\end{figure}

\begin{figure}[h!]
    \centering
    \includegraphics[width=0.92\textwidth]{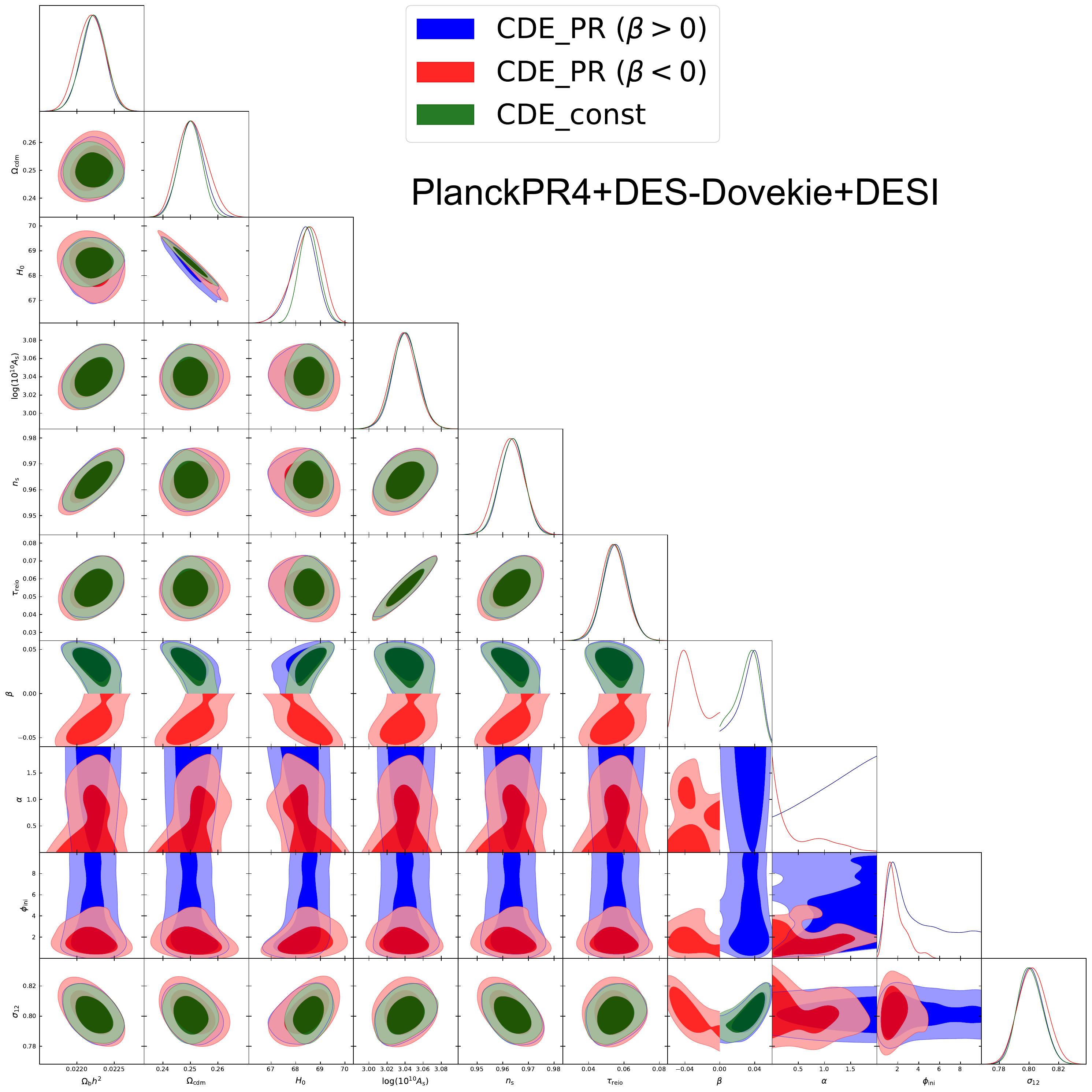}
    \caption{
     Triangle plot for the main parameters of the CDE model with constant potential and with the PR potential both with positive and negative values of the coupling $\beta$, using the data set PlanckPR4+DES-Dovekie+DESI.}
 \label{fig:CP3}
\end{figure}

\begin{table*}[h!] 
\centering
\begin{tabular}{|c ||c |c | c | c|    }
 \multicolumn{1}{c}{} & \multicolumn{1}{c}{} & \multicolumn{1}{c}{} & \multicolumn{1}{c}{} & \multicolumn{1}{c}{} \\
\multicolumn{1}{c}{} &  \multicolumn{4}{c}{PlanckPR4+PantheonPlus+DESI} 
\\\hline
{\small Dataset} & {\small $\Lambda$CDM} & {\small CDE\_const}  & {\small CDE\_PR ($\beta>0$)} & {\small CDE\_PR ($\beta<0$)} 
\\\hline
$\chi^2_{\rm min, PlanckPR4}$ & $10973.93$  & $10974.65$  & $10976.51$  & $10978.27$ \\\hline
$\chi^2_{\rm min, PantheonPlus}$ & $1405.29$ & $1406.74$ & $1403.47$ & $1402.55$ \\\hline
$\chi^2_{\rm min, DESI}$  & $13.92$    &  $10.61$  &  $9.94$  &   $9.19$      \\\hline
$\chi^2_{\rm min, total}$ & $12393.14$ & $12392.01$ & $12389.92$  &  $12390.01$
\\\hline
\end{tabular}
\caption{$\chi^2_{\rm min}$ values for each dataset from the fitting results using the combined dataset PlanckPR4+PantheonPlus+DESI, for the $\Lambda$CDM and CDE models.}
 \label{tab:chi2_Pan}
\end{table*}

\begin{table*}[h!] 
\centering
\begin{tabular}{|c ||c |c | c | c|     }
 \multicolumn{1}{c}{} & \multicolumn{1}{c}{} & \multicolumn{1}{c}{} & \multicolumn{1}{c}{} & \multicolumn{1}{c}{} \\
\multicolumn{1}{c}{} &  \multicolumn{4}{c}{PlanckPR4+DES-Dovekie+DESI} 
\\\hline
{\small Dataset} & {\small $\Lambda$CDM} & {\small CDE\_const}  & {\small CDE\_PR ($\beta>0$)} & {\small CDE\_PR ($\beta<0$)} 
\\\hline
$\chi^2_{\rm min, PlanckPR4}$ & $10976.83$  & 10973.14  & $10978.38$  & $10976.15$ \\\hline
$\chi^2_{\rm min, DES-Dovekie}$ & $1634.81$ & 1635.45  & $1630.89$ & $1631.03$ \\\hline
$\chi^2_{\rm min, DESI}$  &  $12.53$   & 12.18   &  $9.65$  & $9.31$      \\\hline
$\chi^2_{\rm min}$ & $12624.17$ & $12620.79$ & $12618.92$ & $12616.49$
\\\hline
\end{tabular}
\caption{Same as Table. \ref{tab:chi2_Pan}, but using the dataset PlanckPR4+DES-Dovekie+DESI.}\label{tab:chi2_DESDovekie}
\end{table*}

\begin{table*}[t!]
\centering
\begin{tabular}{|c ||c |c | c | c|  c|   }
 \multicolumn{1}{c}{} & \multicolumn{1}{c}{} & \multicolumn{1}{c}{} & \multicolumn{1}{c}{} & \multicolumn{1}{c}{} & \multicolumn{1}{c}{} \\
\multicolumn{1}{c}{} &  \multicolumn{4}{c}{PlanckPR4+PantheonPlus+DESI} 
\\\hline
{\small Parameter} & {\small $\Lambda$CDM} & {\small PR} & {\small CDE\_const}  & {\small CDE\_PR ($\beta>0$)} & {\small CDE\_PR ($\beta<0$)} 
\\\hline
$10^2\omega_b$ & $2.234$ & $2.236$ & $2.212$  & $2.242$ & $2.232$ \\\hline
$\omega_{\rm dm}$ & $0.1180$ & $0.1172$ & $0.1171$ & $0.1163$  & $0.1185$ \\\hline
$H_0$ & $68.07$ & $67.81$ & $68.69$ & $68.17$ & $67.75$ \\\hline
$\beta$ & $-$ & $-$ & $0.038$ &  $0.039$ & $-0.051$ \\\hline
$\alpha$ & $-$ & $0.05$ & $-$ & $1.08$ & $0.17$ \\\hline
${\varphi}_{\rm ini}$ & $-$ & $-$ & $-$ & $0.976$ & $0.381$
\\\hline
$\tilde{w}_0$ & $-1$ & $-0.985 $ & $-1.002$ & $-0.960$ & $-0.934$ \\\hline
$\tilde{w}_a$ & $0$ & $+0.010$ & $-0.005$ & $-0.034$ & $+0.034$ \\\hline
\end{tabular}
\caption{Best-fit values of the parameters controlling the background dynamics obtained with the PlanckPR4+PantheonPlus+DESI dataset. The final two rows provide the effective CPL parameters, $\tilde{w}_0\equiv w_{\rm de}(a=1)$ and $\tilde{w}_a\equiv -dw_{\rm de}/da \big|_{a=1}$, derived from the best-fit results of each respective model.}
\label{tab:BF_PP}
\end{table*}

\begin{table*}[t!]
\centering
\begin{tabular}{|c ||c |c | c | c|  c|   }
 \multicolumn{1}{c}{} & \multicolumn{1}{c}{} & \multicolumn{1}{c}{} & \multicolumn{1}{c}{} & \multicolumn{1}{c}{} & \multicolumn{1}{c}{} \\
\multicolumn{1}{c}{} &  \multicolumn{4}{c}{PlanckPR4+DES-Dovekie+DESI} 
\\\hline
{\small Parameter} & {\small $\Lambda$CDM} & {\small PR} & {\small CDE\_const}  & {\small CDE\_PR ($\beta>0$)} & {\small CDE\_PR ($\beta<0$)} 
\\\hline
$10^2\omega_b$ & $2.229$ & $2.232$ & $2.223$  & $2.236$ & $2.234$ \\\hline
$\omega_{\rm dm}$ & $0.1180$ & $0.1176$ & $0.1177$ & $0.1158$  & $0.1172$ \\\hline
$H_0$ & $68.00$ & $67.69$ & $68.47$ & $67.79$ & $67.85$ \\\hline
$\beta$ & $-$ & $-$ & $0.036$ &  $0.056$ & $-0.029$ \\\hline
$\alpha$ & $-$ & $0.04$ & $-$ & $1.65$ & $1.58$ \\\hline
${\varphi}_{\rm ini}$ & $-$ & $-$ & $-$ & $1.287$ & $1.517$
\\\hline
$\tilde{w}_0$ & $-1$ & $ -0.987$ & $-1.001$ & $-0.943 $ & $-0.942 $ \\\hline
$\tilde{w}_a$ & $0$ & $+0.009 $ & $-0.002$ & $-0.060$ & $-0.093$ \\\hline
\end{tabular}
\caption{Best-fit values of the parameters controlling the background dynamics and the effective CPL parameters obtained with the PlanckPR4+DES-Dovekie+DESI dataset.}
\label{tab:BF_DES}
\end{table*}


\section{Results obtained with DESy5}\label{app:tables_DESy5}

In this appendix, we display the fitting results obtained with PlanckPR4+DESI in combination with the SNIa from DESy5, which were affected by calibration issues that have been recently corrected in the new DES-Dovekie sample. DESy5 pointed to a larger evidence for dynamical dark energy and for the effective crossing of the phantom divide than that obtained with Pantheon+ and DES-Dovekie (cf. Tables \ref{tab:tableCDE} and \ref{tab:tableCDEII}). The signal reached the $\sim 4\sigma$ CL in the context of the CPL parametrization (see, e.g., \cite{DESI:2025zgx}). Our results of Tables \ref{tab:tableCDEIII} and \ref{tab:chi2_DESY5} fully agree with this trend. With DESy5 we find a decrease in $\chi^2$ of more than 14 units with respect to the $\Lambda$CDM and a strong evidence for a non-null coupling in the negative $\beta$ case in terms of $\Delta$AIC, which translates into an exclusion of the standard model at $3\sigma$ CL. This is consistent with the results reported in \cite{Wang:2025znm}.

\begin{table*}[h!]
\centering
\begin{tabular}{|c ||c |c | c | c|     }
 \multicolumn{1}{c}{} & \multicolumn{1}{c}{} & \multicolumn{1}{c}{} & \multicolumn{1}{c}{} & \multicolumn{1}{c}{} \\
\multicolumn{1}{c}{} &  \multicolumn{4}{c}{PlanckPR4+DESy5+DESI} 
\\\hline
{\small Parameter} & {\small $\Lambda$CDM} & {\small CDE\_const}  & {\small CDE\_PR ($\beta>0$)} & {\small CDE\_PR ($\beta<0$)} 
\\\hline
$10^2\omega_b$ & $2.233\pm 0.0014$ & $2.222^{+0.015}_{-0.016}$  & $2.224^{+0.018}_{-0.016}$ & $2.221^{+0.016}_{-0.017}$ \\\hline
$\omega_{\rm dm}$ & $0.1178\pm 0.0006$ & $0.1176\pm 0.0007$ & $0.1168\pm 0.0008$ & $0.1178\pm 0.0007$ \\\hline
$\ln(10^{10}A_s)$ & $3.045^{+0.013}_{-0.014}$ & $3.040\pm 0.014$ & $3.042\pm 0.015$ & $3.040^{+0.013}_{-0.015}$\\\hline
$n_s$ & $0.967\pm 0.004$ & $0.963^{+0.005}_{-0.004}$ & $0.964^{+0.006}_{-0.004}$  & $0.963\pm 0.004$ \\\hline
$\tau_{\rm reio}$ & $0.058^{+0.006}_{-0.007}$ & $0.055\pm 0.007$  & $0.056^{+0.007}_{-0.008}$ & $0.055^{+0.006}_{-0.008}$ \\\hline
$H_0$ & $68.09^{+0.30}_{-0.31}$ & $68.43^{+0.38}_{-0.43}$ & $67.91^{+0.59}_{-0.55}$ & $67.56^{+0.74}_{-0.71}$ \\\hline
$\beta$ & $-$ & $0.029^{+0.017}_{-0.011}$ & $0.031^{+0.016}_{-0.011}$  & $-0.042^{+0.008}_{-0.012}$ \\\hline
$\alpha$ & $-$ & $-$ & $>0.31$  & $<0.77$ \\\hline
${\varphi}_{\rm ini}$ & $-$ & $-$ & $2.70^{+0.02}_{-2.38}$ & $0.68^{+0.12}_{-0.24}$ \\\hline\hline
$r_d$ & $147.72\pm 0.22$ & $147.55^{+0.27}_{-0.24}$ & $147.59^{+0.25}_{-0.27}$ & $147.51^{+0.37}_{-0.24}$ \\\hline
$M$ & $-19.392\pm 0.008$ & $-19.383\pm 0.011$  & $-19.389\pm 0.012$ & $-19.388\pm 0.012$ \\\hline
$\sigma_{12}$ & $0.793^{+0.006}_{-0.005}$ & $0.801^{+0.007}_{-0.009}$  & $0.800^{+0.007}_{-0.009}$ & $0.802\pm 0.009$ \\\hline
$S_{8}$ & $0.808\pm 0.008$ & $0.813\pm 0.009$ & $0.812^{+0.008}_{-0.009}$  &  $0.819\pm 0.009$ \\\hline
$\chi^2_{\rm min}$ & $12637.57$ & $12635.92$ & $12629.24$ & $12623.45$
\\\hline
$\Delta{\rm AIC}$ & $-$  & $-0.35$  & $+2.33$ & $+8.12$
\\\hline
$E_{\Lambda{\rm CDM}}$ & $-$  & $1.29\sigma$  & $2.06\sigma$ & $3.00\sigma$
\\\hline
\end{tabular}
\caption{As in Tables \ref{tab:tableCDE} and \ref{tab:tableCDEII}, but employing the SNIa from DESy5 \cite{DES:2024jxu}.}
\label{tab:tableCDEIII}
\end{table*}

\begin{table*}[h!]
\centering
\begin{tabular}{|c ||c |c | c | c|     }
 \multicolumn{1}{c}{} & \multicolumn{1}{c}{} & \multicolumn{1}{c}{} & \multicolumn{1}{c}{} & \multicolumn{1}{c}{} \\
\multicolumn{1}{c}{} &  \multicolumn{4}{c}{PlanckPR4+DESy5+DESI} 
\\\hline
{\small Dataset} & {\small $\Lambda$CDM} & {\small CDE\_const}  & {\small CDE\_PR ($\beta>0$)} & {\small CDE\_PR ($\beta<0$)} 
\\\hline
$\chi^2_{\rm min, PlanckPR4}$ & $10974.37$  & $10973.68$  & $10977.01$  & $10975.65$ \\\hline
$\chi^2_{\rm min, DESy5}$ & $1648.71$ & $1651.48$ &  $1640.32$& $1639.03$ \\\hline
$\chi^2_{\rm min, DESI}$  &  $14.19$   & $10.76$   &  $11.91$  &   $8.77$      \\\hline
$\chi^2_{\rm min}$ & $12637.57$ & $12635.92$ & $12629.24$ & $12623.45$
\\\hline
\end{tabular}
\caption{Same as Tables \ref{tab:chi2_Pan} and \ref{tab:chi2_DESDovekie}, but using the dataset PlanckPR4+DESy5+DESI.}\label{tab:chi2_DESY5}
\end{table*}


\vspace{5.5cm}

\bibliographystyle{ieeetr} 
\addcontentsline{toc}{section}{References}

\bibliography{references}

@article{Gomez-Valent:2020mqn,
    author = "G{\'o}mez-Valent, Adri{\`a} and Pettorino, Valeria and Amendola, Luca",
    title = "{Update on coupled dark energy and the $H_0$ tension}",
    eprint = "2004.00610",
    archivePrefix = "arXiv",
    primaryClass = "astro-ph.CO",
    doi = "10.1103/PhysRevD.101.123513",
    journal = "Phys. Rev. D",
    volume = "101",
    number = "12",
    pages = "123513",
    year = "2020"
}

@article{Wetterich:1987fm,
    author = "Wetterich, C.",
    title = "{Cosmology and the Fate of Dilatation Symmetry}",
    eprint = "1711.03844",
    archivePrefix = "arXiv",
    primaryClass = "hep-th",
    reportNumber = "PRINT-87-0756, DESY-87-123",
    doi = "10.1016/0550-3213(88)90193-9",
    journal = "Nucl. Phys. B",
    volume = "302",
    pages = "668--696",
    year = "1988"
}

@article{DESI:2025zgx,
    author = "Abdul Karim, M. and others",
    collaboration = "DESI",
    title = "{DESI DR2 results. II. Measurements of baryon acoustic oscillations and cosmological constraints}",
    eprint = "2503.14738",
    archivePrefix = "arXiv",
    primaryClass = "astro-ph.CO",
    reportNumber = "FERMILAB-PUB-25-0169-PPD",
    doi = "10.1103/tr6y-kpc6",
    journal = "Phys. Rev. D",
    volume = "112",
    number = "8",
    pages = "083515",
    year = "2025"
}

@article{Ong:2026tta,
    author = "Ong, Dily Duan Yi and Yallup, David and Handley, Will",
    title = "{The Bayesian view of DESI DR2: Evidence and tension in a combined analysis with CMB and supernovae across cosmological models}",
    eprint = "2603.05472",
    archivePrefix = "arXiv",
    primaryClass = "astro-ph.CO",
    month = "3",
    year = "2026",
    Note = "arXiv:2603.05472"
}

@article{Chakraborty:2025syu,
    author = "Chakraborty, Amlan and Chanda, Prolay K. and Das, Subinoy and Dutta, Koushik",
    title = "{DESI results: hint towards coupled dark matter and dark energy}",
    eprint = "2503.10806",
    archivePrefix = "arXiv",
    primaryClass = "astro-ph.CO",
    reportNumber = "JCAP11(2025)047",
    doi = "10.1088/1475-7516/2025/11/047",
    journal = "JCAP",
    volume = "11",
    pages = "047",
    year = "2025"
}

@article{Chevallier:2000qy,
    author = "Chevallier, Michel and Polarski, David",
    title = "{Accelerating universes with scaling dark matter}",
    eprint = "gr-qc/0009008",
    archivePrefix = "arXiv",
    doi = "10.1142/S0218271801000822",
    journal = "Int. J. Mod. Phys. D",
    volume = "10",
    pages = "213--224",
    year = "2001"
}

@article{Wolf:2024stt,
    author = "Wolf, William J. and Ferreira, Pedro G. and Garc{\'\i}a-Garc{\'\i}a, Carlos",
    title = "{Matching current observational constraints with nonminimally coupled dark energy}",
    eprint = "2409.17019",
    archivePrefix = "arXiv",
    primaryClass = "astro-ph.CO",
    doi = "10.1103/PhysRevD.111.L041303",
    journal = "Phys. Rev. D",
    volume = "111",
    number = "4",
    pages = "L041303",
    year = "2025"
}

@article{Pan:2025psn,
    author = "Pan, Jiaming and Ye, Gen",
    title = "{Nonminimally coupled gravity constraints from DESI DR2 data}",
    eprint = "2503.19898",
    archivePrefix = "arXiv",
    primaryClass = "astro-ph.CO",
    doi = "10.1103/hqwq-m19h",
    journal = "Phys. Rev. D",
    volume = "113",
    number = "4",
    pages = "L041304",
    year = "2026"
}

@article{LaPenna:2026avs,
    author = "La Penna, Lorenzo and Notari, Alessio and Redi, Michele",
    title = "{Mimicking Phantom Dark Energy with Evolving Dark Matter Mass}",
    eprint = "2601.05235",
    archivePrefix = "arXiv",
    primaryClass = "astro-ph.CO",
    month = "1",
    year = "2026",
    Note = "arXiv:2601.05235"
}

@article{Sola:2016our,
    author = "Sol\`a, Joan and Karimkhani, Elahe and Khodam-Mohammadi, A.",
    title = "{Higgs potential from extended Brans{\textendash}Dicke theory and the time-evolution of the fundamental constants}",
    eprint = "1609.00350",
    archivePrefix = "arXiv",
    primaryClass = "gr-qc",
    doi = "10.1088/1361-6382/34/2/025006",
    journal = "Class. Quant. Grav.",
    volume = "34",
    number = "2",
    pages = "025006",
    year = "2017"
}

@article{Benisty:2024lmj,
    author = "Benisty, David and Pan, Supriya and Staicova, Denitsa and Di Valentino, Eleonora and Nunes, Rafael C.",
    title = "{Late-time constraints on interacting dark energy: Analysis independent of H0, rd, and MB}",
    eprint = "2403.00056",
    archivePrefix = "arXiv",
    primaryClass = "astro-ph.CO",
    doi = "10.1051/0004-6361/202449883",
    journal = "Astron. Astrophys.",
    volume = "688",
    pages = "A156",
    year = "2024"
}

@article{Li:2025ops,
    author = "Li, Jun-Xian and Wang, Shuang",
    title = "{Reconstructing dark energy with model independent methods after DESI DR2}",
    eprint = "2506.22953",
    archivePrefix = "arXiv",
    primaryClass = "astro-ph.CO",
    doi = "10.1140/epjc/s10052-025-15065-1",
    journal = "Eur. Phys. J. C",
    volume = "85",
    number = "11",
    pages = "1308",
    year = "2025"
}

@article{Afroz:2025iwo,
    author = "Afroz, Samsuzzaman and Mukherjee, Suvodip",
    title = "{Hint toward an inconsistency between BAO and supernovae datasets: The evidence of redshift evolving dark energy from DESI DR2 is absent}",
    eprint = "2504.16868",
    archivePrefix = "arXiv",
    primaryClass = "astro-ph.CO",
    doi = "10.1103/k59d-l795",
    journal = "Phys. Rev. D",
    volume = "113",
    number = "8",
    pages = "083514",
    year = "2026"
}

@article{Fritzsch:2012qc,
    author = "Fritzsch, Harald and Sol\`a, Joan",
    title = "{Matter Non-conservation in the Universe and Dynamical Dark Energy}",
    eprint = "1202.5097",
    archivePrefix = "arXiv",
    primaryClass = "hep-ph",
    doi = "10.1088/0264-9381/29/21/215002",
    journal = "Class. Quant. Grav.",
    volume = "29",
    pages = "215002",
    year = "2012"
}

@article{Tudes:2024jpg,
    author = {T{\"u}des, Bilal and Amendola, Luca},
    title = "{Non-linear cosmological perturbations for coupled dark energy}",
    eprint = "2411.06014",
    archivePrefix = "arXiv",
    primaryClass = "astro-ph.CO",
    doi = "10.1088/1475-7516/2025/09/001",
    journal = "JCAP",
    volume = "09",
    pages = "001",
    year = "2025"
}

@article{Zheng:2025owb,
    author = {Zheng, Ziyang and Jia, Hanqiong and T{\"u}des, Bilal and Chudaykin, Anton and Kunz, Martin and Amendola, Luca},
    title = "{One-loop kernels in scale-dependent Horndeski theory}",
    eprint = "2505.16767",
    archivePrefix = "arXiv",
    primaryClass = "astro-ph.CO",
    doi = "10.1088/1475-7516/2025/11/035",
    journal = "JCAP",
    volume = "11",
    pages = "035",
    year = "2025"
}

@article{Silva:2025bnn,
    author = "Silva, Emanuelly and Hartmann, Gabriel and Nunes, Rafael C.",
    title = "{One-loop power spectrum corrections in interacting dark energy cosmologies}",
    eprint = "2512.11678",
    archivePrefix = "arXiv",
    primaryClass = "astro-ph.CO",
    doi = "10.1103/mjkg-skbt",
    journal = "Phys. Rev. D",
    volume = "113",
    number = "6",
    pages = "063548",
    year = "2026"
}

@article{deCruzPerez:2025dni,
    author = "de Cruz P{\'e}rez, Javier and G{\'o}mez-Valent, Adri{\`a} and Sol{\`a} Peracaula, Joan",
    title = "{Dynamical dark energy models in light of the latest observations}",
    eprint = "2512.20616",
    archivePrefix = "arXiv",
    primaryClass = "astro-ph.CO",
    doi = "10.1103/xchk-xlk1",
    journal = "Phys. Rev. D",
    volume = "113",
    number = "8",
    pages = "083521",
    year = "2026"
}

@article{Chiba:1999wt,
    author = "Chiba, Takeshi",
    title = "{Quintessence, the gravitational constant, and gravity}",
    eprint = "gr-qc/9903094",
    archivePrefix = "arXiv",
    reportNumber = "UTAP-321",
    doi = "10.1103/PhysRevD.60.083508",
    journal = "Phys. Rev. D",
    volume = "60",
    pages = "083508",
    year = "1999"
}

@article{Bartolo:1999sq,
    author = "Bartolo, Nicola and Pietroni, Massimo",
    title = "{Scalar tensor gravity and quintessence}",
    eprint = "hep-ph/9908521",
    archivePrefix = "arXiv",
    reportNumber = "DFPD-99-TH-40",
    doi = "10.1103/PhysRevD.61.023518",
    journal = "Phys. Rev. D",
    volume = "61",
    pages = "023518",
    year = "2000"
}

@article{Holden:1999hm,
    author = "Holden, Damien J. and Wands, David",
    title = "{Selfsimilar cosmological solutions with a nonminimally coupled scalar field}",
    eprint = "gr-qc/9908026",
    archivePrefix = "arXiv",
    reportNumber = "PU-RCG-99-8",
    doi = "10.1103/PhysRevD.61.043506",
    journal = "Phys. Rev. D",
    volume = "61",
    pages = "043506",
    year = "2000"
}

@article{H0DN:2025lyy,
    author = "Casertano, Stefano and others",
    collaboration = "H0DN",
    title = "{The Local Distance Network: a community consensus report on the measurement of the Hubble constant at 1{\%} precision}",
    eprint = "2510.23823",
    archivePrefix = "arXiv",
    primaryClass = "astro-ph.CO",
    month = "10",
    year = "2025",
    Note = "arXiv:2510.23823"
}

@article{Riess:2021jrx,
    author = "Riess, Adam G. and others",
    title = "{A Comprehensive Measurement of the Local Value of the Hubble Constant with 1 km/s/Mpc Uncertainty from the Hubble Space Telescope and the SH0ES Team}",
    eprint = "2112.04510",
    archivePrefix = "arXiv",
    primaryClass = "astro-ph.CO",
    doi = "10.3847/2041-8213/ac5c5b",
    journal = "Astrophys. J. Lett.",
    volume = "934",
    number = "1",
    pages = "L7",
    year = "2022"
}

@article{Amendola:1999qq,
    author = "Amendola, Luca",
    title = "{Scaling solutions in general nonminimal coupling theories}",
    eprint = "astro-ph/9904120",
    archivePrefix = "arXiv",
    doi = "10.1103/PhysRevD.60.043501",
    journal = "Phys. Rev. D",
    volume = "60",
    pages = "043501",
    year = "1999"
}

@article{Castello:2024lhl,
    author = "Castello, Sveva and Zheng, Ziyang and Bonvin, Camille and Amendola, Luca",
    title = "{Testing the equivalence principle across the Universe: A model-independent approach with galaxy multitracing}",
    eprint = "2412.08627",
    archivePrefix = "arXiv",
    primaryClass = "astro-ph.CO",
    doi = "10.1103/1my7-zklj",
    journal = "Phys. Rev. D",
    volume = "111",
    number = "12",
    pages = "123559",
    year = "2025"
}

@article{Zheng:2025oiq,
    author = "Zheng, Ziyang and Schneider, Malte and Amendola, Luca",
    title = "{Testing the cosmological Euler equation: Viscosity, equivalence principle, and gravity beyond general relativity}",
    eprint = "2511.11554",
    archivePrefix = "arXiv",
    primaryClass = "astro-ph.CO",
    doi = "10.1103/nth3-y54x",
    journal = "Phys. Rev. D",
    volume = "113",
    number = "8",
    pages = "083520",
    year = "2026"
}

@article{Esteban:2024eli,
    author = "Esteban, Ivan and Gonzalez-Garcia, M. C. and Maltoni, Michele and Martinez-Soler, Ivan and Pinheiro, Jo{\~a}o Paulo and Schwetz, Thomas",
    title = "{NuFit-6.0: updated global analysis of three-flavor neutrino oscillations}",
    eprint = "2410.05380",
    archivePrefix = "arXiv",
    primaryClass = "hep-ph",
    reportNumber = "IFT-UAM/CSIC-24-140, YITP-SB-2024-24, IPPP/24/64, IPPP/24/64, IFT-UAM/CSIC-24-140, YITP-SB-2024-24",
    doi = "10.1007/JHEP12(2024)216",
    journal = "JHEP",
    volume = "12",
    pages = "216",
    year = "2024"
}

@article{Savastano:2019zpr,
    author = "Savastano, Stefano and Amendola, Luca and Rubio, Javier and Wetterich, Christof",
    title = "{Primordial dark matter halos from fifth forces}",
    eprint = "1906.05300",
    archivePrefix = "arXiv",
    primaryClass = "astro-ph.CO",
    reportNumber = "HIP-2019-18/TH",
    doi = "10.1103/PhysRevD.100.083518",
    journal = "Phys. Rev. D",
    volume = "100",
    number = "8",
    pages = "083518",
    year = "2019"
}

@article{Amendola:2017xhl,
    author = "Amendola, Luca and Rubio, Javier and Wetterich, Christof",
    title = "{Primordial black holes from fifth forces}",
    eprint = "1711.09915",
    archivePrefix = "arXiv",
    primaryClass = "astro-ph.CO",
    doi = "10.1103/PhysRevD.97.081302",
    journal = "Phys. Rev. D",
    volume = "97",
    number = "8",
    pages = "081302",
    year = "2018"
}

@article{Gonzalez-Fuentes:2026rgu,
    author = "Gonz{\'a}lez-Fuentes, Alex and G{\'o}mez-Valent, Adri{\`a}",
    title = "{Exploring the interplay of late-time dynamical dark energy and new physics before recombination}",
    eprint = "2603.26560",
    archivePrefix = "arXiv",
    primaryClass = "astro-ph.CO",
    month = "3",
    year = "2026",
    Note = "accepted for publication in JCAP, arXiv:2603.26560"
}

@article{DESI:2024aqx,
    author = "Calder\'on, R. and others",
    collaboration = "DESI",
    title = "{DESI 2024: reconstructing dark energy using crossing statistics with DESI DR1 BAO data}",
    eprint = "2405.04216",
    archivePrefix = "arXiv",
    primaryClass = "astro-ph.CO",
    doi = "10.1088/1475-7516/2024/10/048",
    journal = "JCAP",
    volume = "10",
    pages = "048",
    year = "2024"
}

@article{Yuan:2025twx,
    author = "Yuan, Guan-Wen and Davis, Anne-Christine and Giannotti, Maurizio and Vagnozzi, Sunny and Visinelli, Luca and Vogel, Julia K.",
    title = "{Direct detection of solar chameleons with electron recoil data from XENONnT}",
    eprint = "2511.01655",
    archivePrefix = "arXiv",
    primaryClass = "hep-ph",
    doi = "10.1103/8mpf-5s3k",
    journal = "Phys. Rev. D",
    volume = "113",
    number = "12",
    pages = "123024",
    year = "2026"
}

@article{Feleppa:2025clx,
    author = "Feleppa, Fabiano and Lambiase, Gaetano and Vagnozzi, Sunny",
    title = "{Imprints of screened dark energy on nonlocal quantum correlations}",
    eprint = "2508.18448",
    archivePrefix = "arXiv",
    primaryClass = "gr-qc",
    doi = "10.1103/y314-4x4s",
    journal = "Phys. Rev. D",
    volume = "112",
    number = "8",
    pages = "084011",
    year = "2025"
}

@article{Tsai:2021irw,
    author = "Tsai, Yu-Dai and Wu, Youjia and Vagnozzi, Sunny and Visinelli, Luca",
    title = "{Novel constraints on fifth forces and ultralight dark sector with asteroidal data}",
    eprint = "2107.04038",
    archivePrefix = "arXiv",
    primaryClass = "hep-ph",
    reportNumber = "FERMILAB-PUB-21-298-AE-T, LCTP-21-17",
    doi = "10.1088/1475-7516/2023/04/031",
    journal = "JCAP",
    volume = "04",
    pages = "031",
    year = "2023"
}

@article{Vagnozzi:2021quy,
    author = "Vagnozzi, Sunny and Visinelli, Luca and Brax, Philippe and Davis, Anne-Christine and Sakstein, Jeremy",
    title = "{Direct detection of dark energy: The XENON1T excess and future prospects}",
    eprint = "2103.15834",
    archivePrefix = "arXiv",
    primaryClass = "hep-ph",
    doi = "10.1103/PhysRevD.104.063023",
    journal = "Phys. Rev. D",
    volume = "104",
    number = "6",
    pages = "063023",
    year = "2021"
}

@article{Brax:2022olf,
    author = "Brax, Philippe and Davis, Anne-Christine and Elder, Benjamin",
    title = "{Screened scalar fields in hydrogen and muonium}",
    eprint = "2207.11633",
    archivePrefix = "arXiv",
    primaryClass = "hep-ph",
    doi = "10.1103/PhysRevD.107.044008",
    journal = "Phys. Rev. D",
    volume = "107",
    number = "4",
    pages = "044008",
    year = "2023"
}

@article{Elder:2019yyp,
    author = "Elder, Benjamin and Vardanyan, Valeri and Akrami, Yashar and Brax, Philippe and Davis, Anne-Christine and Decca, Ricardo S.",
    title = "{Classical symmetron force in Casimir experiments}",
    eprint = "1912.10015",
    archivePrefix = "arXiv",
    primaryClass = "gr-qc",
    doi = "10.1103/PhysRevD.101.064065",
    journal = "Phys. Rev. D",
    volume = "101",
    number = "6",
    pages = "064065",
    year = "2020"
}

@article{Brax:2018zfb,
    author = "Brax, Philippe and Davis, Anne-Christine and Elder, Benjamin and Wong, Leong Khim",
    title = "{Constraining screened fifth forces with the electron magnetic moment}",
    eprint = "1802.05545",
    archivePrefix = "arXiv",
    primaryClass = "hep-ph",
    doi = "10.1103/PhysRevD.97.084050",
    journal = "Phys. Rev. D",
    volume = "97",
    number = "8",
    pages = "084050",
    year = "2018"
}

@article{March:2021mqu,
    author = "March, Riccardo and Bertolami, Orfeu and Muccino, Marco and Gomes, Claudio and Dell'Agnello, Simone",
    title = "{Cassini and extra force constraints to nonminimally coupled gravity with a screening mechanism}",
    eprint = "2111.13270",
    archivePrefix = "arXiv",
    primaryClass = "gr-qc",
    doi = "10.1103/PhysRevD.105.044048",
    journal = "Phys. Rev. D",
    volume = "105",
    number = "4",
    pages = "044048",
    year = "2022"
}

@article{Amendola:1999er,
    author = "Amendola, Luca",
    title = "{Coupled quintessence}",
    eprint = "astro-ph/9908023",
    archivePrefix = "arXiv",
    doi = "10.1103/PhysRevD.62.043511",
    journal = "Phys. Rev. D",
    volume = "62",
    pages = "043511",
    year = "2000"
}

@article{Berti:2025phi,
    author = "Berti, Maria and Bellini, Emilio and Bonvin, Camille and Kunz, Martin and Viel, Matteo and Zumalac\'arregui, Miguel",
    title = "{Reconstructing the dark energy density in light of DESI BAO observations}",
    eprint = "2503.13198",
    archivePrefix = "arXiv",
    primaryClass = "astro-ph.CO",
    doi = "10.1103/dj3k-84v4",
    journal = "Phys. Rev. D",
    volume = "112",
    number = "2",
    pages = "023518",
    year = "2025"
}

@article{Jiang:2024xnu,
    author = "Jiang, Jun-Qian and Pedrotti, Davide and da Costa, Simony Santos and Vagnozzi, Sunny",
    title = "{Nonparametric late-time expansion history reconstruction and implications for the Hubble tension in light of recent DESI and type Ia supernovae data}",
    eprint = "2408.02365",
    archivePrefix = "arXiv",
    primaryClass = "astro-ph.CO",
    doi = "10.1103/PhysRevD.110.123519",
    journal = "Phys. Rev. D",
    volume = "110",
    number = "12",
    pages = "123519",
    year = "2024"
}

@article{Perrotta:1999am,
    author = "Perrotta, Francesca and Baccigalupi, Carlo and Matarrese, Sabino",
    title = "{Extended quintessence}",
    eprint = "astro-ph/9906066",
    archivePrefix = "arXiv",
    doi = "10.1103/PhysRevD.61.023507",
    journal = "Phys. Rev. D",
    volume = "61",
    pages = "023507",
    year = "1999"
}

@article{Tsai:2023zza,
    author = "Tsai, Yu-Dai and Farnocchia, Davide and Micheli, Marco and Vagnozzi, Sunny and Visinelli, Luca",
    title = "{Constraints on fifth forces and ultralight dark matter from OSIRIS-REx target asteroid Bennu}",
    eprint = "2309.13106",
    archivePrefix = "arXiv",
    primaryClass = "hep-ph",
    reportNumber = "UCI-HEP-TR-2023-04, FERMILAB-PUB-23-538-T-V",
    doi = "10.1038/s42005-024-01779-3",
    journal = "Commun. Phys.",
    volume = "7",
    number = "1",
    pages = "311",
    year = "2024"
}

@article{Uzan:1999ch,
    author = "Uzan, Jean-Philippe",
    title = "{Cosmological scaling solutions of nonminimally coupled scalar fields}",
    eprint = "gr-qc/9903004",
    archivePrefix = "arXiv",
    reportNumber = "UGVA-DPT-1998-12-1022",
    doi = "10.1103/PhysRevD.59.123510",
    journal = "Phys. Rev. D",
    volume = "59",
    pages = "123510",
    year = "1999"
}

@article{Linder:2002et,
    author = "Linder, Eric V.",
    title = "{Exploring the expansion history of the universe}",
    eprint = "astro-ph/0208512",
    archivePrefix = "arXiv",
    doi = "10.1103/PhysRevLett.90.091301",
    journal = "Phys. Rev. Lett.",
    volume = "90",
    pages = "091301",
    year = "2003"
}

@article{Gannouji:2006jm,
    author = "Gannouji, Radouane and Polarski, David and Ranquet, Andre and Starobinsky, Alexei A.",
    title = "{Scalar-Tensor Models of Normal and Phantom Dark Energy}",
    eprint = "astro-ph/0606287",
    archivePrefix = "arXiv",
    doi = "10.1088/1475-7516/2006/09/016",
    journal = "JCAP",
    volume = "09",
    pages = "016",
    year = "2006"
}

@article{Li:2026xaz,
    author = "Li, Tian-Nuo and Giar{\`e}, William and Du, Guo-Hong and Li, Yun-He and Di Valentino, Eleonora and Zhang, Jing-Fei and Zhang, Xin",
    title = "{Strong Evidence for Dark Sector Interactions}",
    eprint = "2601.07361",
    archivePrefix = "arXiv",
    primaryClass = "astro-ph.CO",
    month = "1",
    year = "2026",
    Note = "arXiv:2601.07361"
}

@article{Goh:2023mau,
    author = "Goh, Lisa W. K. and Bachs-Esteban, Joan and G{\'o}mez-Valent, Adri{\`a} and Pettorino, Valeria and Rubio, Javier",
    title = "{Observational constraints on early coupled quintessence}",
    eprint = "2308.06406",
    archivePrefix = "arXiv",
    primaryClass = "astro-ph.CO",
    doi = "10.1103/PhysRevD.109.023530",
    journal = "Phys. Rev. D",
    volume = "109",
    number = "2",
    pages = "023530",
    year = "2024"
}

@article{Goh:2022gxo,
    author = "Goh, Lisa W. K. and G{\'o}mez-Valent, Adri{\`a} and Pettorino, Valeria and Kilbinger, Martin",
    title = "{Constraining constant and tomographic coupled dark energy with low-redshift and high-redshift probes}",
    eprint = "2211.13588",
    archivePrefix = "arXiv",
    primaryClass = "astro-ph.CO",
    doi = "10.1103/PhysRevD.107.083503",
    journal = "Phys. Rev. D",
    volume = "107",
    number = "8",
    pages = "083503",
    year = "2023"
}

@article{Gomez-Valent:2022hkb,
    author = "G{\'o}mez-Valent, Adri{\`a}",
    title = "{Fast test to assess the impact of marginalization in Monte~Carlo analyses and its application to cosmology}",
    eprint = "2203.16285",
    archivePrefix = "arXiv",
    primaryClass = "astro-ph.CO",
    doi = "10.1103/PhysRevD.106.063506",
    journal = "Phys. Rev. D",
    volume = "106",
    number = "6",
    pages = "063506",
    year = "2022"
}

@article{Planck:2015bue,
    author = "Ade, P. A. R. and others",
    collaboration = "Planck",
    title = "{Planck 2015 results. XIV. Dark energy and modified gravity}",
    eprint = "1502.01590",
    archivePrefix = "arXiv",
    primaryClass = "astro-ph.CO",
    doi = "10.1051/0004-6361/201525814",
    journal = "Astron. Astrophys.",
    volume = "594",
    pages = "A14",
    year = "2016"
}

@article{Bansal:2026axl,
    author = "Bansal, Prakhar and Huterer, Dragan",
    title = "{On the Difficulties with Late-Time Solutions for the Hubble Tension}",
    eprint = "2602.06293",
    archivePrefix = "arXiv",
    primaryClass = "astro-ph.CO",
    month = "2",
    year = "2026", 
    Note = "arXiv:2602.06293"
}

@article{DES:2021wwk,
    author = "Abbott, T. M. C. and others",
    collaboration = "DES",
    title = "{Dark Energy Survey Year 3 results: Cosmological constraints from galaxy clustering and weak lensing}",
    eprint = "2105.13549",
    archivePrefix = "arXiv",
    primaryClass = "astro-ph.CO",
    reportNumber = "FERMILAB-PUB-21-221-AE, DES-2020-0617",
    doi = "10.1103/PhysRevD.105.023520",
    journal = "Phys. Rev. D",
    volume = "105",
    number = "2",
    pages = "023520",
    year = "2022"
}

@article{Camarena:2025upt,
    author = "Camarena, David and Greene, Kylar and Houghteling, John and Cyr-Racine, Francis-Yan",
    title = "{Designing concordant distances in the age of precision cosmology: The impact of density fluctuations}",
    eprint = "2507.17969",
    archivePrefix = "arXiv",
    primaryClass = "astro-ph.CO",
    doi = "10.1103/1gj5-3x4m",
    journal = "Phys. Rev. D",
    volume = "112",
    number = "8",
    pages = "083526",
    year = "2025"
}

@article{Miyatake:2023njf,
    author = "Miyatake, Hironao and others",
    title = "{Hyper Suprime-Cam Year 3 results: Cosmology from galaxy clustering and weak lensing with HSC and SDSS using the emulator based halo model}",
    eprint = "2304.00704",
    archivePrefix = "arXiv",
    primaryClass = "astro-ph.CO",
    doi = "10.1103/PhysRevD.108.123517",
    journal = "Phys. Rev. D",
    volume = "108",
    number = "12",
    pages = "123517",
    year = "2023"
}

@article{Wright:2025xka,
    author = "Wright, Angus H. and others",
    title = "{KiDS-Legacy: Cosmological constraints from cosmic shear with the complete Kilo-Degree Survey}",
    eprint = "2503.19441",
    archivePrefix = "arXiv",
    primaryClass = "astro-ph.CO",
    doi = "10.1051/0004-6361/202554908",
    journal = "Astron. Astrophys.",
    volume = "703",
    pages = "A158",
    year = "2025"
}

@article{Pedrotti:2025ccw,
    author = "Pedrotti, Davide and Escamilla, Luis A. and Marra, Valerio and Perivolaropoulos, Leandros and Vagnozzi, Sunny",
    title = "{BAO miscalibration cannot rescue late-time solutions to the Hubble tension}",
    eprint = "2510.01974",
    archivePrefix = "arXiv",
    primaryClass = "astro-ph.CO",
    doi = "10.1103/pn9j-8whx",
    journal = "Phys. Rev. D",
    volume = "113",
    number = "4",
    pages = "043507",
    year = "2026"
}

@article{Gomez-Valent:2023uof,
    author = "G{\'o}mez-Valent, Adri{\`a} and Favale, Arianna and Migliaccio, Marina and Sen, Anjan A.",
    title = "{Late-time phenomenology required to solve the H0 tension in view of the cosmic ladders and the anisotropic and angular BAO datasets}",
    eprint = "2309.07795",
    archivePrefix = "arXiv",
    primaryClass = "astro-ph.CO",
    doi = "10.1103/PhysRevD.109.023525",
    journal = "Phys. Rev. D",
    volume = "109",
    number = "2",
    pages = "023525",
    year = "2024"
}

@article{Lee:2022cyh,
    author = "Lee, Bum-Hoon and Lee, Wonwoo and Colg{\'a}in, Eoin {\'O}. and Sheikh-Jabbari, M. M. and Thakur, Somyadip",
    title = "{Is local H $_{0}$ at odds with dark energy EFT?}",
    eprint = "2202.03906",
    archivePrefix = "arXiv",
    primaryClass = "astro-ph.CO",
    doi = "10.1088/1475-7516/2022/04/004",
    journal = "JCAP",
    volume = "04",
    number = "04",
    pages = "004",
    year = "2022"
}

@article{Krishnan:2021dyb,
    author = "Krishnan, Chethan and Mohayaee, Roya and Colg{\'a}in, Eoin {\'O}. and Sheikh-Jabbari, M. M. and Yin, Lu",
    title = "{Does Hubble tension signal a breakdown in FLRW cosmology?}",
    eprint = "2105.09790",
    archivePrefix = "arXiv",
    primaryClass = "astro-ph.CO",
    doi = "10.1088/1361-6382/ac1a81",
    journal = "Class. Quant. Grav.",
    volume = "38",
    number = "18",
    pages = "184001",
    year = "2021"
}

@article{Sola:2017znb,
    author = "Sol{\`a}, Joan and G{\'o}mez-Valent, Adri{\`a} and de Cruz P{\'e}rez, Javier",
    title = "{The $H_0$ tension in light of vacuum dynamics in the Universe}",
    eprint = "1705.06723",
    archivePrefix = "arXiv",
    primaryClass = "astro-ph.CO",
    doi = "10.1016/j.physletb.2017.09.073",
    journal = "Phys. Lett. B",
    volume = "774",
    pages = "317--324",
    year = "2017"
}

@article{Knox:2019rjx,
    author = "Knox, Lloyd and Millea, Marius",
    title = "{Hubble constant hunter{\textquoteright}s guide}",
    eprint = "1908.03663",
    archivePrefix = "arXiv",
    primaryClass = "astro-ph.CO",
    doi = "10.1103/PhysRevD.101.043533",
    journal = "Phys. Rev. D",
    volume = "101",
    number = "4",
    pages = "043533",
    year = "2020"
}

@article{Freedman:2024eph,
    author = "Freedman, Wendy L. and Madore, Barry F. and Hoyt, Taylor J. and Jang, In Sung and Lee, Abigail J. and Owens, Kayla A.",
    title = "{Status Report on the Chicago-Carnegie Hubble Program (CCHP): Measurement of the Hubble Constant Using the Hubble and James Webb Space Telescopes}",
    eprint = "2408.06153",
    archivePrefix = "arXiv",
    primaryClass = "astro-ph.CO",
    doi = "10.3847/1538-4357/adce78",
    journal = "Astrophys. J.",
    volume = "985",
    number = "2",
    pages = "203",
    year = "2025",
    note = "[Erratum: Astrophys.J. 993, 252 (2025)]"
}

@article{Pettorino:2013oxa,
    author = "Pettorino, Valeria",
    title = "{Testing modified gravity with Planck: the case of coupled dark energy}",
    eprint = "1305.7457",
    archivePrefix = "arXiv",
    primaryClass = "astro-ph.CO",
    doi = "10.1103/PhysRevD.88.063519",
    journal = "Phys. Rev. D",
    volume = "88",
    pages = "063519",
    year = "2013"
}

@article{Pettorino:2012ts,
    author = "Pettorino, Valeria and Amendola, Luca and Baccigalupi, Carlo and Quercellini, Claudia",
    title = "{Constraints on coupled dark energy using CMB data from WMAP and SPT}",
    eprint = "1207.3293",
    archivePrefix = "arXiv",
    primaryClass = "astro-ph.CO",
    doi = "10.1103/PhysRevD.86.103507",
    journal = "Phys. Rev. D",
    volume = "86",
    pages = "103507",
    year = "2012"
}

@article{Keeley:2025rlg,
    author = "Keeley, Ryan E. and Shafieloo, Arman and Matthewson, William L.",
    title = "{Could We Be Fooled about Phantom Crossing?}",
    eprint = "2506.15091",
    archivePrefix = "arXiv",
    primaryClass = "astro-ph.CO",
    month = "6",
    year = "2025",
    Note= "arXiv:2506.15091"
}

@article{Blas:2011rf,
    author = "Blas, Diego and Lesgourgues, Julien and Tram, Thomas",
    title = "{The Cosmic Linear Anisotropy Solving System (CLASS) II: Approximation schemes}",
    eprint = "1104.2933",
    archivePrefix = "arXiv",
    primaryClass = "astro-ph.CO",
    reportNumber = "CERN-PH-TH-2011-082, LAPTH-010-11",
    doi = "10.1088/1475-7516/2011/07/034",
    journal = "JCAP",
    volume = "07",
    pages = "034",
    year = "2011"
}

@article{Torrado:2020dgo,
    author = "Torrado, Jesus and Lewis, Antony",
    title = "{Cobaya: Code for Bayesian Analysis of hierarchical physical models}",
    eprint = "2005.05290",
    archivePrefix = "arXiv",
    primaryClass = "astro-ph.IM",
    reportNumber = "TTK-20-15",
    doi = "10.1088/1475-7516/2021/05/057",
    journal = "JCAP",
    volume = "05",
    pages = "057",
    year = "2021"
}

@article{Lewis:2019xzd,
    author = "Lewis, Antony",
    title = "{GetDist: a Python package for analysing Monte Carlo samples}",
    eprint = "1910.13970",
    archivePrefix = "arXiv",
    primaryClass = "astro-ph.IM",
    doi = "10.1088/1475-7516/2025/08/025",
    journal = "JCAP",
    volume = "08",
    pages = "025",
    year = "2025"
}

@article{Lesgourgues:2011re,
    author = "Lesgourgues, Julien",
    title = "{The Cosmic Linear Anisotropy Solving System (CLASS) I: Overview}",
    eprint = "1104.2932",
    archivePrefix = "arXiv",
    primaryClass = "astro-ph.IM",
    month = "4",
    year = "2011",
    Note = "arXiv:1104.2932"
}

@article{Barros:2018efl,
    author = "Barros, Bruno J. and Amendola, Luca and Barreiro, Tiago and Nunes, Nelson J.",
    title = "{Coupled quintessence with a $\Lambda$CDM background: removing the $\sigma_8$ tension}",
    eprint = "1802.09216",
    archivePrefix = "arXiv",
    primaryClass = "astro-ph.CO",
    doi = "10.1088/1475-7516/2019/01/007",
    journal = "JCAP",
    volume = "01",
    pages = "007",
    year = "2019"
}

@article{Cortes:2025joz,
    author = "Cort{\^e}s, Marina and Liddle, Andrew R.",
    title = "{On DESI's DR2 exclusion of LCDM}",
    eprint = "2504.15336",
    archivePrefix = "arXiv",
    primaryClass = "astro-ph.CO",
    doi = "10.1093/mnrasl/slaf108",
    journal = "Mon. Not. Roy. Astron. Soc.",
    volume = "544",
    pages = "L121--L125",
    year = "2025"
}

@article{Gomez-Valent:2024tdb,
    author = "G\'omez-Valent, Adri\`a and Sol{\`a} Peracaula, Joan",
    title = "{Phantom Matter: A Challenging Solution to the Cosmological Tensions}",
    eprint = "2404.18845",
    archivePrefix = "arXiv",
    primaryClass = "astro-ph.CO",
    doi = "10.3847/1538-4357/ad7a62",
    journal = "Astrophys. J.",
    volume = "975",
    number = "1",
    pages = "64",
    year = "2024"
}

@article{Gomez-Valent:2024ejh,
    author = "G{\'o}mez-Valent, Adria and Sol{\`a} Peracaula, Joan",
    title = "{Composite dark energy and the cosmological tensions}",
    eprint = "2412.15124",
    archivePrefix = "arXiv",
    primaryClass = "astro-ph.CO",
    doi = "10.1016/j.physletb.2025.139391",
    journal = "Phys. Lett. B",
    volume = "864",
    pages = "139391",
    year = "2025"
}

@article{Efstratiou:2025iqi,
    author = "Efstratiou, Dimitrios and Paraskevas, Evangelos Achilleas and Perivolaropoulos, Leandros",
    title = "{Addressing the DESI DR2 phantom-crossing anomaly and enhanced H0 tension with reconstructed scalar-tensor gravity}",
    eprint = "2511.04610",
    archivePrefix = "arXiv",
    primaryClass = "astro-ph.CO",
    doi = "10.1103/zdcg-4sdf",
    journal = "Phys. Rev. D",
    volume = "113",
    number = "12",
    pages = "123525",
    year = "2026"
}

@article{Wolf:2025acj,
    author = "Wolf, William J. and Ferreira, Pedro G. and Garc{\'\i}a-Garc{\'\i}a, Carlos",
    title = "{Cosmological constraints on Galileon dark energy with broken shift symmetry}",
    eprint = "2509.17586",
    archivePrefix = "arXiv",
    primaryClass = "astro-ph.CO",
    doi = "10.1103/bxvj-bsv1",
    journal = "Phys. Rev. D",
    volume = "113",
    number = "2",
    pages = "023551",
    year = "2026"
}

@article{Yao:2025wlx,
    author = "Yao, Zhibang and Ye, Gen and Silvestri, Alessandra",
    title = "{A general model for dark energy crossing the phantom divide}",
    eprint = "2508.01378",
    archivePrefix = "arXiv",
    primaryClass = "gr-qc",
    doi = "10.1088/1475-7516/2025/10/078",
    journal = "JCAP",
    volume = "10",
    pages = "078",
    year = "2025"
}

@article{Tsujikawa:2025wca,
    author = "Tsujikawa, Shinji",
    title = "{Crossing the phantom divide in scalar-tensor and vector-tensor theories}",
    eprint = "2508.17231",
    archivePrefix = "arXiv",
    primaryClass = "astro-ph.CO",
    reportNumber = "WUCG-25-09",
    doi = "10.1103/y858-4swl",
    journal = "Phys. Rev. D",
    volume = "113",
    number = "4",
    pages = "L041301",
    year = "2026"
}

@article{Braglia:2025gdo,
    author = "Braglia, Matteo and Chen, Xingang and Loeb, Abraham",
    title = "{Exotic dark matter and the DESI anomaly}",
    eprint = "2507.13925",
    archivePrefix = "arXiv",
    primaryClass = "astro-ph.CO",
    doi = "10.1088/1475-7516/2025/11/064",
    journal = "JCAP",
    volume = "11",
    pages = "064",
    year = "2025"
}

@article{Odintsov:2024woi,
    author = "Odintsov, Sergei D. and S{\'a}ez-Chill{\'o}n G{\'o}mez, Diego and Sharov, German S.",
    title = "{Modified gravity/dynamical dark energy vs $\Lambda $CDM: is the game over?}",
    eprint = "2412.09409",
    archivePrefix = "arXiv",
    primaryClass = "gr-qc",
    doi = "10.1140/epjc/s10052-025-14013-3",
    journal = "Eur. Phys. J. C",
    volume = "85",
    number = "3",
    pages = "298",
    year = "2025"
}

@article{Mishra:2025goj,
    author = "Mishra, Swagat S. and Matthewson, William L. and Sahni, Varun and Shafieloo, Arman and Shtanov, Yuri",
    title = "{Braneworld dark energy in light of DESI~DR2}",
    eprint = "2507.07193",
    archivePrefix = "arXiv",
    primaryClass = "astro-ph.CO",
    doi = "10.1088/1475-7516/2025/11/018",
    journal = "JCAP",
    volume = "11",
    pages = "018",
    year = "2025"
}

@article{Nojiri:2025low,
    author = "Nojiri, Shin'ichi and Odintsov, S. D. and Oikonomou, V. K.",
    title = "{Phantom crossing and oscillating dark energy with F(R) gravity}",
    eprint = "2506.21010",
    archivePrefix = "arXiv",
    primaryClass = "gr-qc",
    reportNumber = "KEK-TH-2734, KEK-Cosmo-0383",
    doi = "10.1103/16yg-966k",
    journal = "Phys. Rev. D",
    volume = "112",
    number = "10",
    pages = "104035",
    year = "2025"
}

@article{Giani:2025hhs,
    author = "Giani, Leonardo and Von Marttens, Rodrigo and Piattella, Oliver Fabio",
    title = "{The matter with(in) CPL}",
    eprint = "2505.08467",
    archivePrefix = "arXiv",
    primaryClass = "astro-ph.CO",
    doi = "10.33232/001c.142699",
    month = "5",
    year = "2025",
    Note = "arXiv:2505.08467"
}

@article{Cai:2025mas,
    author = "Cai, Yifu and Ren, Xin and Qiu, Taotao and Li, Mingzhe and Zhang, Xinmin",
    title = "{The Quintom theory of dark energy after DESI DR2}",
    eprint = "2505.24732",
    archivePrefix = "arXiv",
    primaryClass = "astro-ph.CO",
    doi = "10.1093/nsr/nwag115",
    month = "5",
    year = "2025",
    Note = "arXiv: 2505.24732"
}

@article{Chen:2025wwn,
    author = "Chen, Xingang and Loeb, Abraham",
    title = "{Evolving dark energy or dark matter with an evolving equation-of-state?}",
    eprint = "2505.02645",
    archivePrefix = "arXiv",
    primaryClass = "astro-ph.CO",
    doi = "10.1088/1475-7516/2025/07/059",
    journal = "JCAP",
    volume = "07",
    pages = "059",
    year = "2025"
}

@article{Yang:2025mws,
    author = "Yang, Yuhang and Wang, Qingqing and Ren, Xin and Saridakis, Emmanuel N. and Cai, Yi-Fu",
    title = "{Modified Gravity Realizations of Quintom Dark Energy after DESI DR2}",
    eprint = "2504.06784",
    archivePrefix = "arXiv",
    primaryClass = "astro-ph.CO",
    doi = "10.3847/1538-4357/ade43f",
    journal = "Astrophys. J.",
    volume = "988",
    number = "1",
    pages = "123",
    year = "2025"
}

@article{Wolf:2025jed,
    author = "Wolf, William J. and Garc{\'\i}a-Garc{\'\i}a, Carlos and Anton, Theodore and Ferreira, Pedro G.",
    title = "{Assessing Cosmological Evidence for Nonminimal Coupling}",
    eprint = "2504.07679",
    archivePrefix = "arXiv",
    primaryClass = "astro-ph.CO",
    doi = "10.1103/jysf-k72m",
    journal = "Phys. Rev. Lett.",
    volume = "135",
    number = "8",
    pages = "081001",
    year = "2025"
}

@article{Akarsu:2026anp,
    author = {Akarsu, {\"O}zg{\"u}r and Caruana, Maria and Dialektopoulos, Konstantinos F. and Escamilla, Luis A. and Kahya, Emre O. and Levi Said, Jackson},
    title = "{Hints of sign-changing scalar field energy density and a transient acceleration phase at $z\sim 2$ from model-agnostic reconstructions}",
    eprint = "2602.08928",
    archivePrefix = "arXiv",
    primaryClass = "astro-ph.CO",
    month = "2",
    year = "2026",
    Note = "arXiv:2602.08928"
}

@article{Fardon:2003eh,
    author = "Fardon, Rob and Nelson, Ann E. and Weiner, Neal",
    title = "{Dark energy from mass varying neutrinos}",
    eprint = "astro-ph/0309800",
    archivePrefix = "arXiv",
    reportNumber = "UW-PT-03-22",
    doi = "10.1088/1475-7516/2004/10/005",
    journal = "JCAP",
    volume = "10",
    pages = "005",
    year = "2004"
}

@article{Cheng:2025yue,
    author = "Cheng, Hanyu and Pan, Supriya and Di Valentino, Eleonora",
    title = "{Beyond Two Parameters: Revisiting Dark Energy with the Latest Cosmic Probes}",
    eprint = "2512.09866",
    archivePrefix = "arXiv",
    primaryClass = "astro-ph.CO",
    doi = "10.3847/1538-4357/ae3a8f",
    journal = "Astrophys. J.",
    volume = "999",
    number = "2",
    pages = "190",
    year = "2026"
}

@article{Artola:2025zzb,
    author = "Artola, Mikel and Ayuso, Ismael and Lazkoz, Ruth and Salzano, Vincenzo",
    title = "{Is Chevallier-Polarski-Linder dark energy a mirage?}",
    eprint = "2510.04191",
    archivePrefix = "arXiv",
    primaryClass = "astro-ph.CO",
    doi = "10.1103/rz3s-zz61",
    journal = "Phys. Rev. D",
    volume = "113",
    number = "2",
    pages = "023513",
    year = "2026"
}

@article{Ibarra-Uriondo:2026zbp,
    author = "Ibarra-Uriondo, Be{\~n}at and Bouhmadi-L{\'o}pez, Mariam",
    title = "{Sign-Switching Dark Energy: Smooth Transitions with Recent {\textbackslash}textit{DESI DR2} Observations}",
    eprint = "2602.12347",
    archivePrefix = "arXiv",
    primaryClass = "astro-ph.CO",
    month = "2",
    year = "2026",
    Note ="arXiv:2602.12347"
}

@article{Park:2026iqa,
    author = "Park, Chan-Gyung and de Cruz P\'erez, Javier and Ratra, Bharat",
    title = "{Is the $w_0w_a$CDM cosmological parameterization evidence for dark energy dynamics partially caused by the excess smoothing of Planck PR4 CMB anisotropy data?}",
    eprint = "2604.03756",
    archivePrefix = "arXiv",
    primaryClass = "astro-ph.CO",
    month = "4",
    year = "2026",
    Note = "arXiv:2604.03756"
}

@article{Khoury:2025txd,
    author = "Khoury, Justin and Lin, Meng-Xiang and Trodden, Mark",
    title = "{Apparent w{\ensuremath{<}}-1 and a Lower S8 from Dark Axion and Dark Baryons Interactions}",
    eprint = "2503.16415",
    archivePrefix = "arXiv",
    primaryClass = "astro-ph.CO",
    doi = "10.1103/w4qb-plk8",
    journal = "Phys. Rev. Lett.",
    volume = "135",
    number = "18",
    pages = "181001",
    year = "2025"
}

@article{Giare:2025pzu,
    author = "Giar{\`e}, William and Mahassen, Tariq and Di Valentino, Eleonora and Pan, Supriya",
    title = "{An overview of what current data can (and cannot yet) say about evolving dark energy}",
    eprint = "2502.10264",
    archivePrefix = "arXiv",
    primaryClass = "astro-ph.CO",
    doi = "10.1016/j.dark.2025.101906",
    journal = "Phys. Dark Univ.",
    volume = "48",
    pages = "101906",
    year = "2025"
}

@article{Wolf:2024eph,
    author = "Wolf, William J. and Garc{\'\i}a-Garc{\'\i}a, Carlos and Bartlett, Deaglan J. and Ferreira, Pedro G.",
    title = "{Scant evidence for thawing quintessence}",
    eprint = "2408.17318",
    archivePrefix = "arXiv",
    primaryClass = "astro-ph.CO",
    doi = "10.1103/PhysRevD.110.083528",
    journal = "Phys. Rev. D",
    volume = "110",
    number = "8",
    pages = "083528",
    year = "2024"
}

@article{Ye:2024ywg,
    author = "Ye, Gen and Martinelli, Matteo and Hu, Bin and Silvestri, Alessandra",
    title = "{Hints of Nonminimally Coupled Gravity in DESI 2024 Baryon Acoustic Oscillation Measurements}",
    eprint = "2407.15832",
    archivePrefix = "arXiv",
    primaryClass = "astro-ph.CO",
    doi = "10.1103/PhysRevLett.134.181002",
    journal = "Phys. Rev. Lett.",
    volume = "134",
    number = "18",
    pages = "181002",
    year = "2025"
}

@article{Gomez-Valent:2025mfl,
    author = "G{\'o}mez-Valent, Adri{\`a} and Gonz{\'a}lez-Fuentes, Alex",
    title = "{Effective phantom divide crossing with standard and negative quintessence}",
    eprint = "2508.00621",
    archivePrefix = "arXiv",
    primaryClass = "astro-ph.CO",
    doi = "10.1016/j.physletb.2025.140096",
    journal = "Phys. Lett. B",
    volume = "872",
    pages = "140096",
    year = "2026"
}

@article{Peebles:1987ek,
    author = "Peebles, P. J. E. and Ratra, Bharat",
    title = "{Cosmology with a Time Variable Cosmological Constant}",
    reportNumber = "PUPT-1069",
    doi = "10.1086/185100",
    journal = "Astrophys. J. Lett.",
    volume = "325",
    pages = "L17",
    year = "1988"
}

@article{Ratra:1987rm,
    author = "Ratra, Bharat and Peebles, P. J. E.",
    title = "{Cosmological Consequences of a Rolling Homogeneous Scalar Field}",
    reportNumber = "PUPT-1072",
    doi = "10.1103/PhysRevD.37.3406",
    journal = "Phys. Rev. D",
    volume = "37",
    pages = "3406",
    year = "1988"
}

@article{SolaPeracaula:2018wwm,
    author = "Sol\`a Peracaula, Joan and G\'omez-Valent, Adri\`a and de Cruz P{\'e}rez, Javier",
    title = "{Signs of Dynamical Dark Energy in Current Observations}",
    eprint = "1811.03505",
    archivePrefix = "arXiv",
    primaryClass = "astro-ph.CO",
    doi = "10.1016/j.dark.2019.100311",
    journal = "Phys. Dark Univ.",
    volume = "25",
    pages = "100311",
    year = "2019"
}

@article{Goh:2025upc,
    author = "Goh, L. W. K. and Taylor, A. N.",
    title = "{Phantom Crossing with Quintom Models}",
    eprint = "2509.12335",
    archivePrefix = "arXiv",
    primaryClass = "astro-ph.CO",
    doi = "10.1093/mnras/staf1927",
    journal = "Mon. Not. Roy. Astron. Soc.",
    volume = "3142",
    pages = "3157",
    year = "2025"
}

@article{DESI:2025fii,
    author = "Lodha, K. and others",
    collaboration = "DESI",
    title = "{Extended dark energy analysis using DESI DR2 BAO measurements}",
    eprint = "2503.14743",
    archivePrefix = "arXiv",
    primaryClass = "astro-ph.CO",
    reportNumber = "FERMILAB-PUB-25-0164-PPD",
    doi = "10.1103/w4c6-1r5j",
    journal = "Phys. Rev. D",
    volume = "112",
    number = "8",
    pages = "083511",
    year = "2025"
}

@article{Chakraborty:2024xas,
    author = "Chakraborty, Amlan and Ray, Tulip and Das, Subinoy and Banerjee, Arka and Ganesan, Vidhya",
    title = "{Hint of Dark Matter{\textendash}Dark Energy Interaction in DESI DR2 and Current Cosmological Dataset?}",
    eprint = "2403.14247",
    archivePrefix = "arXiv",
    primaryClass = "astro-ph.CO",
    doi = "10.3847/1538-4357/ae2ff8",
    journal = "Astrophys. J.",
    volume = "998",
    number = "1",
    pages = "83",
    year = "2026"
}

@article{Carron:2022eyg,
    author = "Carron, Julien and Mirmelstein, Mark and Lewis, Antony",
    title = "{CMB lensing from Planck PR4~maps}",
    eprint = "2206.07773",
    archivePrefix = "arXiv",
    primaryClass = "astro-ph.CO",
    doi = "10.1088/1475-7516/2022/09/039",
    journal = "JCAP",
    volume = "09",
    pages = "039",
    year = "2022"
}

@article{Feleppa:2025vop,
    author = "Feleppa, Fabiano and de Graaf, Welmoed Marit and Brax, Philippe and Lambiase, Gaetano",
    title = "{Bounds on screened dark energy from near-Earth space-based measurements}",
    eprint = "2511.08448",
    archivePrefix = "arXiv",
    primaryClass = "gr-qc",
    doi = "10.1103/gss2-qpp1",
    journal = "Phys. Rev. Lett.",
    volume = "136",
    number = "10",
    pages = "101002",
    year = "2026"
}

@article{Sola:2005et,
    author = "Sol\`a, Joan and Stefancic, Hrvoje",
    title = "{Effective equation of state for dark energy: Mimicking quintessence and phantom energy through a variable lambda}",
    eprint = "astro-ph/0505133",
    archivePrefix = "arXiv",
    reportNumber = "UB-ECM-PF-05-11",
    doi = "10.1016/j.physletb.2005.08.051",
    journal = "Phys. Lett. B",
    volume = "624",
    pages = "147--157",
    year = "2005"
}

@article{Akaike,
  author={Akaike, H.},
  journal={IEEE Transactions on Automatic Control}, 
  title={A new look at the statistical model identification}, 
  year={1974},
  volume={19},
  number={6},
  pages={716-723},
  doi={10.1109/TAC.1974.1100705}}

@article{Sanchez:2020vvb,
    author = "S\'anchez, Ariel G.",
    title = "{Arguments against using $h^{-1}{\rm Mpc}$ units in observational cosmology}",
    eprint = "2002.07829",
    archivePrefix = "arXiv",
    primaryClass = "astro-ph.CO",
    doi = "10.1103/PhysRevD.102.123511",
    journal = "Phys. Rev. D",
    volume = "102",
    number = "12",
    pages = "123511",
    year = "2020"
}

@article{Forconi:2025cwp,
    author = "Forconi, Matteo and Favale, Arianna and G{\'o}mez-Valent, Adri{\`a}",
    title = "{Illustrating the consequences of a misuse of {\ensuremath{\sigma}}8 in cosmology}",
    eprint = "2501.11571",
    archivePrefix = "arXiv",
    primaryClass = "astro-ph.CO",
    doi = "10.1103/rpf5-ldks",
    journal = "Phys. Rev. D",
    volume = "112",
    number = "2",
    pages = "023517",
    year = "2025"
}

@article{Rosenberg:2022sdy,
    author = "Rosenberg, Erik and Gratton, Steven and Efstathiou, George",
    title = "{CMB power spectra and cosmological parameters from Planck PR4 with CamSpec}",
    eprint = "2205.10869",
    archivePrefix = "arXiv",
    primaryClass = "astro-ph.CO",
    doi = "10.1093/mnras/stac2744",
    journal = "Mon. Not. Roy. Astron. Soc.",
    volume = "517",
    number = "3",
    pages = "4620--4636",
    year = "2022"
}

@article{Efstathiou:2019mdh,
    author = "Efstathiou, George and Gratton, Steven",
    title = "{A Detailed Description of the CamSpec Likelihood Pipeline and a Reanalysis of the Planck High Frequency Maps}",
    eprint = "1910.00483",
    archivePrefix = "arXiv",
    primaryClass = "astro-ph.CO",
    doi = "10.21105/astro.1910.00483",
    month = "10",
    year = "2019", 
    Note = "arXiv:1910.00483"
}

@article{DES:2024jxu,
    author = "Abbott, T. M. C. and others",
    collaboration = "DES",
    title = "{The Dark Energy Survey: Cosmology Results with {\ensuremath{\sim}}1500 New High-redshift Type Ia Supernovae Using the Full 5 yr Data Set}",
    eprint = "2401.02929",
    archivePrefix = "arXiv",
    primaryClass = "astro-ph.CO",
    reportNumber = "FERMILAB-PUB-23-0821-PPD, DES-2023-805",
    doi = "10.3847/2041-8213/ad6f9f",
    journal = "Astrophys. J. Lett.",
    volume = "973",
    number = "1",
    pages = "L14",
    year = "2024"
}

@article{DES:2025sig,
    author = "Popovic, B. and others",
    collaboration = "DES",
    title = "{The Dark Energy Survey Supernova Program: A Reanalysis Of Cosmology Results And Evidence For Evolving Dark Energy With An Updated Type Ia Supernova Calibration}",
    eprint = "2511.07517",
    archivePrefix = "arXiv",
    primaryClass = "astro-ph.CO",
    reportNumber = "FERMILAB-PUB-25-0842-CSAID-PPD",
    doi = "10.1093/mnras/stag632",
    journal = "Mon. Not. Roy. Astron. Soc.",
    volume = "548",
    pages = "stag632",
    year = "2026"
}

@article{Gomez-Valent:2022bku,
    author = "G{\'o}mez-Valent, Adri{\`a} and Zheng, Ziyang and Amendola, Luca and Wetterich, Christof and Pettorino, Valeria",
    title = "{Coupled and uncoupled early dark energy, massive neutrinos, and the cosmological tensions}",
    eprint = "2207.14487",
    archivePrefix = "arXiv",
    primaryClass = "astro-ph.CO",
    doi = "10.1103/PhysRevD.106.103522",
    journal = "Phys. Rev. D",
    volume = "106",
    number = "10",
    pages = "103522",
    year = "2022"
}

@article{Gonzalez-Fuentes:2025lei,
    author = "Gonz{\'a}lez-Fuentes, Alex and G{\'o}mez-Valent, Adri{\`a}",
    title = "{Reconstruction of dark energy and late-time cosmic expansion using the Weighted Function Regression method}",
    eprint = "2506.11758",
    archivePrefix = "arXiv",
    primaryClass = "astro-ph.CO",
    doi = "10.1088/1475-7516/2025/12/049",
    journal = "JCAP",
    volume = "12",
    pages = "049",
    year = "2025"
}

@article{Hergt:2026moc,
    author = "Hergt, Lukas Tobias and Henrot-Versill{\'e}, Sophie and Tristram, Matthieu and Scott, Douglas",
    title = "{Consistency of standard cosmologies using Bayesian model comparison and tension quantification}",
    eprint = "2602.06115",
    archivePrefix = "arXiv",
    primaryClass = "astro-ph.CO",
    month = "2",
    year = "2026",
    Note = "arXiv:2602.06115"
}

@article{Jhaveri:2026bla,
    author = "Jhaveri, Tanisha and Karwal, Tanvi and Crawford, Thomas and Hu, Wayne and Khalife, Ali Rida and Balkenhol, Lennart and Ge, Fei",
    title = "{Disentangling cosmic distance tensions with early and late dark energy}",
    eprint = "2604.08530",
    archivePrefix = "arXiv",
    primaryClass = "astro-ph.CO",
    month = "4",
    year = "2026", 
    Note = "arXiv:2604.08530"
}

@article{Semenaite:2025ohg,
    author = "Semenaite, A. and others",
    title = "{Joint cosmological fits to DESI-DR1 full-shape clustering and weak gravitational lensing in configuration space}",
    eprint = "2512.15961",
    archivePrefix = "arXiv",
    primaryClass = "astro-ph.CO",
    reportNumber = "FERMILAB-PUB-25-0957-PPD",
    doi = "10.5281/zenodo.17934989",
    month = "12",
    year = "2025", 
    Note = "arXiv:2512.15961"
}

@article{Poulin:2025nfb,
    author = "Poulin, Vivian and Smith, Tristan L. and Calder{\'o}n, Rodrigo and Simon, Th{\'e}o",
    title = "{Impact of ACT DR6 and DESI DR2 for early dark energy and the Hubble tension}",
    eprint = "2505.08051",
    archivePrefix = "arXiv",
    primaryClass = "astro-ph.CO",
    doi = "10.1103/bx25-1g5d",
    journal = "Phys. Rev. D",
    volume = "113",
    number = "6",
    pages = "063519",
    year = "2026"
}

@article{Adi:2025hyj,
    author = "Adi, Tal",
    title = "{Lowering the horizon on Dark Energy: A late-time response to early solutions for the Hubble tension}",
    eprint = "2509.12331",
    archivePrefix = "arXiv",
    primaryClass = "astro-ph.CO",
    doi = "10.1088/1475-7516/2026/03/015",
    journal = "JCAP",
    volume = "03",
    pages = "015",
    year = "2026"
}

@article{Chaussidon:2025npr,
    author = "Chaussidon, E. and others",
    title = "{Early time solution as an alternative to the late time evolving dark energy with DESI DR2 BAO}",
    eprint = "2503.24343",
    archivePrefix = "arXiv",
    primaryClass = "astro-ph.CO",
    reportNumber = "FERMILAB-PUB-25-0241-PPD",
    doi = "10.1103/xtql-wh3h",
    journal = "Phys. Rev. D",
    volume = "112",
    number = "6",
    pages = "063548",
    year = "2025"
}

@article{Sharma:2025iux,
    author = "Sharma, Ravi Kumar and Lesgourgues, Julien",
    title = "{Constraints on neutrino mass and dark energy agnostic to the sound horizon}",
    eprint = "2510.15835",
    archivePrefix = "arXiv",
    primaryClass = "astro-ph.CO",
    reportNumber = "TTK-25-31",
    doi = "10.1088/1475-7516/2026/02/034",
    journal = "JCAP",
    volume = "02",
    pages = "034",
    year = "2026"
}

@article{Poulin:2024ken,
    author = "Poulin, Vivian and Smith, Tristan L. and Calder{\'o}n, Rodrigo and Simon, Th{\'e}o",
    title = "{Implications of the cosmic calibration tension beyond H0 and the synergy between early- and late-time new physics}",
    eprint = "2407.18292",
    archivePrefix = "arXiv",
    primaryClass = "astro-ph.CO",
    doi = "10.1103/PhysRevD.111.083552",
    journal = "Phys. Rev. D",
    volume = "111",
    number = "8",
    pages = "083552",
    year = "2025"
}

@article{SolaPeracaula:2022hpd,
    author = "Sol\`a Peracaula, Joan",
    title = "{The cosmological constant problem and running vacuum in the expanding universe}",
    eprint = "2203.13757",
    archivePrefix = "arXiv",
    primaryClass = "gr-qc",
    doi = "10.1098/rsta.2021.0182",
    journal = "Phil. Trans. Roy. Soc. Lond. A",
    volume = "380",
    pages = "20210182",
    year = "2022"
}

@article{Ong:2025utx,
    author = "Ong, Dily Duan Yi and Yallup, David and Handley, Will",
    title = "{A Bayesian Perspective on Evidence for Evolving Dark Energy}",
    eprint = "2511.10631",
    archivePrefix = "arXiv",
    primaryClass = "astro-ph.CO",
    month = "11",
    year = "2025",
    Note = "arXiv:2511.10631"
}

@article{Patel:2024odo,
    author = "Patel, Vrund and Chakraborty, Amlan and Amendola, Luca",
    title = "{The prior dependence of the DESI results}",
    eprint = "2407.06586",
    archivePrefix = "arXiv",
    primaryClass = "astro-ph.CO",
    month = "7",
    year = "2024",
    Note = "arXiv:2407.06586"
}

@article{Das:2005yj,
    author = "Das, Subinoy and Corasaniti, Pier Stefano and Khoury, Justin",
    title = "{Super-acceleration as signature of dark sector interaction}",
    eprint = "astro-ph/0510628",
    archivePrefix = "arXiv",
    doi = "10.1103/PhysRevD.73.083509",
    journal = "Phys. Rev. D",
    volume = "73",
    pages = "083509",
    year = "2006"
}

@article{DES:2024hip,
    collaboration = "DES",
    title = "{The Dark Energy Survey Supernova Program: Cosmological Analysis and Systematic Uncertainties}",
    eprint = "2401.02945",
    archivePrefix = "arXiv",
    primaryClass = "astro-ph.CO",
    reportNumber = "FERMILAB-PUB-23-693-PPD",
    doi = "10.3847/1538-4357/ad5e6c",
    journal = "Astrophys. J.",
    volume = "975",
    number = "1",
    pages = "86",
    year = "2024"
}

@article{Scolnic:2021amr,
    author = "Scolnic, Dan and others",
    title = "{The Pantheon+ Analysis: The Full Data Set and Light-curve Release}",
    eprint = "2112.03863",
    archivePrefix = "arXiv",
    primaryClass = "astro-ph.CO",
    doi = "10.3847/1538-4357/ac8b7a",
    journal = "Astrophys. J.",
    volume = "938",
    number = "2",
    pages = "113",
    year = "2022"
}

@article{Park:2024vrw,
    author = "Park, Chan-Gyung and de Cruz P{\'e}rez, Javier and Ratra, Bharat",
    title = "{Using non-DESI data to confirm and strengthen the DESI 2024 spatially flat w0waCDM cosmological parametrization result}",
    eprint = "2405.00502",
    archivePrefix = "arXiv",
    primaryClass = "astro-ph.CO",
    doi = "10.1103/PhysRevD.110.123533",
    journal = "Phys. Rev. D",
    volume = "110",
    number = "12",
    pages = "123533",
    year = "2024"
}

@article{Yang:2025uyv,
    author = "Yang, Weiqiang and Zhang, Sibo and Mena, Olga and Pan, Supriya and Di Valentino, Eleonora",
    title = "{Dark Energy Is Not That Into You: Variable Couplings after DESI DR2 BAO}",
    eprint = "2508.19109",
    archivePrefix = "arXiv",
    primaryClass = "astro-ph.CO",
    month = "8",
    year = "2025",
    Note = "arXiv:2508.19109"
}

@article{Ghedini:2025epp,
    author = "Ghedini, Pietro and Hajjar, Rasmi and Mena, Olga",
    title = "{Dark energy and neutrinos along the cosmic expansion history}",
    eprint = "2512.16781",
    archivePrefix = "arXiv",
    primaryClass = "astro-ph.CO",
    doi = "10.1016/j.dark.2026.102237",
    journal = "Phys. Dark Univ.",
    volume = "52",
    pages = "102237",
    year = "2026"
}

@article{Bottaro:2024pcb,
    author = "Bottaro, Salvatore and Castorina, Emanuele and Costa, Marco and Redigolo, Diego and Salvioni, Ennio",
    title = "{From 100~kpc to 10~Gpc: Dark matter self-interactions before and after DESI observations}",
    eprint = "2407.18252",
    archivePrefix = "arXiv",
    primaryClass = "astro-ph.CO",
    doi = "10.1103/gc78-96l5",
    journal = "Phys. Rev. D",
    volume = "112",
    number = "2",
    pages = "023525",
    year = "2025"
}

@article{Bottaro:2023wkd,
    author = "Bottaro, Salvatore and Castorina, Emanuele and Costa, Marco and Redigolo, Diego and Salvioni, Ennio",
    title = "{Unveiling Dark Forces with Measurements of the Large Scale Structure of the Universe}",
    eprint = "2309.11496",
    archivePrefix = "arXiv",
    primaryClass = "astro-ph.CO",
    doi = "10.1103/PhysRevLett.132.201002",
    journal = "Phys. Rev. Lett.",
    volume = "132",
    number = "20",
    pages = "201002",
    year = "2024"
}

@article{Archidiacono:2022iuu,
    author = "Archidiacono, Maria and Castorina, Emanuele and Redigolo, Diego and Salvioni, Ennio",
    title = "{Unveiling dark fifth forces with linear cosmology}",
    eprint = "2204.08484",
    archivePrefix = "arXiv",
    primaryClass = "astro-ph.CO",
    reportNumber = "CERN-TH-2022-066",
    doi = "10.1088/1475-7516/2022/10/074",
    journal = "JCAP",
    volume = "10",
    pages = "074",
    year = "2022"
}

@article{Costa:2025kwt,
    author = "Costa, Marco and Creque-Sarbinowski, Cyril and Simon, Olivier and Weiner, Zachary J.",
    title = "{Dark forces suppress structure growth}",
    eprint = "2510.00098",
    archivePrefix = "arXiv",
    primaryClass = "astro-ph.CO",
    doi = "10.1088/1475-7516/2026/06/055",
    journal = "JCAP",
    volume = "06",
    pages = "055",
    year = "2026"
}

@article{Wang:2026wrk,
    author = "Wang, Jia-Qi and Cai, Rong-Gen and Guo, Zong-Kuan and Li, Yun-He and Wang, Shao-Jiang and Zhang, Xin",
    title = "{Non-minimally coupled quintessence with sign-switching interaction}",
    eprint = "2604.02204",
    archivePrefix = "arXiv",
    primaryClass = "astro-ph.CO",
    month = "4",
    year = "2026",
    Note = "arXiv:2604.02204"
}

@article{Wang:2025znm,
    author = "Wang, Jia-Qi and Cai, Rong-Gen and Guo, Zong-Kuan and Wang, Shao-Jiang",
    title = "{Resolving the Planck-DESI tension by non-minimally coupled quintessence}",
    eprint = "2508.01759",
    archivePrefix = "arXiv",
    primaryClass = "astro-ph.CO",
    month = "8",
    year = "2025", 
    Note = "arXiv:2508.01759"
}

@article{Park:2025fbl,
    author = "Park, Chan-Gyung and Ratra, Bharat",
    title = "{Updated observational constraints on $\phi$CDM dynamical dark energy cosmological models}",
    eprint = "2509.25812",
    archivePrefix = "arXiv",
    primaryClass = "astro-ph.CO",
    month = "9",
    year = "2025", 
    Note = "arXiv:2509.25812"
}
\end{document}